\documentclass[preprint,3p,times,twocolumn]{elsarticle}

\usepackage{amssymb}
\usepackage{amsmath}
\usepackage{graphicx}
\usepackage[ruled,vlined]{algorithm2e}
\usepackage{array}
\usepackage{booktabs}
\usepackage{hyperref}
\usepackage{enumitem}
\usepackage[table,xcdraw]{xcolor}
\usepackage{listings}

\journal{Additive Manufacturing}

\begin{document}
\begin{frontmatter}

\title{Implicit Toolpath Generation for Functionally Graded Additive Manufacturing via Gradient-Informed Slicing}
\affiliation[1]{
  organization={University of Colorado},
  city={Boulder},
  state={Colorado},
  postcode={80309},
  country={USA}
}

\affiliation[2]{
  organization={Draper Scholars, The Charles Stark Draper Laboratory, Inc},
  city={Cambridge},
  state={Massachusetts},
  postcode={02139},
  country={USA}
}

\affiliation[3]{
  organization={The Charles Stark Draper Laboratory, Inc},
  city={Cambridge},
  state={Massachusetts},
  postcode={02139},
  country={USA}
}

\author[1,2]{Charles Wade\corref{cor1}}
\author[3]{Devon Beck}
\author[1]{Robert MacCurdy \corref{cor2}}
\cortext[cor1]{charles.wade@colorado.edu}
\cortext[cor2]{Corresponding author, maccurdy@colorado.edu}

\begin{abstract}
This paper presents a novel gradient-informed slicing method for functionally graded additive manufacturing (FGM) that overcomes the limitations of conventional toolpath planning approaches, which struggle to produce truly continuous gradients. By integrating multi-material gradients into the toolpath generation process, our method enables the fabrication of FGMs with complex gradients that vary seamlessly in any direction.  We leverage OpenVCAD's implicit representation of geometry and material fields to directly extract iso-contours, enabling accurate, controlled gradient toolpaths. Two novel strategies are introduced to integrate these gradients into the toolpath planning process. The first strategy maintains traditional perimeter, skin, and infill structures subdivided by mixture ratios, with automated 'zippering' to mitigate stress concentrations. The second strategy fills iso-contoured regions densely, printing directly against gradients to eliminate purging and reduce waste. Both strategies accommodate gradually changing printing parameters, such as mixed filament ratios, toolhead switching, and variable nozzle temperatures for foaming materials. This capability allows for controlled variation of composition, density, and other properties within a single build, expanding the design space for functionally graded parts. Experimental results demonstrate the fabrication of high-quality FGMs with complex, multi-axis gradients, highlighting the versatility of our method. We showcase the successful implementation of both strategies on a range of geometries and material combinations, demonstrating the potential of our approach to produce intricate and functional FGMs. Although we demonstrate our methodology with material extrusion, it is applicable to any g-code based system. This work provides a robust, open-source, and automated framework for designing and fabricating advanced FGMs, accelerating research in multi-material additive manufacturing.
\end{abstract}




\end{frontmatter}


\section{Introduction}
The field of additive manufacturing (AM) has rapidly expanded to encompass objects composed of multiple materials, each conferring distinct properties. Conventional material extrusion (MEX) multi-material 3D printing strategies typically rely on discrete transitions between a set of base materials, achieved through hardware configurations like multiple nozzles or filament-switching systems \cite{prusa_mmu3, bambu_ams}. Although these approaches can produce objects with spatial changes in color, stiffness, or strength, their slicing methods result in abrupt material interfaces. This limits the full exploitation of material combinations and prohibits the creation of continuous, functionally graded materials (FGMs). In this work we present the first fully automated, functionally graded materials slicing framework that converts arbitrarily complex three-dimensional material fields directly into printer-ready G-code. By eliminating the need for hand-crafted gradient G-code, which has been a main bottleneck limiting design and application complexity, our method provides a direct translation from volumetric gradient design to practical fabrication on any G-code–based additive manufacturing platform.

FGMs are increasingly attractive for advanced engineering applications, as their composition and associated properties vary smoothly and continuously throughout their volume \cite{li_review_2020}. Continuous gradients can enhance mechanical performance, optimize weight distribution, and introduce novel functionalities while eliminating sharp interfaces that could otherwise lead to stress concentrations. However, achieving high-fidelity FGMs using toolpath-based AM processes, such as MEX, Direct Ink Write (DIW), and Directed Energy Deposition (DED), presents significant challenges compared to Material Jetting methods that support voxel-based design specifications.

Voxel-based AM methods, like Material Jetting, realize FGMs by selectively interdigitating multiple base materials within individual voxels during a single pass of the print head \cite{wade_openvcad_2024}. Voxel material jetting printers typically accept image stacks with a finite color palette as input. Each image represents a z-slice, while each pixel's color is mapped to a discrete material channel within the printer. Given the high-resolution of material jetting systems, two discrete materials can be interdigitated at different ratios to create detailed gradients when viewed at the macro-scale. Since voxel composition usually does not influence the toolpath, no additional planning is required beyond specifying the region to be printed. In contrast, toolpath-based methods face substantial obstacles due to their inherent design for homogeneous materials, traditionally represented by a handful of discrete boundary surfaces with singular material assignments. Existing slicing algorithms and toolpath planning techniques do not accommodate complex, smoothly varying gradients. Despite these limitations, there is substantial interest in extending toolpath-based systems to produce objects with spatially varying properties, encompassing both multi-material composites and single-material structures whose properties vary according to adjustable process parameters.

Process parameters influencing material properties in AM can be categorized into instantaneous and non-instantaneous changes. Instantaneous parameters, such as laser power or print speed, can be modified without significant delay. Conversely, non-instantaneous parameters require stabilization time or print material usage (e.g. dead extrusion volume) due to inherent physical constraints. Examples of non-instantaneous parameters include filament mixing ratios in mixing hotends, filament or toolhead switches, and nozzle temperature adjustments. Table~\ref{tab:process_parameters} summarizes examples of these parameters across different 3D printing modalities.

\begin{figure*}[t]
    \centering
    \includegraphics[width=0.9\linewidth]{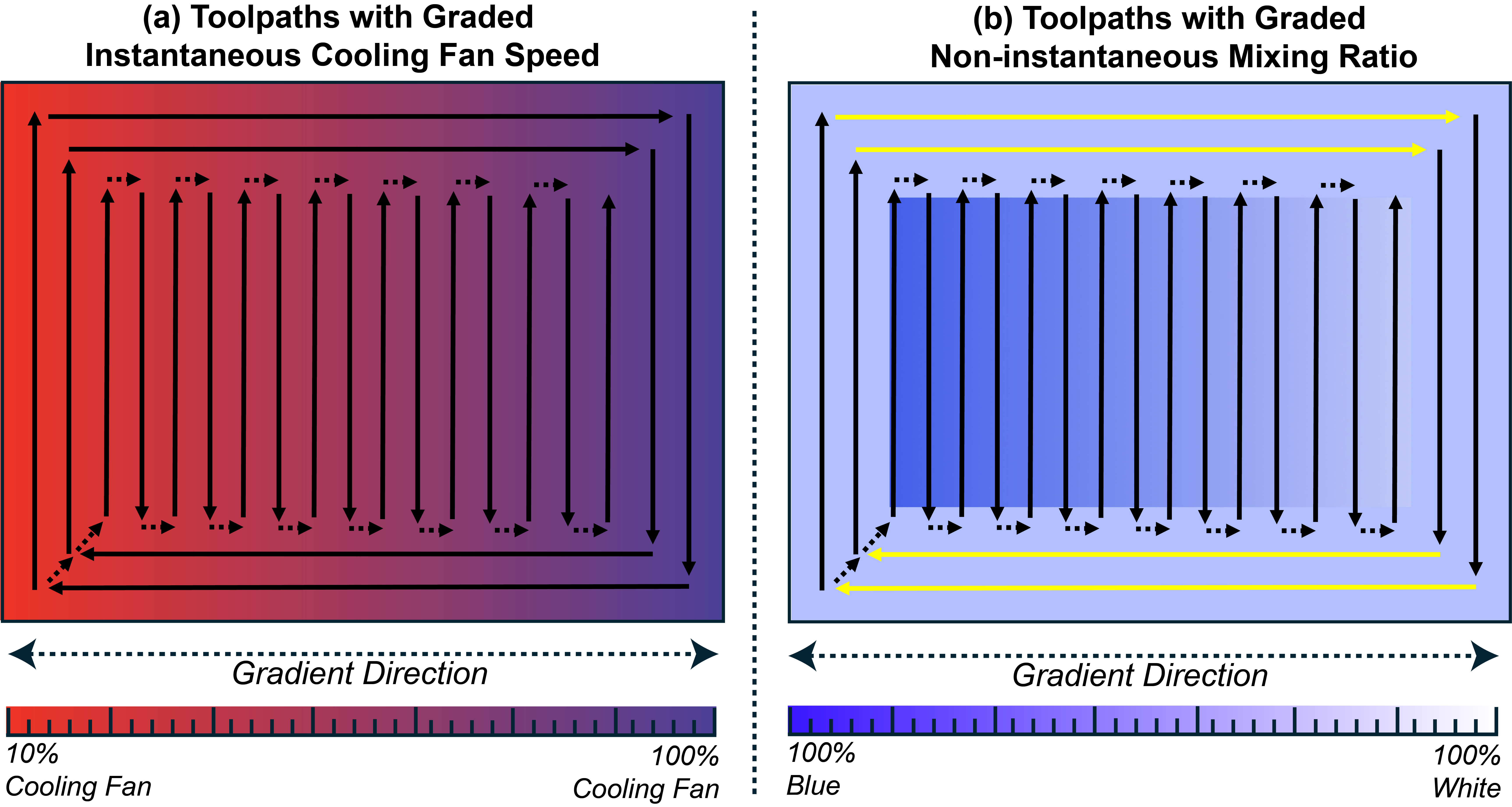}
    \caption{Comparison of instantaneous and non-instantaneous process parameters on a single-layer toolpath for a rectangular object. Solid arrows indicate printing paths, dashed arrows represent non-extrusion travel moves. (a) An instantaneous parameter (cooling fan speed) is linearly graded from 100\% to 50\% along the $x$-axis. Each pre-planned toolpath segment directly samples the intended gradient without any transition issues. (b) A non-instantaneous parameter (two-material mixing ratio) graded linearly from 100\% blue to 100\% white along the $x$-axis. Toolpaths aligned with the gradient (highlighted in yellow) traverse the entire gradient in a single continuous segment, exceeding the mixing capabilities of the printer and resulting in an unintended intermediate color. Conversely, toolpaths oriented perpendicular to the gradient direction have less variation between consecutive segments, allowing the printer sufficient time to stabilize and accurately reproduce the intended gradient.}
    \label{fig:instant_vs_non_instant}
\end{figure*}

Figure \ref{fig:instant_vs_non_instant} illustrates the challenges associated with a naive gradient implementation approach that applies gradients directly to pre-planned G-code. When toolpaths are pre-planned without considering the intended gradient, instantaneous parameters (e.g. cooling fan speed) accurately reflect gradient changes through simple sampling . In contrast, applying the same approach to non-instantaneous parameters (e.g., multi-material mixing ratios) results in unintended intermediate states due to insufficient transition time. This observation underscores the necessity for gradient-informed slicers to generate toolpaths that constrain changes in non-instantaneous parameters to physically achievable transitions \cite{leoni_functionally_2023}.

\begin{table*}[h]
    \centering
    \begin{tabular}{@{}p{0.45\textwidth}p{0.45\textwidth}@{}}
        \toprule
        \textbf{Instantaneous} & \textbf{Non-Instantaneous} \\ 
        \midrule
        Laser Power (DED) \cite{borish_automated_2022, gibson_beyond_2019} & Filament Mixing (MEX) \cite{green_local_2023} \\
        Extrusion Rate (MEX) \cite{geng_effects_2019} & Nozzle Temperature (MEX) \cite{ozdemir_xpandables_2023}\\
        Cooling Fan Speed (MEX) \cite{lee_influence_2019} & Multi-Filament Paths (MEX) \cite{prusa_xl_toolchanger, bambu_ams}\\
        Print Speed (MEX) \cite{loskot_influence_2023, geng_effects_2019} & Material Mixing (DIW) \cite{duncan_lowloss_2023} \\
        Viscous Thread Instability Parameters (MEX) \cite{lipton_3d_2016, emery_foams_2024} & Material Swapping (MEX, VPP) \cite{prusa_mmu3, shaukat_review_2022, shaukat_dual-vat_2024}\\
        Beam Focus (PBF) \cite{yuan_effect_2022} & \\
        \bottomrule
    \end{tabular}
    \caption{Examples of instantaneous and non-instantaneous 3D printing process parameters. Instantaneous parameters can change immediately during printing without significant delay, whereas non-instantaneous parameters require a finite stabilization time or material flow before fully transitioning to a new state. Processes listed include Directed Energy Deposition (DED), Direct Ink Write (DIW), Powder Bed Fusion (PBF), Material Extrusion (MEX), and Vat Photopolymerization (VPP).}
    \label{tab:process_parameters}
\end{table*}

To achieve functionally graded materials, researchers have primarily explored hardware-based solutions, such as mixing hotends or multiple deposition heads, to physically implement FGMs from several base materials \cite{green_hotend_nodate}. Other approaches have leveraged variations in process parameters on single-material systems, such as modulating extrusion temperatures with foaming filaments, to achieve graded material behaviors \cite{ozdemir_xpandables_2023}. Nonetheless, these advances largely rely on manual design workflows, approximate mesh representations, or ad-hoc adjustments to generated toolpaths. A comprehensive solution demands an integrated design and slicing strategy, wherein geometry and material distributions are precisely defined during the design stage and accurately translated into toolpaths capable of faithfully reproducing intended gradients.

In this paper, we introduce a novel slicing method explicitly designed for the fabrication of FGMs using non-instantaneous process parameters. We validate our method across three distinct MEX systems: (1) a two-material mixing extruder, (2) a five-head extruder system with a tool-changer mechanism, and (3) a single-material system using temperature-responsive foaming filaments. Our method builds upon OpenVCAD \cite{wade_openvcad_2024}, an implicit design framework capable of representing both geometry and material distributions as continuous spatial fields rather than discrete surfaces.  Leveraging this implicit representation, we propose two complementary strategies for toolpath generation. Our first strategy preserves traditional slicing paradigms, like walls, skins, and infills, and subdivides them into discrete process parameter state regions, intelligently ordering these regions and adding purge towers where necessary to accommodate gradual composition changes. Our second strategy forgoes conventional toolpath structures, to generate toolpaths that “print against the gradient” and densely cover regions defined by material iso-contours, thereby achieving smoother transitions without purge towers. Although we demonstrate our method with MEX, it is generalizable and can be applied to other process parameters on any toolpath-based system, such as direct ink write or directed energy deposition. Our method also replaces workflows that require users to manually specify G-code on a per-design basis. 

\begin{figure*}[h!]
    \centering
    \includegraphics[width=\linewidth]{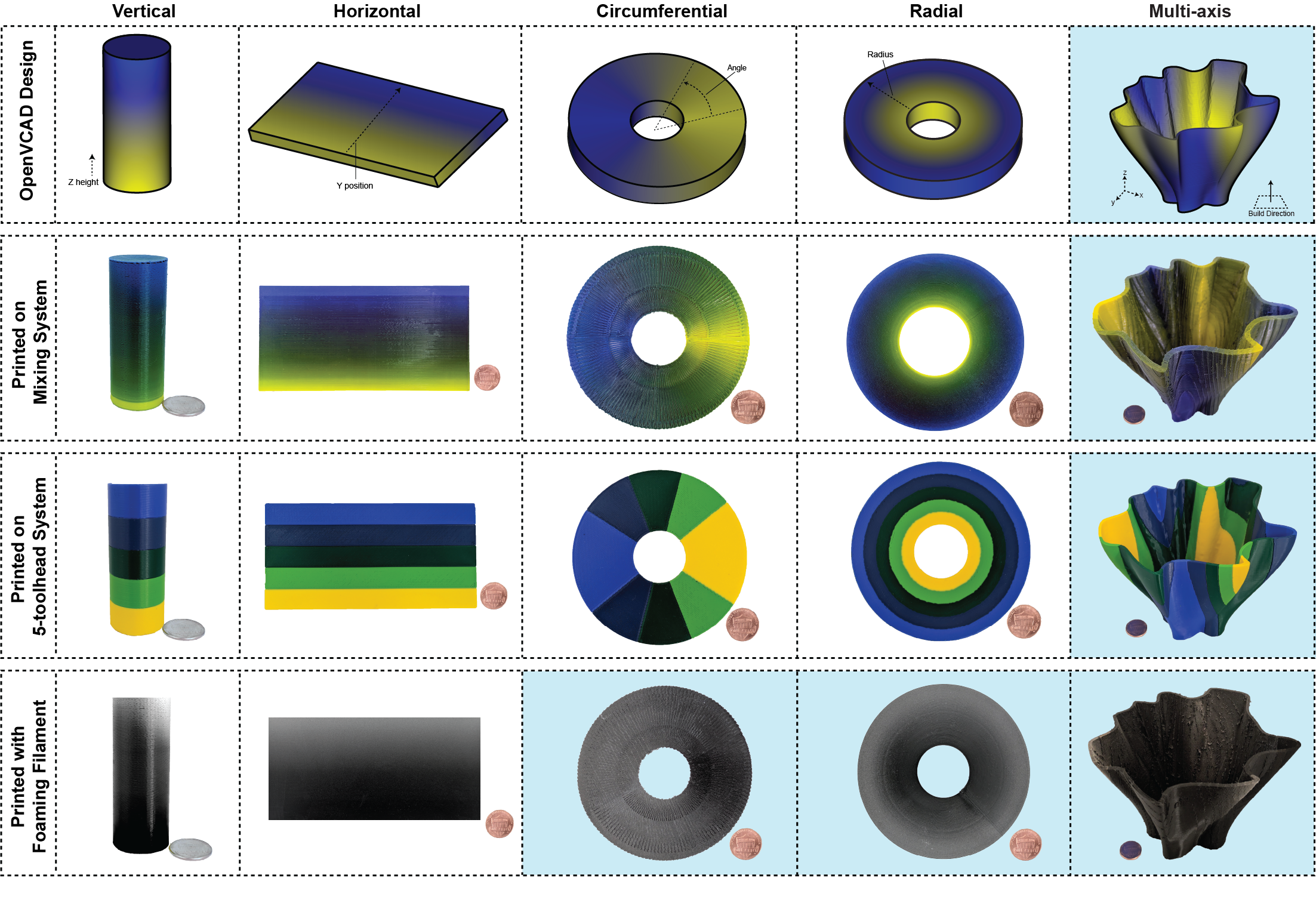}
    \caption{Comparison of gradient complexity achievable by different slicing methods across three distinct MEX systems: two-material mixing, multi-head tool-changer, and temperature-responsive foaming filament. OpenVCAD design scripts are given along with a render of each object. White boxes indicate gradient capabilities previously demonstrated by Green et al., Garland \& Fadel, and Ozdemir \& Doubrovski~\cite{green_local_2023, garland_design_2015, ozdemir_xpandables_2023}, while light blue boxes represent novel gradient configurations made possible through our proposed method. Our approach generalizes to enable slicing of both complex 3D gradients and simpler single-axis gradients across all three systems, and introduces the first demonstration of radial and circumferential gradients with foaming filaments.}
    \label{fig:big_comparison}
\end{figure*}

Through experiments, printed examples, and analysis of complex geometries and gradients, we show that our method improves gradient fidelity, works across multiple printing modalities, and provides an automated method for joining adjacent material regions by “zippering” toolpaths at interfaces. From figure \ref{fig:big_comparison}, we see how our method enables systems to print gradients regardless of their spatial complexity. The results illustrate that automated, gradient-driven slicing can produce intricate, functionally graded parts while minimizing the manual overhead traditionally associated with functionally graded 3D printing.

The remainder of this paper is organized as follows: Section 2 reviews related work in multi-material AM, gradient-based strategies, and toolpath planning. Section 3 details our methods, including the use of OpenVCAD, iso-contouring, and the two proposed toolpath strategies for FGMs. Section 4 presents the results, including test prints, gradient analysis, and limitations. Finally, Section 5 concludes with a summary of contributions and potential directions for future research in automated gradient toolpath planning.

\section{Related Work}
In this section, we summarize recent advancements in hardware and planning methods for functionally grading process parameters across additive manufacturing domains. We begin by contrasting homogeneous and heterogeneous multi-material 3D printing, focusing on the respective strategies for constructing objects with multiple materials. Next, we examine work on mixing systems, including fused filament fabrication (FFF) and Direct Ink Write technologies, to achieve functionally graded designs. Beyond mixing printers, we consider other process parameters that can mimic multi-material behavior, such as printing with temperature-responsive filaments containing foaming agents to alter mechanical properties. Finally, we highlight the need for novel toolpath planning strategies for functionally grading process parameters, including methods for local composition control and the mechanical implications of heterogeneous printing strategies.

\subsection{Homogeneous Multi-material 3D Printing}
Homogeneous design refers to the use of individual base materials to construct multi-material objects \cite{fayolle_modeling_2021}. This approach is a cornerstone of multi-material 3D printing in consumer and industrial contexts. One of the most common implementations uses multiple extrusion nozzles, where each nozzle deposits a specific material in turn, layer by layer \cite{khan_review_2024, e3d_toolchanger_review, prusa_xl_toolchanger, abilgaziyev_design_2015, ali_multi-nozzle_2016}. Another strategy employs filament multiplexing, where a single extruder dynamically switches between materials during a layer \cite{bambu_ams, prusa_mmu3, mosaic_palette_3}. While effective for creating objects with a limited number of base materials, these techniques fall short in capturing gradients.

Recent advancements have introduced techniques to produce visual gradients in 3D-printed parts, primarily focusing on aesthetic enhancements rather than functional material properties. Littler et al. propose an automated filament inking system for multi-color FFF 3D printing \cite{littler_automated_2022}. Their approach allows for low-cost, customizable multi-color printing by pre-processing filament with ink rather than switching feedstock or nozzles during printing. A similar commercial printer that uses inks to dye filament is made by \textit{da Vinci} \cite{davinci}. Other methods apply halftoning principles adapted from 2D printing to create the appearance of continuous tones in multi-material FFF. Reiner et al. introduced a dual-color method that creates grayscale tones using sinusoidal contour modulations but is limited by resolution and viewing angles \cite{reiner_dualcolor_2014}. Kuipers et al. extend this with 3D hatching and halftoning techniques to improve grayscale gradients across various orientations, leveraging sagging effects for consistency \cite{kuipers_3d_2017,kuipers_hatching_2018}. Song et al. proposed a method for color gradients using a mixing FFF printer, dividing layers into strata with controlled thickness to achieve smooth transitions \cite{song_colored_2019}. In contrast, our work expands toolpath planning to functionally grade material properties, such as density, stiffness, and thermal responsiveness, moving beyond visual gradients.

Another approach for creating gradients on homogeneous printers focuses on achieving spatially varying mechanical properties. Takahashi et al. introduced programmable filaments for multi-material 3D printing, where distinct material regions are pre-fabricated within the filament itself  \cite{takahashi_programmable_2020}. This is achieved by first printing a coil geometry composed of different materials, which is then used as the input filament for subsequent prints. Their method facilitates controlled material distribution, enabling functional gradients on single-material printers. Building on this, Ahn et al. extend the concept by embedding multiple base materials directly within the filament, enabling properties like mechanical strength and electrical conductivity to be spatially programmed \cite{ahn_3d_2024}. However, both approaches face challenges in achieving smooth transitions between material segments and in accurately translating complex, user-defined gradients into printable filaments. Prefabricating filaments requires precise and sustained alignment between the filament feedstock and the G-code instructions, leaving little room for alignment errors between the embedded gradients and the intended design. This alignment becomes particularly problematic during extruder purging and material swapping processes, where precise synchronization of filament gradients with the print geometry is difficult to achieve reliably.

\subsection{Mechanical Performance of Multi-material Prints}
A key challenge in multi-material 3D printing is the bonding strength between materials, particularly when adjacent regions are printed at different times within a layer. Extensive investigations have explored post-processing techniques to enhance the adhesion of multi-material MEX parts \cite{tiwary_overview_2021}. Dairabayeva et al. examines seven interface designs, such as overlap, encapsulation, and mechanical interlocking, to assess their impact on tensile strength, strain at break, Young’s modulus, and yield strength \cite{dairabayeva_investigation_2023}. Their results show that mechanical locking mechanisms significantly improved adhesion and structural integrity, especially in multi-material specimens combining Tough PLA and TPU. This highlights the importance of interface design in optimizing the mechanical performance of multi-material parts.

Sardinha et al. investigate the effects of seam overlap and ironing on the mechanical performance of parts fabricated with Multi-Extrusion Filament Fabrication. Using PLA tensile specimens, the study demonstrates that increasing seam overlap improves mechanical strength through enhanced polymer diffusion across adjacent layers \cite{sardinha2024multi_extrusion}. While ironing marginally improves seam strength, it was less effective than overlap and carries risks of thermal degradation. This work highlights the importance of seam placement and overlap optimization for improving the mechanical properties of large-format parts. Although seam overlapping improves strength, this method is underutilized because toolpath generation is currently a manual process. In our work, we leverage the findings of Sardinha et al. and address the gap of manual toolpath generation by proposing an algorithm for automatically determining optimal overlap regions.

\subsection{Heterogeneous Multi-material 3D Printing}
Heterogeneous design represents a significant departure from homogeneous approaches by focusing on real-time material mixing during the printing process to create new composite materials. Unlike homogeneous designs, which combine distinct base materials into discrete regions or employ visual tricks like halftoning, heterogeneous designs physically blend materials to achieve functionally graded properties. This distinction enables heterogeneous designs to achieve more complex gradients, such as compositional, mechanical, or thermal variations, which are otherwise unattainable.

In our work, we focus on heterogeneous design achieved through mixing systems and temperature-responsive materials. Mixing systems, can be categorized into two primary modalities: Fused Filament Fabrication and Direct Ink Write. As we discuss in the following sections, both modalities face similar limitations regarding toolpath planning and these challenges underscore the need for novel strategies to fully realize the potential of heterogeneous design.

\subsubsection{Fused Filament Fabrication Mixing Systems}
Mixing hotend 3D printers are designed to enable multi-material printing by combining different filament inputs within a single nozzle. These systems are either active mixing or non-active mixing. Non-active mixing systems combine filaments without mechanical agitation, relying on the natural flow of the materials through predefined channels in the hotend. In these systems, transitions between materials often involves purging the entire hotend to clear out the previous mixing ratio before stabilizing a new mixture. This process can result in material waste and visible transitions in the print where one mixture ratio ends and the next begins. Active mixing hotends, on the other hand, use mechanical components such as augers or mixing impellers to blend the filaments inside the nozzle chamber in real-time. These systems enable finer control over the mixture ratios and allow for smoother transitions between materials. Active mixing improves on some of the challenges faced by non-active systems, such as long purge times and abrupt changes in material composition. However, both systems face limitations in how quickly they can adjust the mixture ratios, which must be considered in the toolpath planning process to ensure the highest print quality.

Han et al. optimized nozzle geometry to improve color mixing in multi-filament FFF systems  \cite{han_design_2017}. Their work, however, focuses solely on aesthetic applications and highlights the difficulty of achieving consistent mixing due to laminar flow constraints. Similarly, Kennedy and Christ introduce active mixing hardware to blend polymer filaments, demonstrating improved control over material properties \cite{kennedy_printing_2020}. While effective for predefined polymer blends, they do not present a method for continuous gradients. These works underscore the need for automated and scalable planning methods to support complex, multi-material graded designs.

Adapa and Jagadish focus on the design and fabrication of an internal mixer and filament extruder for producing hybrid composite filaments for FFF, addressing challenges in material blending and filament quality \cite{adapa_design_2023}. Similarly, Green et al. present a method for local composition control in FFF using an active mixing hotend \cite{green_local_2023, green_hotend_nodate}. They demonstrate the ability to control mixture ratios dynamically to produce complex gradients both within and across layers. Their work highlights how graded structures can be realized with various orientations, such as horizontal, vertical, radial, and circumferential gradients. Columns one through four in Figure \ref{fig:big_comparison}, which use a mixing system, were adapted from Green et al. \cite{green_local_2023}. Moreover, they discuss the limitations of instantaneous composition changes due to the non-zero transition time required to purge material from the mixing chamber. Green et al. focus primarily on hardware implementation and demonstrate producing gradients along a single axis. Teng et al. introduce a single-nozzle multi-filament system with active mixing designed for multi-material additive manufacturing \cite{teng_single-nozzle_2024}. Their work achieves high fidelity in both material composition and spatial distribution using an embedded auger system. However, their work focuses on hardware advancements and does not provide a comprehensive slicing strategy for mixing systems. Both Green et al. and Teng et al. report that instantaneous composition changes are not feasible due to the time required for materials to purge from the mixing chamber.  These hardware advances create a new challenge in planning methods to account for delayed material transitions.

Leoni et al. \cite{leoni_functionally_2023} propose a toolpath planning method for functionally graded additive manufacturing that bridges the gap between design and material extrusion. Their method involves the manipulation of G-code based on a volumetric model that describes material distributions within an object. While Leoni et al. effectively automate the material distribution process for multi-material parts, they only achieve linear gradients along the z-axis. This constraint simplifies the printing process but limits the complexity of the gradients that can be realized.  In contrast, our proposed method can plan toolpaths for complex gradients that vary spatially in three dimensions simultaneously based on arbitrary user-specified functions.  Leoni et al.  also acknowledge challenges related to material transition delays caused by the volume of the mixing chamber, echoing the limitations discussed in other works. Similarly, they discuss the necessity of anticipating the extrusion amount to mitigate the delay between a commanded mixture ratio change and the actual composition leaving the nozzle caused by the volume of the melting chamber. We address these challenges by introducing a method to accommodate the delayed material transition in mixing hotend systems.

Garland and Fadel propose a method for designing and manufacturing functionally graded material objects using a standard FFF 3D printer \cite{garland_design_2015}. They use a level-set function to model material gradients, which is discretized into boundary surfaces for slicing. However, this approach faces significant challenges, including the loss of smoothness and fidelity during the conversion to triangulated meshes. As the number of extruders increases for finer gradients, aligning and assigning materials to numerous discrete meshes becomes impractical in slicing software. Furthermore, they do not demonstrate their method with complex multi-axis gradients. While their method pre-processes a single object into multiple meshes for standard slicing, our proposed method incorporates gradient information directly into a modified slicing process that operates over a single object.

\subsubsection{Direct Ink Write Mixing Systems}
Direct Ink Write (DIW) mixing systems offer a versatile approach to functionally graded material fabrication by enabling real-time mixing of inks with distinct compositions. These systems leverage active mixing mechanisms to achieve precise control over material properties, such as mechanical stiffness, thermal or electrical conductivity, or reactivity.

Recent advancements in hardware systems for DIW explore various methods for mixing composite materials. Ober et al. established foundational principles for active mixing in complex fluids, emphasizing the role of microfluidic control to achieve homogeneous outputs even with disparate inks \cite{ober_active_2015}. Building on this, Serex et al. demonstrated how microfluidics could enhance control in DIW systems, enabling seamless transitions between materials with contrasting viscosities and properties  \cite{serex_microfluidics_2018}. Expanding on these capabilities, Golobic et al. applied active mixing to reactive inks, showcasing how real-time blending could deposit thermite-based materials \cite{golobic_active_2019}. This work underscored the potential of DIW systems to fabricate functional devices requiring high precision and controlled reactivity. For more general applications, Ortega et al. explored the integration of DIW systems with multi-material deposition strategies, highlighting advancements in flow control to achieve smooth gradients across multiple domains \cite{ortega_active_2019}. Similarly, Liao et al. addressed challenges in mixing high-viscosity inks, particularly for bioengineering applications, where precise material uniformity is critical \cite{liao_active_2024}. Duncan et al. demonstrated the use of DIW systems for fabricating graded dielectrics, achieving low-loss RF components with spatially tailored properties \cite{duncan_lowloss_2023}. Pelz et al. and Li et al. employ a mixing DIW system for ceramics and other structural materials, achieving complex material distributions that enhance mechanical and thermal performance \cite{pelz_multi-material_2021, li_fabricating_2018}. While these works present novel hardware developments and applications they do not directly address the challenges with translating a functionally graded design into instructions for these printers. Instead, these works rely on expert human intuition and intervention in the path planning process to add mixture change commands and optimize the build order. 

While DIW systems offer significant advantages, including precise spatial control and compatibility with a wide range of materials, they face challenges similar to those encountered in mixing FFF systems.  Additionally, the current literature lacks comprehensive approaches to automate toolpath planning for gradient-based designs in DIW systems. Most existing works focus heavily on hardware innovations and material development, leaving toolpath generation reliant on existing slicing approaches, which have limitations described in sections \ref{sec:lcc} and \ref{sec:toolpath_intro}. This gap underscores the need for integrated solutions that seamlessly connect design intent with automated toolpath planning to fully exploit the capabilities of mixing systems.

\subsection{Single Material Multi-function Printing}
Although mixing systems provide an intuitive option for creating functionally graded parts, multi-material-like behavior can be created using other process parameters. The use of foaming filaments in FFF printing has opened the door to creating parts with controllable density, enabling functionally graded materials with varied properties. These filaments, containing chemical foaming agents, can expand at certain temperatures, resulting in a change in density and color. While several studies have explored multi-density regions, the computational design of continuous gradients has yet to be fully addressed. Damanpack et al. demonstrate how printing parameters like temperature and flow rate control the porosity and density of PLA filaments with foaming agents \cite{damanpack_porous_2021}.  They develop a model to predict density and mechanical properties based on temperature. However, the study focused on discrete density regions rather than continuous gradients, leaving room for future computational approaches to create smoother transitions.

Ozdemir and Doubrovski explore the use of a single-filament material capable of varying density based on extrusion temperature \cite{ozdemir_xpandables_2023}. While they highlight the potential for creating multi-property parts and note the aesthetic changes caused by foaming, they do not fabricate continuous gradients. Although they explore the computational design of multiple density regions to create parts with different surface finishes or embedded QR codes, methods to automate seamless property changes throughout a print could advance their findings.

Tammaro et al. and Laureijs both investigate how foaming affects density and mechanical properties. They identify how printing parameters impact foaming but, like earlier work, focus on discrete regions of particular density \cite{laureijs_investigating_nodate, tammaro_microfoamed_2022}. Their results provide insight into the mechanical trade-offs in foamed materials, which could inform future computational models aimed at giving the designer high-level control over part density.

Lalegani Dezaki et al. present an application of foaming filaments in the design and fabrication of soft pneumatic actuators with controllable stiffness \cite{lalegani_dezaki_soft_2023}. The foaming filament’s ability to alter the stiffness of the actuator provides an innovative approach for creating actuators with customizable mechanical properties using a single material. By carefully controlling the printing process, the authors fabricate actuators capable of different bending angles and forces using the same input pressure. This opens the door for further exploration into the computational design of continuous gradients, optimizing these process parameters for even more refined control of multi-density, multi-function actuators.

\subsection{Toolpath Planning for Local Composition Control} \label{sec:lcc}
Current toolpath planning strategies largely follow a common workflow. They begin by taking ``slices'' of the input geometry, which is usually a triangulated mesh, and forming outline polygons. Based on user defined settings, these polygons are offset inward to form walls, filled with hatch or lattice patterns to create skins and infill structures. After this, the toolpaths are optimized for various factors such as minimizing travel (non-extrusion) distance or print time. When the user wishes to produce a print with multi-material regions, they supply multiple input boundary surfaces that each represent discrete homogeneous material regions. When sliced, the resulting cross-sections of these objects are treated as independent printing regions that each receive their own walls, skins, infills, etc. The order is optimized such that each discrete polygon is grouped with other polygons of that same material type. The objects are then printed such that the number of tool or material changes are minimized. However these strategies are insufficient for planning and printing objects that contain functionally graded material regions. In their review of functionally graded additive manufacturing, Loh et al. highlight the need for continual development of slicing and toolpath planning software for functionally graded materials, stating, ``Completely new approaches to slicing, analyzing and preparing FGAM [functionally graded additive manufacturing] fabrication are mandatory. New AM software processes should be able to strategically control the density, directionality and allocation of material substances in a logical distribution throughout the generation of the FGAM model.'' \cite{loh_overview_2018} Our paper will detail one such method that considers the functional gradient while performing the toolpath planning in an automated fashion.

Another branch of related work leverages fields, such as stress or strain distributions, to inform slicing and toolpath planning, with the goal of enhancing structural performance. Steuben et al. introduce an implicit slicing method to generate toolpaths aligned with functional fields such as stress distributions \cite{steuben_implicit_2016}. Their approach integrates design intent directly into toolpath planning, enabling optimized toolpath placement for enhanced mechanical performance. While this method is effective for single-material systems, adapting it to multi-material or functionally graded applications remains an open challenge. Xia et al. and Xiong et al. introduce methods to align deposition paths with principal stress directions derived from Finite Element Analysis \cite{xia_stress-based_2020, xiong_process_2019}. These approaches adaptively adjust toolpath spacing and orientation to reduce material usage while maintaining or enhancing part strength. Chen et al. expand on this by converting stress tensors into scalar fields, enabling automated generation of stress-aligned toolpaths using level sets \cite{chen_field-based_2022}. Liu et al.  further refine this concept for continuous fiber-reinforced polymers, demonstrating significant improvements in stiffness and strength by aligning infill patterns with stress trajectories \cite{liu_stress_2024}. These works collectively show that incorporating stress fields into toolpath planning can substantially improve mechanical performance, particularly for load-bearing components.  Sales et al. extend the application of stress fields by integrating principal stress lines from topology optimization into additive manufacturing workflows \cite{sales_function-aware_2021}. By guiding non-uniform infill patterns along stress lines, they achieve enhanced material efficiency and mechanical robustness. Despite their success in improving part strength, these methods focus exclusively on optimizing toolpaths for single-material systems and are limited to specific applications, such as structural reinforcement. While stress-informed toolpath planning introduces innovative strategies to enhance mechanical properties, it does not address the unique challenges posed by functionally graded materials. Thus, while this body of work lays a foundation for integrating fields into toolpath planning, further work is required to extend these techniques to multi-material or functionally graded contexts.

Although significant progress has been made in leveraging fields to inform toolpath planning, the focus of these works remains on optimizing toolpath directions to enhance part strength or enable continuous fiber printing. These methods do not address the unique challenges of functionally graded 3D printing, where material properties vary continuously across a part. Our proposed method builds on these foundational strategies to meet the demands of functionally graded multi-material systems.

\section{Methods}
\subsection{Toolpath Generation for 3D Printing} \label{sec:toolpath_intro}
While effective for single-material prints, traditional toolpath planning fails to address the complexities introduced by functionally graded prints. In particular, traditional toolpath planning does not support mixing hotends and temperature-responsive materials. In mixing hotends, transitioning between material compositions is not instantaneous; there is a delay as the materials inside the nozzle mix and stabilize, which results in print defects if not managed correctly. Similarly, with temperature-responsive materials such as foaming PLA and TPU, there are limitations in how quickly the nozzle temperature can be adjusted, making it difficult to control material properties in real-time. These inherent delays necessitate careful planning to ensure smooth transitions and to avoid abrupt changes in material properties, which could compromise both mechanical performance and visual aesthetics.

To address these challenges, specialized toolpath strategies are needed for FGM prints that actively consider the gradient of material properties during the planning phase. Rather than focusing purely on geometric accuracy, these strategies must account for how material properties change spatially within the object and generate toolpaths that minimize sharp transitions in material composition or temperature. By ordering the toolpaths to optimize over the rate of change in material properties, it is possible to reduce the need for purge cycles and allow for smoother transitions, enabling more complex multi-material prints that better capture the desired functional gradients and achieve specific mechanical or thermal properties in the final object.

\subsection{Preliminaries}
 These preliminaries establish the theoretical foundation and practical tools necessary to enable functionally graded multi-material designs and their integration into advanced additive manufacturing workflows. A thorough understanding of the concepts introduced in this section is essential for the subsequent development and analysis of the proposed methods.  OpenVCAD's representation framework and the principles of iso-contouring underpin the novel slicing and material assignment strategies that we introduce in this work. We assume readers have a foundational understanding of additive manufacturing slicing architectures, including cross-section generation, typical toolpath categories (walls, skins, infills), travel move optimization, and basic G-code instructions. Readers seeking an in-depth background of slicing methodologies are encouraged to consult the textbook "Motion and Path Planning for Additive Manufacturing" by Roschli et al.~\cite{roschli_motion_2023}.

\subsubsection{OpenVCAD}
OpenVCAD is a method that enables the design and expression of complex, multi-material, functionally graded designs in a compact way \cite{wade_openvcad_2024}. Originally developed for inkjet 3D printing systems, OpenVCAD was created to address the limitations of traditional design tools, which typically rely on surface-based representations like STL or 3MF files as the interface to the 3D printing workflow. These traditional methods are poorly suited for multi-material fabrication because they focus on boundary surfaces, making it difficult to define material distributions throughout the volume of an object.

In contrast, OpenVCAD offers a higher-level design framework that uses implicit representations to define both geometry and material distributions. At any given XYZ location, OpenVCAD provides two essential accessors: the signed distance to the surface, which defines the geometry, and a vector of volume fractions, which defines the material composition. This implicit approach allows for infinite resolution scaling and affine invariance, enabling the user to define complex geometries and material gradients without losing precision. Currently, OpenVCAD is only applicable to 3D printing systems that can accept a voxel-based volumetric input. However, the work presented here extends the use of OpenVCAD by applying its implicit representations to inform toolpath-based systems, guiding both the geometry and material gradient during slicing for multi-material 3D printing.

\subsubsection{Iso-contouring}
Building on the implicit representation provided by OpenVCAD, we leverage iso-contouring algorithms to sample both the geometry and material distribution of a multi-material design. The core idea is to extract polylines that represent specific ``levels'' or contours within the design. These contours provide critical information about both the geometry and the material gradient, which will later be used to inform the toolpath planning. Here, we outline common methods and our modifications to suit our needs in toolpath planning.

\begin{figure*}[h!]
    \centering
    \includegraphics[width=\linewidth]{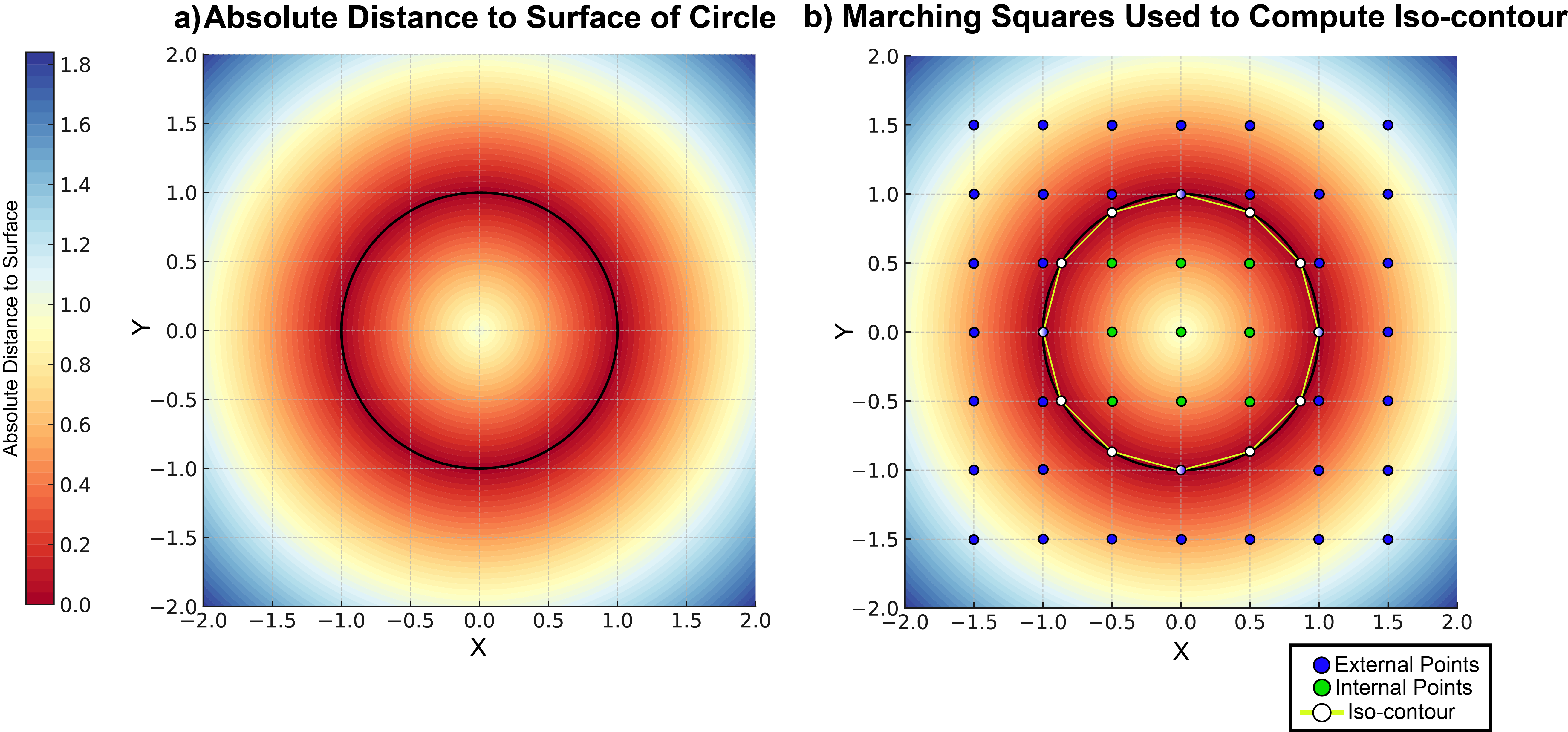}
    \caption{An overview of the Marching Squares algorithm. (a) The OpenVCAD design is sampled and visualized as an absolute value of the signed distance field. The implicit surface is the black line. (b) points are sampled uniformly across space and classified as inside, outside, or on the surface. Using the point classification, a lookup table, and linear interpolation the exact intersection segments are determined. The resulting line segments are stitched into a polygon that approximates the implicit surface.}
    \label{fig:marching_squares}
\end{figure*}

In OpenVCAD, an object's geometry is defined by a signed distance function where a value of zero represents the surface. Using this implicit function, we apply the marching squares algorithm to sample the geometry on a single pre-defined slice plane and generate iso-contour polylines that trace the boundaries of the object. The marching squares algorithm is a natural choice for simplicity, though more advanced algorithms like dual-contouring may offer better results for objects with sharp edges or fine features \cite{maple_geometric_2003, ju_dual_2002}. From Figure \ref{fig:marching_squares}, when given a function $f(x,y)$ that represents an implicit closed surface, the algorithm evaluates grid points and determines whether each is inside, outside, or on the surface. Using the point classification and a lookup table, the line(s) that intersect that grid cell are determined.  The output of the marching squares algorithm is a list of line segments that approximate the outline of the surface. However these line segments are not inherently ordered, and need to be stitched into polylines and polygons. 

Stitching the unordered list of line segments is accomplished with a two-part process. First, an undirected graph is built where the end points of each line are vertices and segments form the connecting edges. A global list of vertices is used to ensure that the same point is not added twice. When two line segments share the same endpoint, they will be represented by a single vertex in the graph. Care must be taken to handle floating point error that occurs as part of the marching squares process. We recommend using a fuzzy comparison with an adjustable tolerance to ensure vertices are merged properly across printing modalities that work at very large or small scales. For our work with desktop filament printers we use a tolerance of $\pm10^{-6} mm$. Once a graph structure is built, a Depth First Search is used to determine the set of connected components that form an ordered polyline or polygon. Care must be taken in this step to handle cycles that might form in the graph structure. Graph cycles represent closed polygons, where non-cyclic connected components represent open polylines. In the case of running marching squares on the geometric evaluator for an OpenVCAD design, we are guaranteed to have closed polygons because OpenVCAD assumes 3D geometries to be specified as watertight and manifold. However, in the next section we will see that this does not hold when we use the marching squares algorithm on the material distribution of an OpenVCAD design.

An alternative approach that was considered is to compute the triangulated mesh using the 3D counterpart to the marching squares algorithm, marching cubes \cite{lorensen_marching_1987}. This algorithm computes a triangulated mesh that is compatible with existing slicing workflows; however, this introduces inaccuracy in the z-direction. Better dimensional accuracy is achieved by computing the outline polygons for each slicing layer directly from the implicit design \cite{wang_universal_2024}. This is because the implicit representation maintains continuity along the z-axis, allowing for precise sampling at any arbitrary layer height without introducing interpolation errors or loss of geometric fidelity.

\begin{table*}[h]
\centering
\renewcommand{\arraystretch}{1.2}
\begin{tabular}{ll}
\rowcolor[HTML]{C0C0C0} 
\textbf{Variable} & \textbf{Description} \\
\(\displaystyle V_{\mathrm{blue}}, V_{\mathrm{yellow}}\) & Volume fraction pair that sum to \(1\). These are ``colors''. Each is defined as a function of \(x, y, z\). \\
\(\displaystyle F_{n}\) & The \(n\)-th face in a 2D arrangement. A face is a closed polygonal region with a uniform mixture ratio. \\
\(\displaystyle \alpha\) & Palette interval bandwidth, expressed as a percentage of the total gradient range (from 0\% to 100\%). \\
\(\displaystyle \beta\) & Zippering overlap, expressed as a percentage of the total gradient length. \\
\(\displaystyle h\) & Layer height ($mm$) used during printing. \\
\(\displaystyle w\) & Extrusion bead width ($mm$). \\
\(\displaystyle V_{\mathrm{melt}}\) & Melt chamber volume ($mm^3$) that must be purged to achieve a stable composition change. \\
\end{tabular}
\caption{Summary of symbols and their definitions used throughout the paper.}
\label{tab:symbols}
\end{table*}

\subsubsection{Connecting Geometry and Material Contours to Discretize Gradients} \label{sec:connecting_geo_mat_contours}
To generate tool pathing, we approximate continuous functional gradients with discrete states. We refer to each discrete state as a ``color''. Although a ``color'' may be the material color  (or another desired material property), it can also denote a set of process parameters, such as material mixture ratio, toolhead index, or extrusion temperature, that achieve some material property. We refer to the set of colors used in a design as the ``palette''. The size of the palette is the number of discrete process parameter states, which we leave as a user-defined setting in this work. A single color in the palette represents a range of colors in the continuous gradient. The palette informs how iso-contours are computed and used for toolpath generation.

We generate iso-contours for both the geometry and material components of OpenVCAD's implicit design.  In generating the iso-contours for the geometry, we only consider an iso-value of zero to find the surface of the geometry.  OpenVCAD guarantees that the iso-contour of the geometry at a given z-height always yields a set of closed polygons. In contrast, the iso-contours of a material distribution may not always produce closed polygons. For instance, a material gradient only varying along the x-axis,would result in open polyline iso-contours.   In generating the iso-contours for the material, we consider many iso-values, corresponding to the number of colors in our palette. 

\begin{figure*}[h!]
    \centering
    \includegraphics[width=\linewidth]{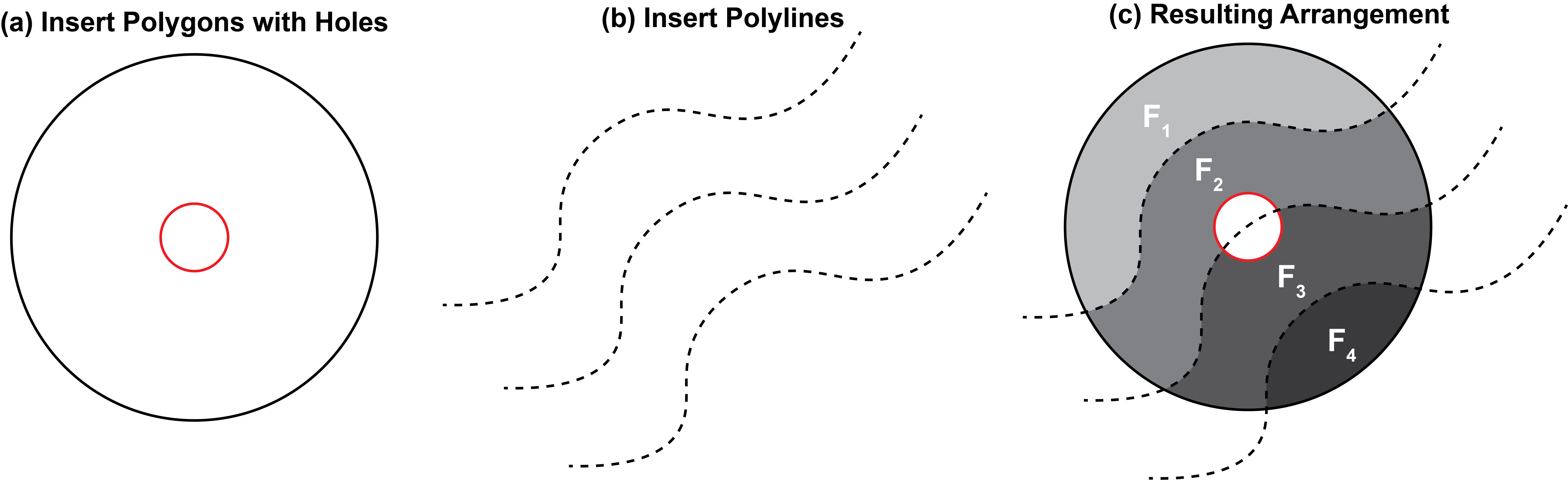}
    \caption{Shows the process of extracting bounded faces using a 2D arrangement data structure. (a) A polygon (black) with a hole (red) is inserted into the arrangement. The polygon with hole is generated from the OpenVCAD signed distance function using Marching Squares. (b) Polylines (dashed) are inserted into the arrangement. Polylines are generated from the OpenVCAD multi-material distribution functions using Marching Squares. (c) The arrangement data structure can be used to extract the bounded faces, $F_1,F_2,F_3$ and $F_4$.}
    \label{fig:arrangement}
\end{figure*}

To use the iso-contours, geometric contours are combined with material contours. The geometric contours define the boundaries of printable area, and material contours subdivide the functional gradient into colors from the palette. The goal is to generate a set of polygonal ``faces," each denoting a region of the layer falling within a specific range in the gradient or a single color of the palette. Algorithm \ref{alg:iso_contours} details the process of generating printable faces from an OpenVCAD design. A 2D arrangement is generated, incorporating both the geometric iso-contours and the material iso-contours \cite{cgal_arrangement_on_surface_2, agarwal_sharir_2000}. The arrangement allows for different polygons and polylines to intersect to form a new set of ``faces" (polygons) that are either bounded or unbounded. Figure \ref{fig:arrangement} shows how the 2D arrangement data structure can be used to extract a list of bounded faces after inserting polygons and polylines. For each bounded ``face" formed in this arrangement, a point within the face is sampled to determine the material at that location. The face is then classified as the appropriate color from the palette and stored for later toolpath generation. This method accommodates both closed and open contours, ensuring each part of the layer is assigned the correct color while maintaining a consistent and manufacturable geometry.

\begin{algorithm}[]
\caption{Generate Printable Faces from OpenVCAD Iso-contours}
\label{alg:iso_contours}
\KwIn{OpenVCAD design, Color Palette, Z-height, XY sampling resolution}
\KwOut{List of colors from the palette with corresponding polygonal faces}
\BlankLine
\SetKwBlock{StepOne}{Step 1: Generate Geometric Iso-contours}{end}
\SetKwBlock{StepTwo}{Step 2: Generate Material Iso-contours}{end}
\SetKwBlock{StepThree}{Step 3: Create 2D Arrangement}{end}
\SetKwBlock{StepFour}{Step 4: Classify Faces}{end}
\SetKwBlock{StepFive}{Step 5: Return List of Faces by Color}{end}

\StepOne{
    Sample the design at Z-height using the signed distance function.\\
    Generate closed polygons defining the object boundary.
}
\StepTwo{
    \ForEach{color in palette}{
        Sample the material distribution at XY resolution.\\
        Extract iso-contours for each color.
    }
}
\StepThree{
    Overlay geometric and material iso-contours.\\
    Generate closed regions where contours intersect.
}
\StepFour{
    \ForEach{closed region in 2D arrangement}{
        Select a random point within the region.\\
        Sample the material at the point.\\
        Classify the region into a color from the palette.
    }
}
\StepFive{
    Return a mapping of colors to corresponding regions.
}
\end{algorithm}

\subsection{Gradient Informed Toolpath Planning}
The goal of gradient-informed toolpath planning is to incorporate both geometric boundaries and material gradients into the toolpath generation process. By using the "faces" generated from the iso-contours of geometry and material distributions, we create specialized toolpaths that not only follow the shape of the object but also respect the material transitions required by the functional gradient. The objective is to minimize sharp transitions in material properties and the amount of purging required, while ensuring a seamless integration between geometry and material distribution.

We propose two strategies to address this challenge, each tailored to different use cases and 3D printing systems. The first approach, which we refer to as ``preserving traditional toolpaths'', builds upon existing tool pathing methods by subdividing standard paths into material regions and adjusting the order of operations to account for material changes. The second approach, which we refer to as ``continuous printing against the gradient", focuses on filling the object as densely as possible while maintaining a high-quality gradient, without relying on traditional toolpath structures. Each method has its own strengths, and the choice of strategy depends on the specific requirements of the print, including material properties, system capabilities, and desired precision in gradient transitions. In both cases, the faces generated from the iso-contours form the foundation of the toolpath planning, ensuring that the material gradients are integrated into the 3D printing process.

\subsection{Strategy 1: Preserving Traditional Toolpaths}
\label{sec:strategy_1}
\begin{figure}[h!]
    \centering
    \includegraphics[width=\linewidth]{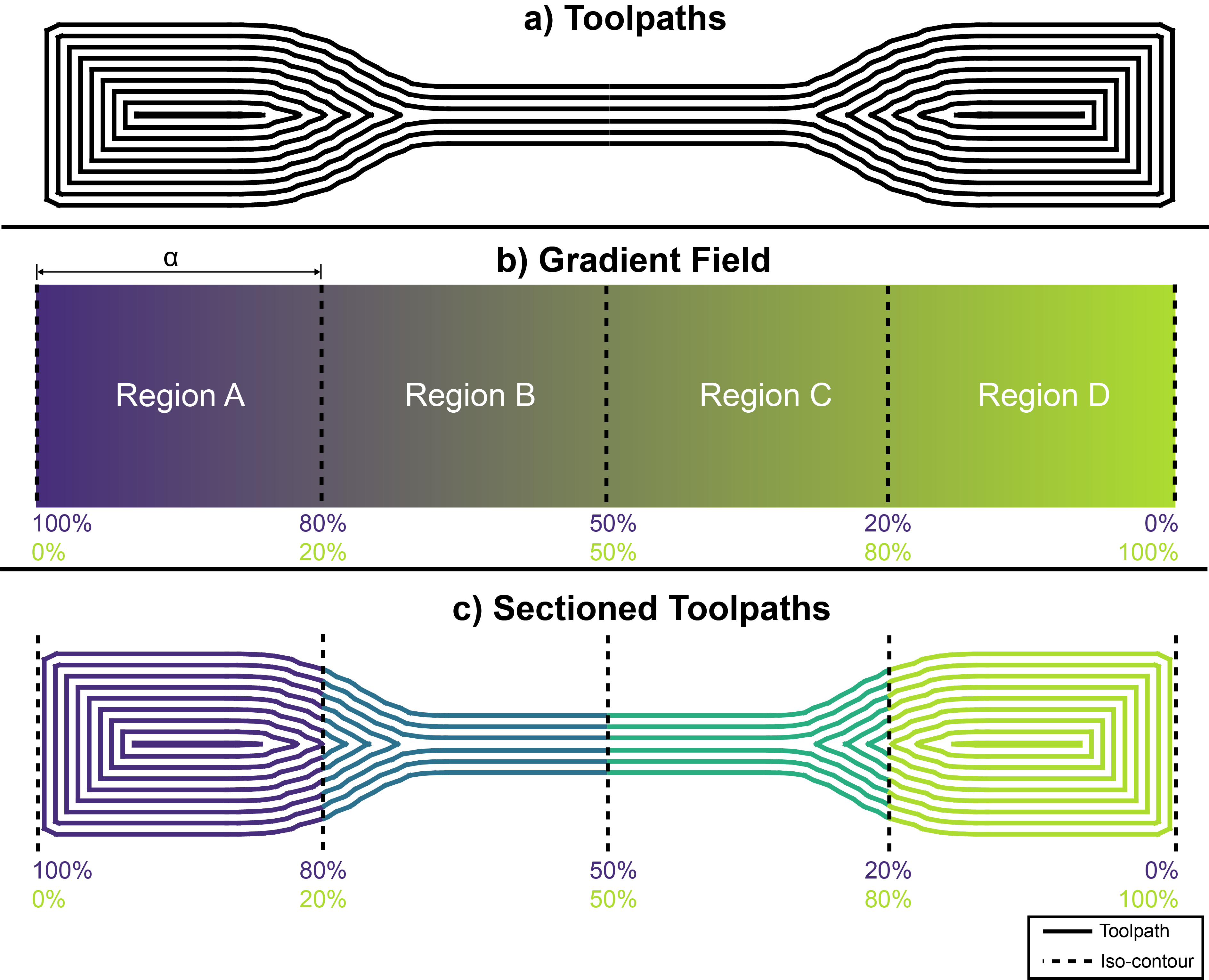}
    \caption{An overview of strategy 1. (a) an OpenVCAD design is sliced into layer outlines given a z-coordinate. These outlines are then toolpath planned (we show an object made entirely of walls for simplicity, but this works with any tool pathing). (b) the OpenVCAD multi-material gradient is sliced into a user defined number of colors to form the palette . The bandwidth of these regions is defined as $\alpha$ and the actual assigned color is the mid-point of the interval. The marching squares algorithm is used to determine the iso-contours that sub-divide the gradient. (c) the iso-contours are used to segment the planned toolpaths into different regions, where each region is a single color. }
    \label{fig:sectioned_paths}
\end{figure}

\begin{figure}[h!]
    \centering
    \includegraphics[width=0.75\linewidth]{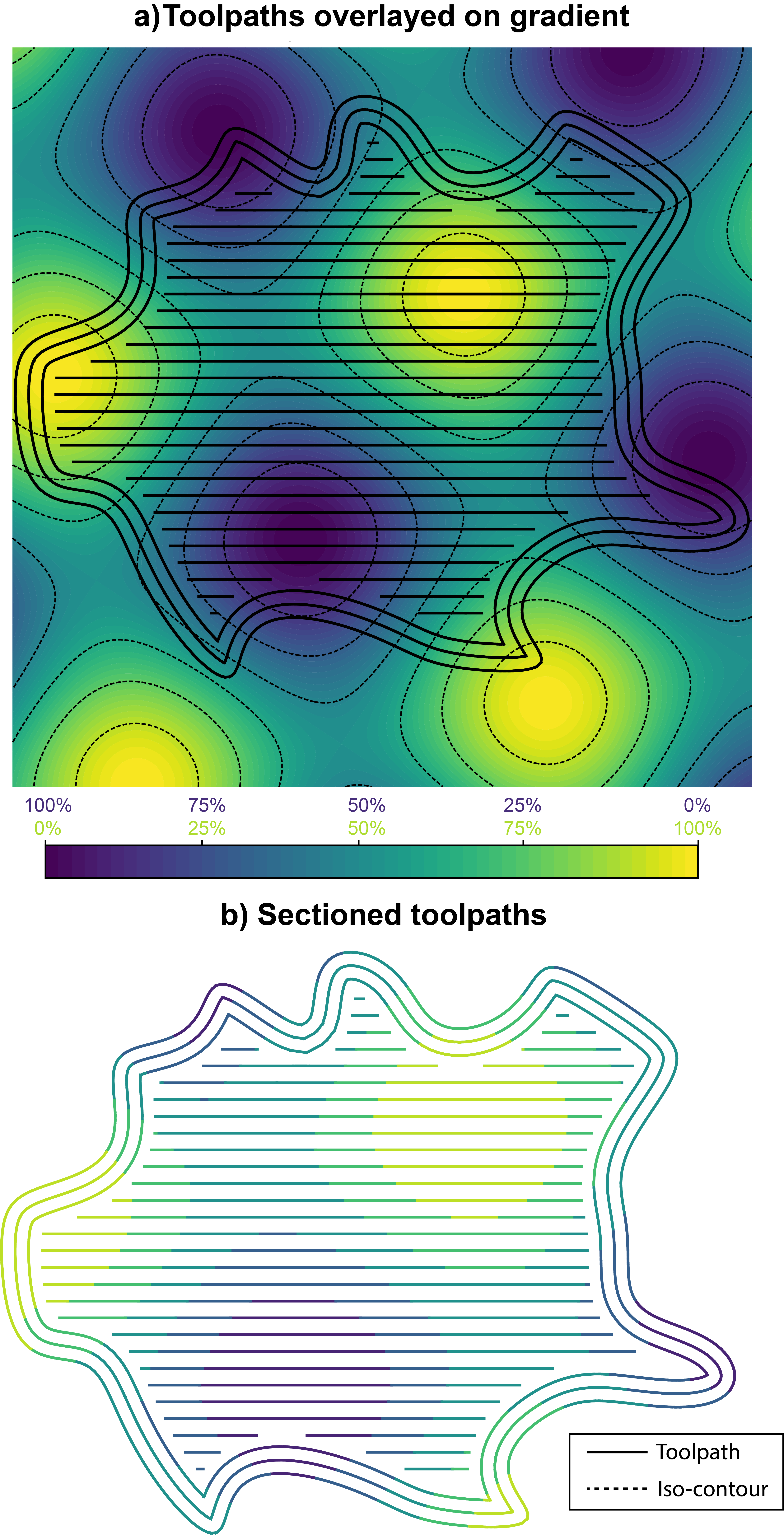}
    \caption{(a) a complex gradient that varies in X and Y is defined in OpenVCAD using volume fraction functions given in Equation \ref{eq:color_distributions}. The gradient is sliced into iso-contours (dashed lined) that sub-divide the space into a user defined number of colors in the color palette. (b) Sliced toolpaths for a layer are segmented and classified into these color palette regions.}
    \label{fig:complex_section}
\end{figure}

The first strategy for gradient-informed toolpath planning, detailed in Algorithm \ref{alg:gradient_toolpath}, focuses on maintaining traditional toolpath generation methods while incorporating material gradient information. The process begins by slicing a layer of an OpenVCAD design's geometry to generate the outlines of the object at the current layer. These outlines serve as the input for a normal toolpath planning process, which generate walls, infills, skins, and other features according to user defined settings. Figure \ref{fig:sectioned_paths} shows a tensile dog bone with a simple linear cross-fade between two materials sliced with this method. Figure \ref{fig:complex_section} further demonstrates this method with a complex gradient that is varying across the \texttt{x} and \texttt{y} axes. 

\subsubsection{Sectioning Toolpaths by Material Gradient}
Once the standard toolpaths have been generated for the layer, the next step is to overlay the gradient field and section the toolpaths according to the user-defined color palette. This is accomplished by using the iso-contours derived from the material gradient to divide the design into distinct faces, each corresponding to a single color from the palette, as described in Section \ref{sec:connecting_geo_mat_contours}.

\begin{algorithm}[]
\caption{Preserve Planned Toolpaths with Gradient Information}
\label{alg:gradient_toolpath}
\KwIn{Geometric outline, Color Palette, List of Faces and their colors}
\KwOut{Ordered toolpaths with color assignments and purge towers}
\BlankLine
\SetKwBlock{StepOne}{Step 1: Generate Traditional Toolpaths}{end}
\SetKwBlock{StepTwo}{Step 2: Section Toolpaths by Color}{end}
\SetKwBlock{StepThree}{Step 3: Optimize Toolpath Order}{end}
\SetKwBlock{StepFour}{Step 4: Insert Purge Towers}{end}
\SetKwBlock{StepFive}{Step 5: Return Ordered Toolpaths}{end}

\StepOne{
    Generate standard toolpaths using the geometric outline.
}
\StepTwo{
    \ForEach{face in faces list}{
        Compute intersections between toolpaths and the face.\\
        Label intersected toolpaths with the face's color.
    }
}
\StepThree{
    Optimize toolpath order to minimize transitions between colors.
}
\StepFour{
    \ForEach{transition between colors}{
        \If{color change exceeds threshold}{
            Insert a purge tower to stabilize the color.
        }
    }
}
\StepFive{
    Return toolpaths with color assignments and purge towers.
}
\end{algorithm}

For each face defined by the material iso-contours, the algorithm computes the intersection of each face with existing toolpaths. The result is a set of segmented toolpaths that are fully contained within each face. Each toolpath segment is labeled with the corresponding color from the palette, as defined by the face it belongs to. This process repeats for all faces across the layer, ensuring that each toolpath is properly segmented and labeled according to the material gradient. This method allows for the preservation of traditional tool pathing while accounting for the material distribution, ensuring that each region of the object is printed with the correct material color.

\begin{equation}
\begin{aligned}
    V_{\text{blue}} &= \frac{1+\sin(0.02x + 0.03y) \cos(0.03x - 0.02y)}{2}, \\
    V_{\text{yellow}} &= 1 - \frac{1+\sin(0.02x + 0.03y) \cos(0.03x - 0.02y)}{2}.
\end{aligned}
\label{eq:color_distributions}
\end{equation}

\subsubsection{Path Order Optimization}\label{sec:path_order_optimizing}
Once the toolpaths have been segmented according to the material gradient, the next step is to order them in a way that minimizes abrupt changes in color or the process parameters that define the ``color''. By ordering the regions to gradually transition from one color to another, we ensure smoother material transitions during the print, reducing the risk of defects or abrupt changes in material properties.

\begin{figure*}[h!]
    \centering
    \includegraphics[width=\linewidth]{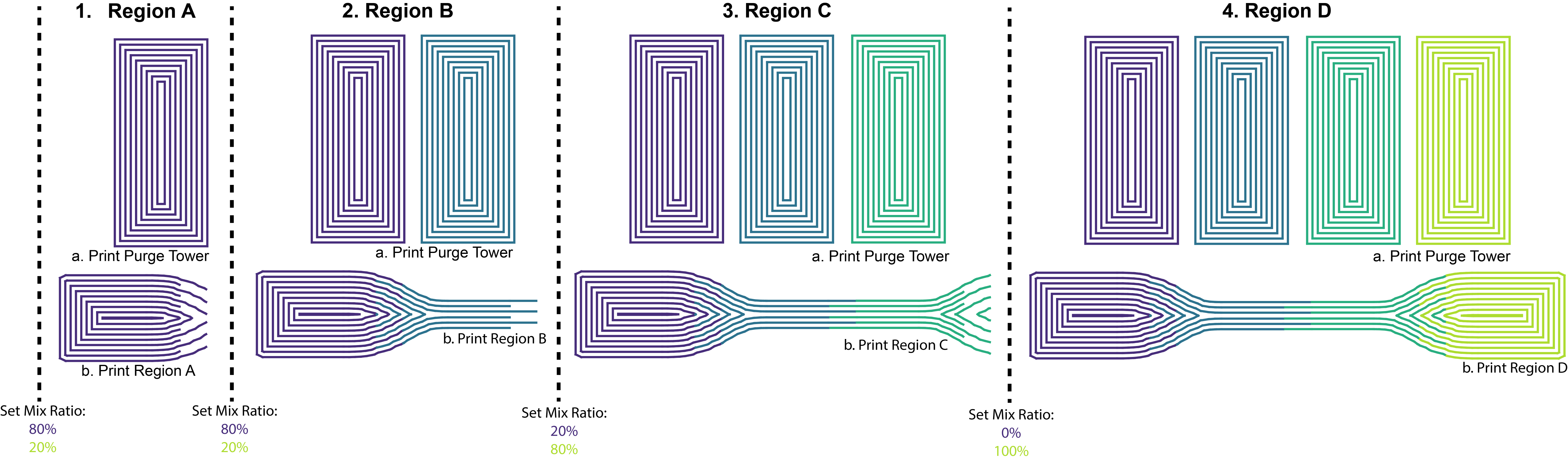}
    \caption{Timeline of the printing process, assuming a mixing hotend system. ``Color'' is determined by the mixture ratio. 1. Mix ratio is set for Region A. Then Region A's purge tower is printed, followed by the paths in that region. the process repeats for Regions B,C, and D. On the following layer, the region ordering will be reversed since the first layer ended with mixture ratio D. However, purge towers are always printed before the paths for that region to ensure a stable mixture ratio has been achieved.}
    \label{fig:interwoven}
\end{figure*}

If a large change in material composition or process parameters occurs between two adjacent regions, a purge tower is printed to stabilize the color before continuing. By reducing the change in process parameters between adjacent regions, the size and frequency of purge towers can be minimized, resulting in a more efficient print. However, dividing the toolpaths into smaller segments may introduce challenges with travel movements and start/stop operations, especially on systems that are sensitive to interruptions. Figure \ref{fig:interwoven} shows the timeline for building a single layer of the tensile dog bone example. Each region is printed successively with purge towers added to stabilize color transitions.

Additionally, the size of purge towers must be calibrated, according to the printing system and the changing process parameters. Ideally, the extruder should fully stabilize to the new color within each region to prevent material bleeding into neighboring regions. For smaller objects or those with fine gradients, larger purge towers may be needed to ensure smooth transitions. With Strategy 1, traditional toolpaths can be preserved while incorporating gradient information, making it suitable for applications where maintaining conventional toolpath structures is beneficial.

\subsubsection{Zippered Toolpaths to Reduce Seam Defects}
\label{sec:strategy_1_zipper}
\begin{figure}[h!]
    \centering
    \includegraphics[width=\linewidth]{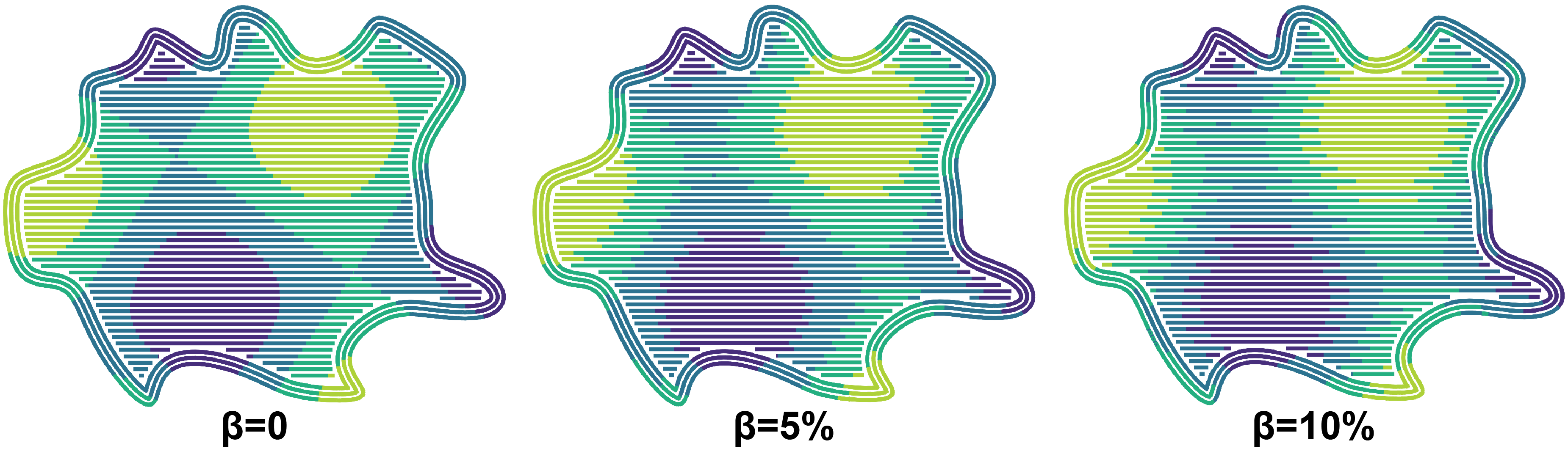}
    \caption{Zippering is added to the $XY$ gradient given in Figure \ref{fig:complex_section} and Equation \ref{eq:color_distributions}. As $\beta$ increases the zippering distance also increases. }
    \label{fig:overlap_complex}
\end{figure}

One of the primary concerns when sectioning toolpaths into distinct material regions is the creation of seams at the boundaries between regions. These seams, while necessary for transitioning between material gradients, can introduce mechanical weaknesses in the printed object \cite{sardinha_influence_2020, taiwo_investigating_2024}. To mitigate these defects, we propose a method of ``zippering" the toolpaths by introducing overlapping regions between adjacent colors. This builds on the work of Sardinha et al. who found that zippering approaches can improve the mechanical performance of objects printed with multiple materials. They have shown that the zippering technique increases the bonding surface area between regions, improving the mechanical integrity of the final print by avoiding sharp, discrete transitions \cite{sardinha2024multi_extrusion}. We present a novel automation of their technique and expand it to support the automatic placement of zippers along any arbitrary material boundary curve (see fig. \ref{fig:overlap_complex}.

\begin{figure}[h!]
    \centering
    \includegraphics[width=\linewidth]{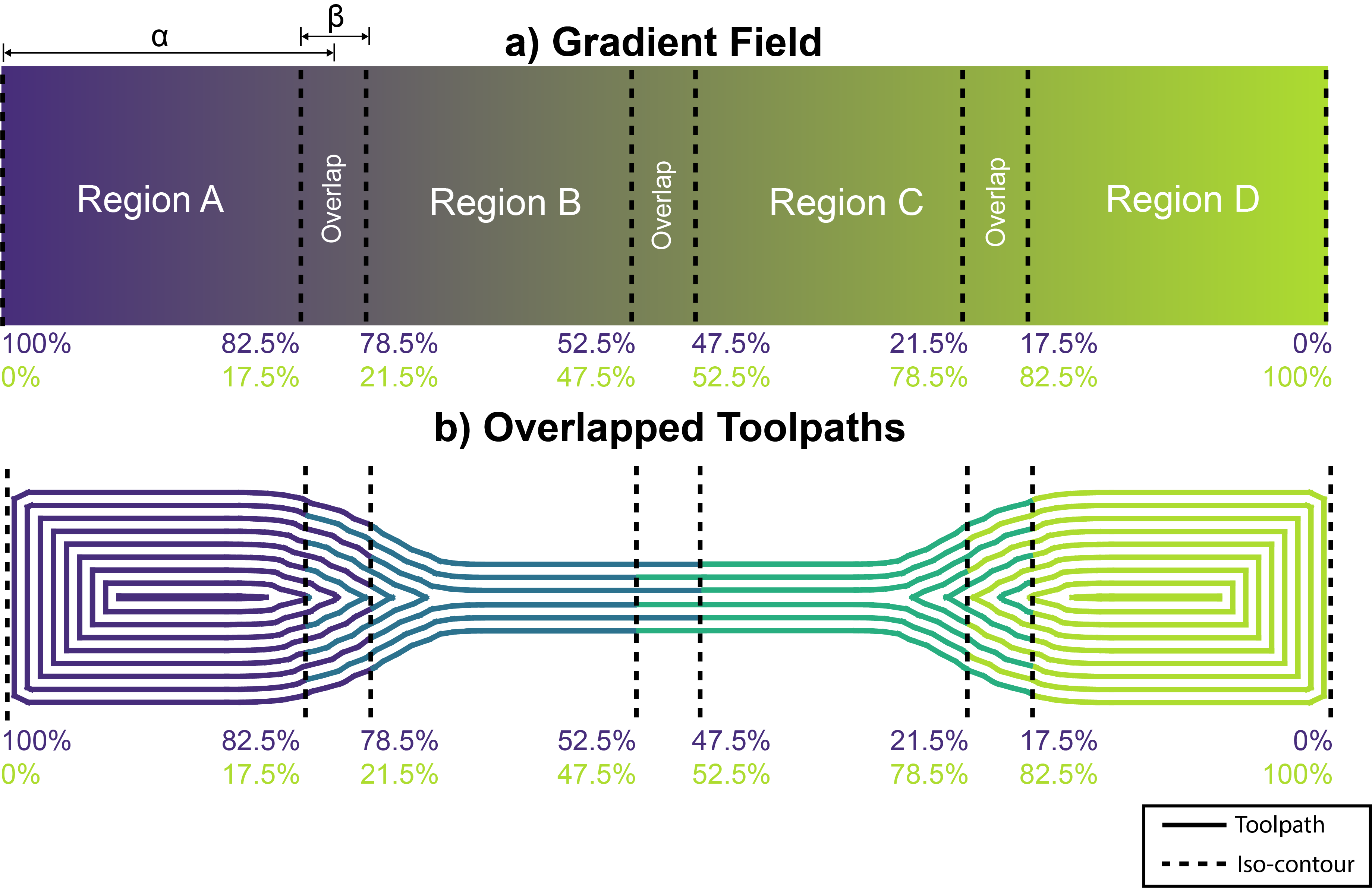}
    \caption{An overview of the zippering process. (a) the color palette from the OpenVCAD design is sub-divided into a user defined number of colors with a bandwidth $\alpha$. Overlap regions are inserted in between the original sub-division with a bandwidth of $\beta$. (b) when the design is segmented into individual colors, paths that are in the overlapping region are alternatively assigned to adjacent colors to create the zipper.}
    \label{fig:overlap}
\end{figure}

As detailed in Algorithm \ref{alg:zippered_toolpaths} and shown in Figure \ref{fig:overlap}, our automated slicing method begins by inserting overlapping regions between each pair of colors in the user-defined palette, and the toolpaths in these regions are divided and assigned to one of the two adjacent colors. The overlap width, denoted as $\beta$, is a user-defined parameter. This region is treated as a transitional zone, where toolpaths are alternately ``stitched" between the adjacent color regions. This effectively blends the materials across the interface, increasing the bonding surface area compared to a sharp seam. While Sardinha et al. demonstrated the efficacy of increasing the surface area between dissimilar material regions to improve strength, their method relied on manual processing to create the zippering effect. Our proposed method not only automates the toolpath generation, but also greatly increases the complexity of the interface geometry that can be zippered. Figure \ref{fig:overlap_complex} shows the toolpaths for an object sliced with the complex two-material gradient given in Algorithm \ref{eq:color_distributions}. We observe that the zippering is correctly placed along the mixture range boundaries when $\beta$ is greater than zero.

When selecting the number of colors in the palette, users must balance gradient smoothness with resulting segmented path length. A greater number of palette colors yields smoother gradients but reduces the length of each individual toolpath segment. While the optimal choice is geometry-dependent, we observe that desktop FFF systems commonly experience print quality degradation when individual path segments become shorter than approximately 5~mm, due to frequent starts and stops in extrusion. Thus, we recommend choosing the palette interval ($\alpha$) such that the number of palette colors ensures that the minimum path segment length remains at or above this 5~mm threshold. Regarding the zippering overlap distance $\beta$, we suggest a default overlap of at least 15 mm for typical desktop-scale prints, informed by findings from Sardinha et al. ~\cite{sardinha2024multi_extrusion}. Practically, users specify $\beta$ as a percentage of the total gradient. Therefore, the absolute overlap distance (in mm) will depend on the physical dimensions of the gradient region within the object.”

\begin{algorithm}[]
\caption{Zippering Toolpaths with Overlapping Regions}
\label{alg:zippered_toolpaths}
\KwIn{Geometric outline,Color Palette, List of Faces and their colors, Overlap width (\(\beta\))}
\KwOut{Zippered toolpaths with overlapping regions and color labels}
\BlankLine
\SetKwBlock{StepOne}{Step 1: Generate Traditional Toolpaths}{end}
\SetKwBlock{StepTwo}{Step 2: Section Toolpaths by Color}{end}
\SetKwBlock{StepThree}{Step 3: Introduce Overlapping Regions}{end}
\SetKwBlock{StepFour}{Step 4: Return Zippered Toolpaths}{end}

\StepOne{
    Generate standard toolpaths with the geometric outline.
}
\StepTwo{
    \ForEach{face in faces list}{
        Compute intersections between toolpaths and the face.\\
        Label intersected toolpaths with the face's color.
    }
}
\StepThree{
    \ForEach{pair of adjacent colors}{
        Define the overlap region with width \(\beta\).\\
        \ForEach{toolpath in the overlap region}{
            Alternate assignment of toolpaths to the first or second color.\\
            Optimize toolpath order to minimize the change in color.
        }
    }
}
\StepFour{
    Return toolpaths organized by color, including overlapping regions.
}
\end{algorithm}

\subsection{Strategy 2: Maximizing Gradient Quality: Continuous Printing Against the Gradient}
\label{sec:strategy_2}
The second approach, termed ``continuous printing against the gradient", diverges from traditional toolpath structures by prioritizing alignment with the material gradient over conventional print elements such as walls, infill, and skin. This method aims to densely cover each region according to the gradient field, thereby enhancing the continuity and quality of material transitions across the printed part.

\begin{figure}[h!]
    \centering
    \includegraphics[width=\linewidth]{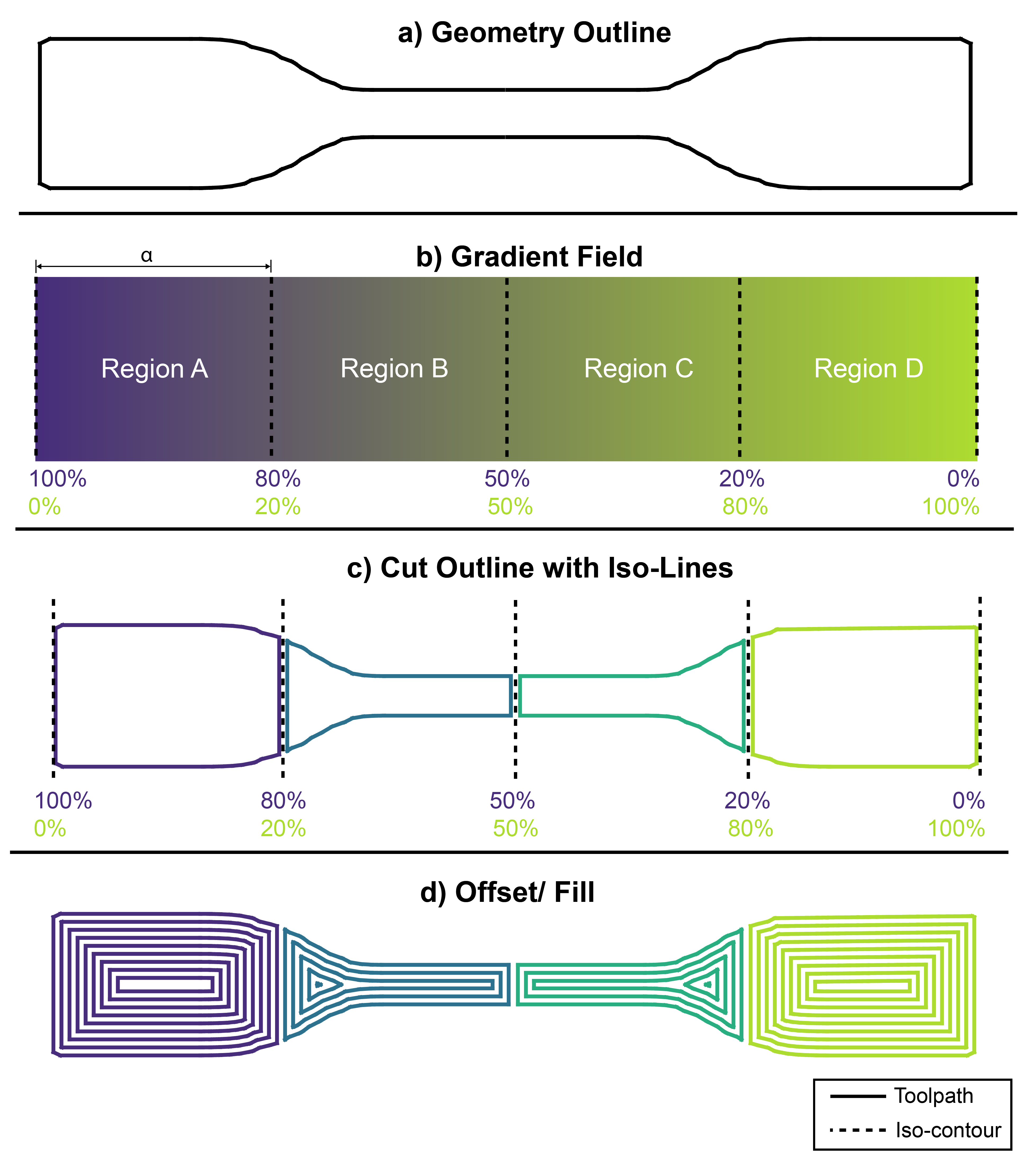}
    \caption{An overview of the Strategy 2 workflow. (a) an OpenVCAD geometry is sliced to determine an outline. The OpenVCAD design's material distribution (b) is sliced into iso-lines that sub-divide the space into colors from the user-defined palette. (c) The iso-lines are used to cut the geometry outline into closed polygons that encapsulate all of the printable area belonging to a specific color. (d) The paths are filled with infill.}
    \label{fig:filled_paths}
\end{figure}

From Figure \ref{fig:filled_paths} and Algorithm \ref{alg:offset_paths}, we see this approach uses the faces generated from iso-contouring the material distribution (see section \ref{sec:connecting_geo_mat_contours}). Since these faces are derived from the gradient field, they are non-overlapping by design, allowing each polygonal face to be offset inwards without intersecting adjacent regions. An initial offset by half a bead width ensures that the toolpaths remain confined within each face, preventing overlap between adjacent mixture regions. Following this offsetting step, each face is filled by either creating successive inward offsets of the polygon, forming concentric polygons that progressively fill the region, or by applying a dense rectilinear pattern with 100\% density to fully cover the area with material. Both approaches achieve complete coverage within each mixture region, a necessity discussed in a later section.

This method offers flexibility for handling complex gradients and supports a greater number of colors in the palette  compared with Strategy 1. In contrast to the first strategy, where additional colors may fragment the toolpaths and introduce starts and stops, additional colors in this strategy enable smoother printing. Also, as the number of colors in the palette increases, the need for purge towers decreases, as gradient transitions are more gradual. In an idealized scenario with a large number of colors, this method generates a single contour set per color, traversing the gradient with the lowest rate of change. Strategy 2 plans toolpaths that ensure the print direction is always perpendicular to the gradient direction, thereby minimizing the change in color and requiring no purge towers.

\begin{algorithm}[h!]
\caption{Continuous Printing Against the Gradient}
\label{alg:offset_paths}
\KwIn{Color Palette, List of Faces and their Color}
\KwOut{Toolpaths organized by color, with optimized path order and color change commands}
\BlankLine
\SetKwBlock{StepOne}{Step 1: Offset Faces Inwards}{end}
\SetKwBlock{StepTwo}{Step 2: Fill Each Face}{end}
\SetKwBlock{StepThree}{Step 3: Optimize Path Order}{end}
\SetKwBlock{StepFour}{Step 4: Return Toolpaths}{end}

\StepOne{
    \ForEach{face in list of faces}{
        Offset the face inwards by half a bead width to prevent overlap.
    }
}
\StepTwo{
    \ForEach{face in list of faces}{
        \uIf{Concentric offset infill}{
            Fill the face with concentric inward offsets.
        }
        \ElseIf{Rectilinear infill}{
            Apply rectilinear infill at 100\% density.\\
            Connect perimeter walls and infill lines with travel moves.
        }
    }
}
\StepThree{
    Order toolpaths by color to minimize mixture transitions between regions.
}
\StepFour{
    Return toolpaths organized by color, with color change commands.
}
\end{algorithm}

\subsubsection{Addressing Non-Linear Gradients with Infill}
\begin{figure}[h!]
    \centering
    \includegraphics[width=0.9\linewidth]{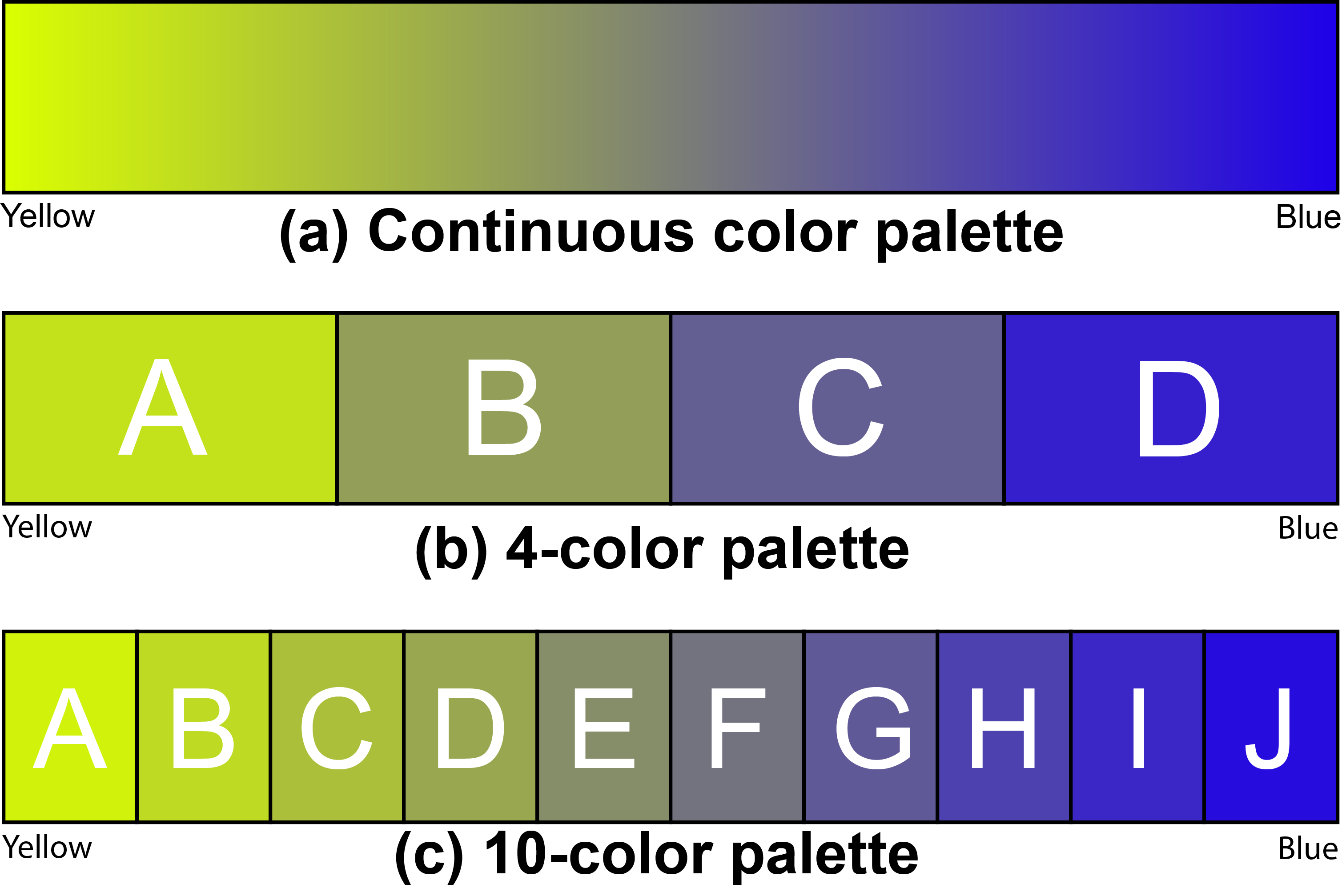}
    \caption{(a) continuous two color palette is sub-divided into (b) 4 and (c) 10 distinct colors. Letters A through J show the order the colors are printed to minimize change in mixture. For each color, the corresponding toolpaths are printed, followed by the next color.}
    \label{fig:color_bar}
\end{figure}

Our method segments the multi-material gradient into regions by computing iso-contours that correspond to specific iso-values. These iso-contours divide the total material mixture space into a user-defined number of equal-sized regions. For example, in a two-material gradient divided into a four-color palette for a mixing hotend printer, the iso-contours delineate regions with mixture ratios of (0–25\%), (25–50\%), (50–75\%), and (75–100\%) for materials A and B, respectively. However, for a given mixture interval, a single ratio is required to be given to the printer. We choose the center point of a mixture interval for this value. This can be extrapolated to other processes parameters as well.  Figure \ref{fig:color_bar} illustrates this linear subdivision, showing both a four-color and a ten-color palette along with a traversal order \textit{A} to \textit{J}. Following this traversal order is crucial to minimize changes in mixture ratio (color) between successively printed regions. Minimizing the change in mixture ratio between regions reduces the required purging. 

\begin{figure}[h!]
    \centering
    \includegraphics[width=\linewidth]{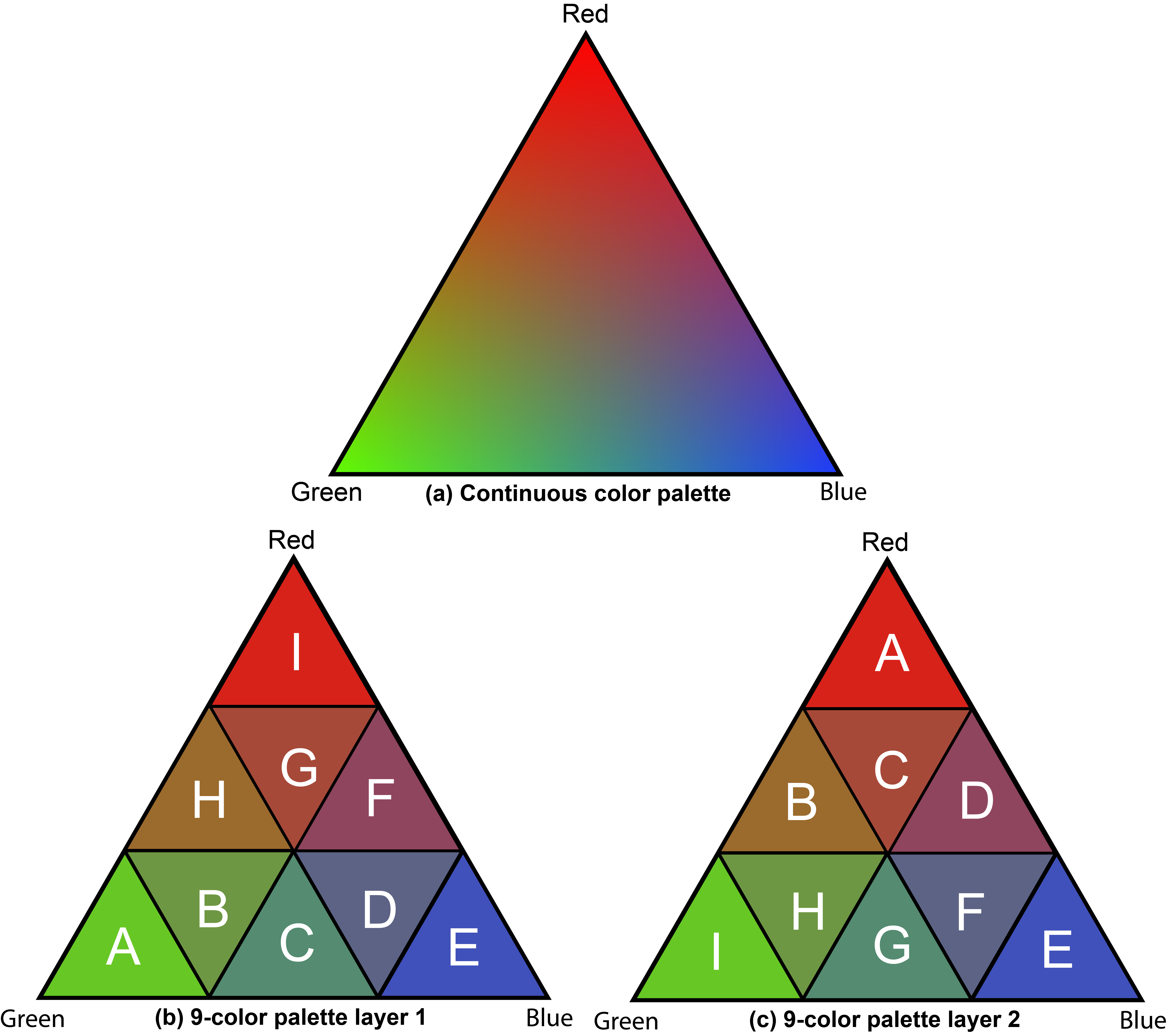}
    \caption{(a) continuous three color palette is sub-divided into (b) and (c) with 9 distinct virtual colors. Letters A through I show the order the colors are printed to minimize change in mixture. (b) and (c) compare two different traversals that minimize the change in mixture ratio between successively printed colors. When printing multiple layers, the traversal order should be flipped on odd numbered layers to match the end mixture state from the previous layer. }
    \label{fig:color_triangle}
\end{figure}

This subdivision process is not limited to two materials; it can be extended to palettes with three or more primary materials. Figure \ref{fig:color_triangle} demonstrates the subdivision of a continuous three-component palette into nine individual ``colors''. While multiple traversal orderings exist for tri-color palettes, the Figure highlights two possible minimal change orders. Additionally, the traversal order alternates between each printed layer to ensure a back-and-forth progression, further minimizing mixture ratio changes across layers. 

\begin{figure}[h!]
    \centering
    \includegraphics[width=\linewidth]{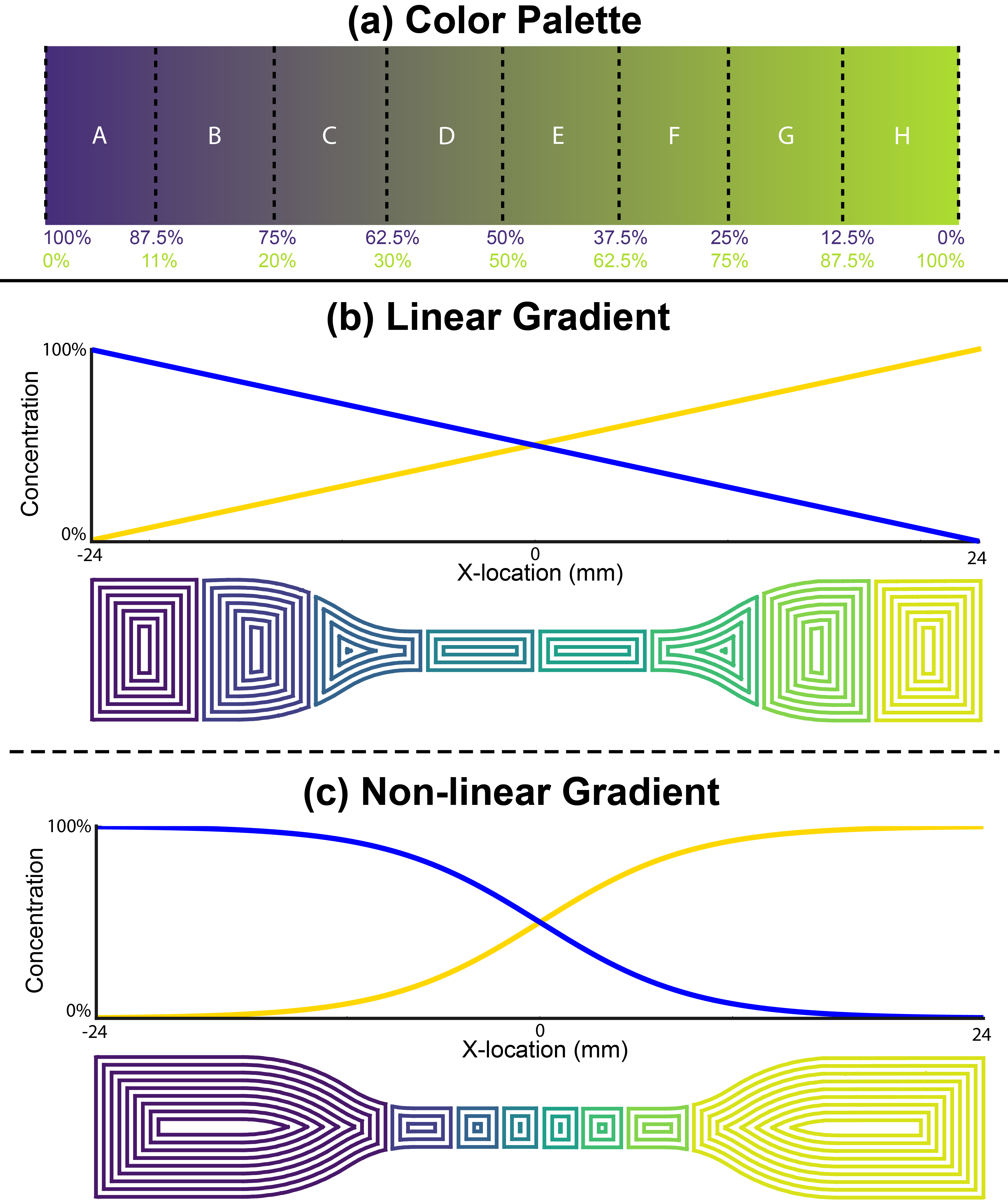}
    \caption{A comparison between OpenVCAD objects with (b) linear and (c) non-linear gradients. (a) shows how the color space is always sub-divided into equal regions. The gradient defined in the OpenVCAD design will determine the size of each segmented region within the print needed to accurately capture the gradient.}
    \label{fig:non-linear}
\end{figure}

We use a linear subdivision in gradient mixture space since OpenVCAD inherently allows for the specification of non-linear gradients in the design. Each color in the palette is assigned an average set of process parameters that serves as the target for printing. Figures \ref{fig:non-linear}(b) and \ref{fig:non-linear}(c) compare OpenVCAD designs with linear and non-linear gradient distributions, respectively, using the same palette subdivision as shown in (a). This comparison highlights that while each palette region has equal divisions in palette space, the actual area of each color may vary, particularly in non-linear designs. In Figure \ref{fig:non-linear}(c), the non-linear mapping in the OpenVCAD design results in varying spatial areas for each region, despite uniform divisions in the color palette. This effect underscores the importance of using an infilling mechanism to cover all of the space within a region, regardless of the area. It should be noted that Strategy 2 has a practical upper bound on the number of colors achievable within a palette. As the number of colors increases, iso-contours become progressively denser, eventually approaching the physical limitation of a single extrusion bead width. Therefore, because printing two colors in the same extrusion bead width is not possible, the practical maximum number of palette colors in Strategy 2 depends explicitly on the object's scale, printer nozzle diameter, and the spatial density of the gradient specified by the design.

\subsubsection{Compensating for Dead Volume}
Dead volume refers to the residual volume within the melt chamber that delays the transition between material states, causing deviations from intended gradient profiles. Purge towers are used in Strategy 1 to mitigate artifacts caused by changes in mixture ratios, ensuring that the hotend's entire dead volume is purged and the new mixture ratio is stabilized before resuming printing. While the larger number of virtual ``colors'' available with Strategy 2 allows for objects to be constructed without purge towers, challenges still arise when attempting to precisely position gradients within a design. This difficulty stems from the delay between commanding a mixture ratio change and the resulting change observed at the nozzle due to the dead volume.

\begin{figure}[]
    \centering
    \includegraphics[width=73mm]{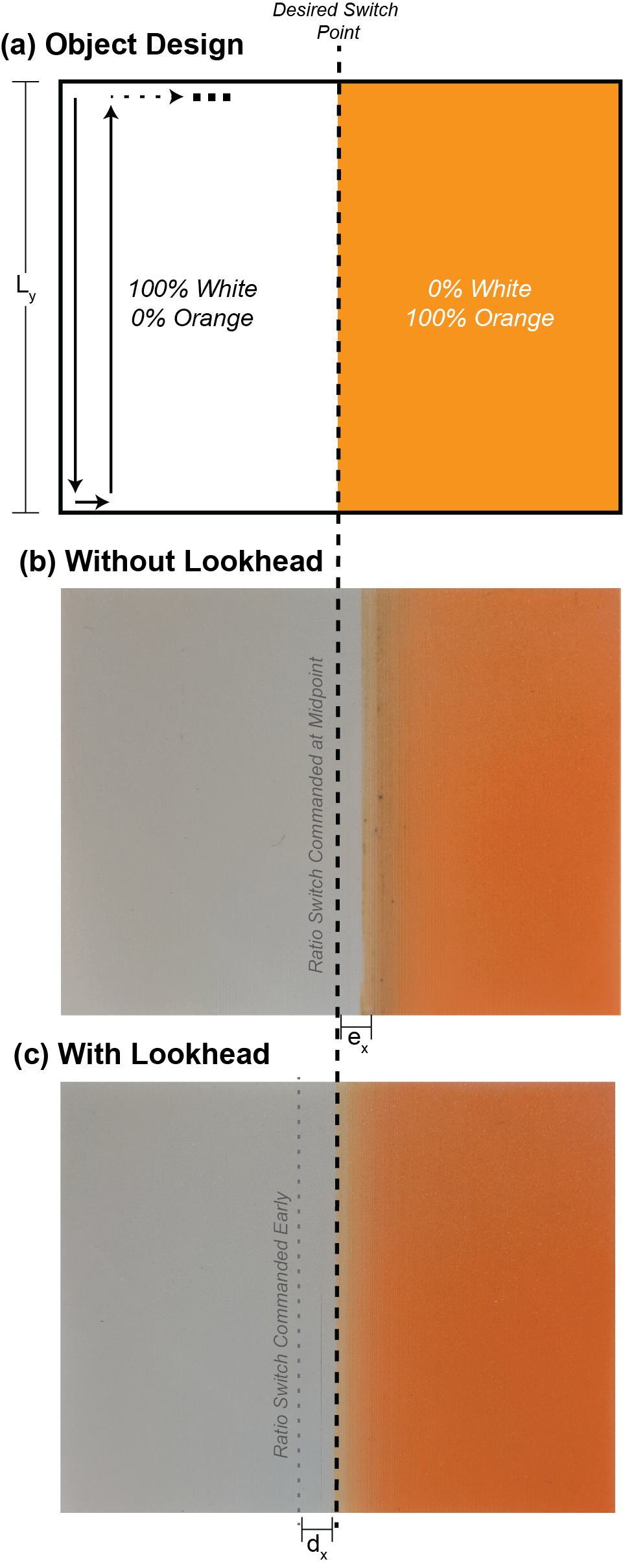}
    \caption{A comparison of the initial calibration print for the Geeetech 3D printer with no look-ahead, the final calibration print with a look-ahead distance of $960\ ~\mathrm{mm}$, and the reference design. The object is printed with a serpentine toolpath pattern, with a material ratio change commanded exactly half way through the printed layer. This Figure is best viewed in color. The same calibration procedure was also used to determine a $450~\mathrm{mm}$ look-ahead distance parameters for the Prusa XL 3D printer. }
    \label{fig:lookadhead}
\end{figure}

Although increasing the number of colors in the palette reduces the observable difference between adjacent printed regions and effectively spreads gradient changes over longer distances, a mechanism is still required to compensate for the inherent delay. Figure \ref{fig:lookadhead} (a) illustrates an object designed with a sharp transition from white to blue at the midpoint of the x-axis. The slicing process generates toolpaths that hatch back and forth along the y-axis, transitioning across the x-axis in short increments. This pathing ensures that each printed segment is exclusively white or blue. 

Figure \ref{fig:lookadhead} (b) demonstrates the artifact caused by the delay: the mixture ratio command is issued precisely at the midpoint of the x-axis, yet the observed transition occurs after this point due to the dead volume. In contrast, Figure \ref{fig:lookadhead} (c) shows the same design printed using a look-ahead mechanism. This mechanism allows the slicer to anticipate mixture ratio changes by analyzing upcoming segments in the toolpath. When a mixture ratio change is detected within a predefined look-ahead distance, the corresponding command is issued earlier to account for the delay introduced by the dead volume. 

The results in Figure \ref{fig:lookadhead} (c) illustrate that the look-ahead mechanism effectively compensates for the delay caused by the dead volume, ensuring the gradient transition occurs at the intended location. A procedure for calculating the lookahead distance is given in Appendix A.

\section{Results}\label{sec:results}
\begin{figure}[h!]
    \centering
    \includegraphics[width=\linewidth]{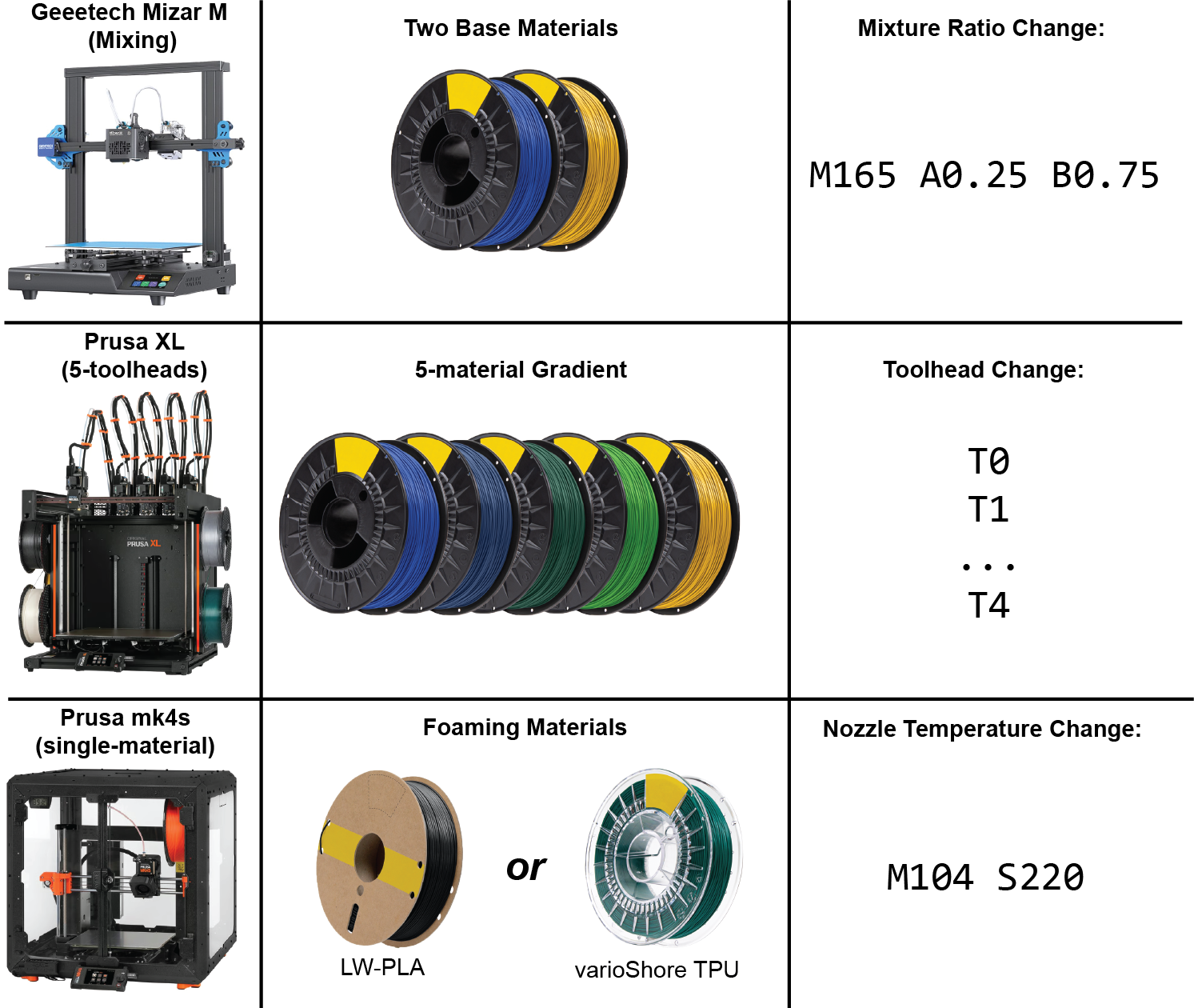}
    \caption{The three modalities employed in this study: a two-material mixing system, a 5-toolhead multi-material MEX printer, and a single-material printer loaded with foaming filament. For the mixing system, two base filaments (blue and yellow) are combined, with the mixture ratio controlled via the Marlin firmware \texttt{M165} command. For the 5-toolhead system, discrete filament colors spanning from blue to yellow are selected using the \texttt{T0, T1, ..., T4} commands. The foaming filament experiments use either LW-PLA or varioShore TPU materials on a single-material Prusa MK4S printer.}
    \label{fig:printers}
\end{figure}

Thus far, the proposed method has been described using abstract ``colors'' within a conceptual ``palette.'' Figure~\ref{fig:printers} illustrates the three additive manufacturing systems used in this work: a two-material mixing hotend, a five-head tool-changer, and a single-material printer using temperature-responsive filaments. Each system employs a different process parameter such as material mixing ratios, discrete toolhead selections, or nozzle temperatures to achieve functional gradients. Full details of the hardware, materials, and G-code parameters are provided in Section~\ref{sec:materials_and_methods}. A central insight demonstrated by these systems is that our proposed slicing method generalizes broadly across diverse printing modalities. The underlying slicing strategy remains consistent across platforms, and only the specific G-code commands used to implement gradient transitions vary. By abstracting these hardware-specific instructions, our method delivers a fully automated, gradient-informed G-code generation process that eliminates manual intervention or the need for post-processing scripts.

\subsection{Color Palettes}
To evaluate the proposed method, an arbitrarily complex geometry with a two-material gradient is selected. Objects are printed with an increasing number of colors in the palette, demonstrating the application of both Strategy 1 and Strategy 2. Both strategies are employed for mixing hotend systems and temperature-responsive foaming PLA materials. Strategy 2 demonstrates the capability of achieving higher color counts without requiring purge towers. 

\begin{figure}[h!]
    \centering
    \includegraphics[width=\linewidth]{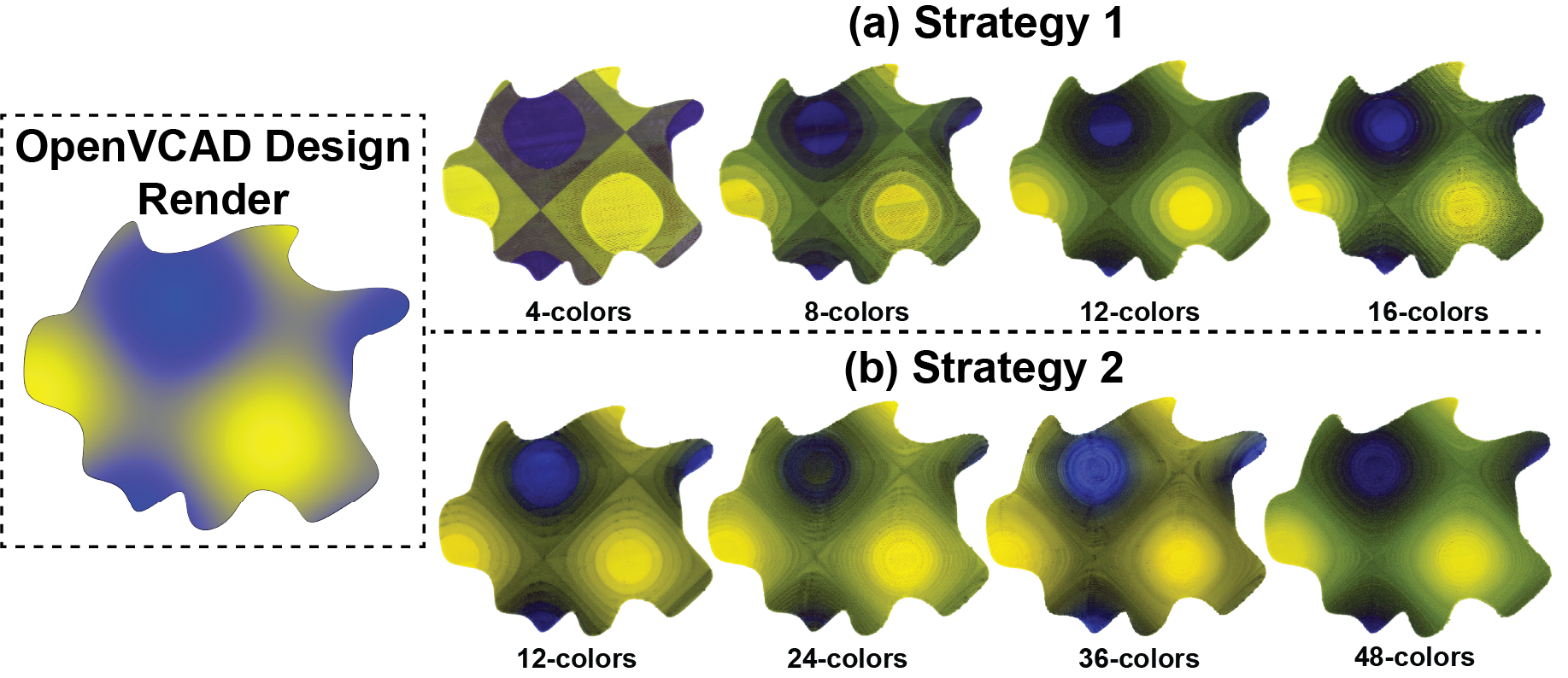}
    \caption{A comparison of the two strategies with increasing numbers of colors in the palette applied to mixing systems. (a) Strategy 1, objects were printed with 4, 8, 12, and 16 colors in the palette. (b) Strategy 2, objects were printed with 12, 24, 36, and 48 colors in the palette.}
    \label{fig:mixing_palette}
\end{figure}

\begin{figure}[h!]
    \centering
    \includegraphics[width=\linewidth]{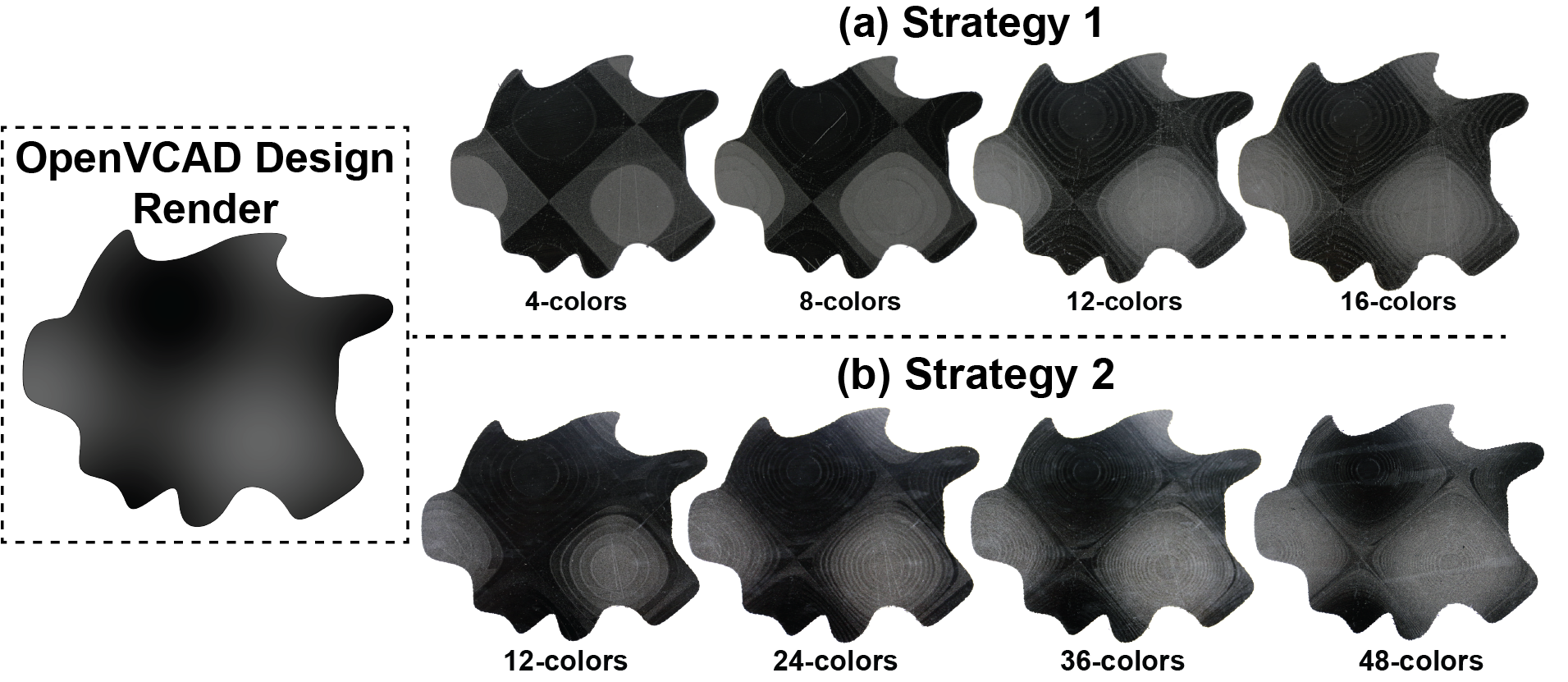}
    \caption{A comparison of the two strategies with increasing numbers of colors in the palette applied to foaming PLA. (a) Strategy 1, objects were printed with 4, 8, 12, and 16 colors in the palette. (b) Strategy 2, objects were printed with 12, 24, 36, and 48 colors in the palette.}
    \label{fig:foaming_palette}
\end{figure}

Figures \ref{fig:mixing_palette} (a) and \ref{fig:foaming_palette} (a) show the results of Strategy 1 applied to the mixing hotend system and foaming PLA, respectively. Objects are printed with 4, 8, 12, and 16 colors in the palette. As observed in these figures, increasing the number of colors enhances the visual quality of the gradient. However, a trade-off exists; the number of purge towers required equals the number of colors in the palette. Sixteen colors were chosen as the upper limit for these examples due to diminishing part quality from excessive segmentation, starts, and stops. Notably, larger objects would permit a greater number of palette colors due to larger individual path segment size in each region. In Figures \ref{fig:mixing_palette} (a) and \ref{fig:foaming_palette} (a) we  emphasize that the perimeter walls and infill structure were preserved, being segmented into different color regions.

Figures \ref{fig:mixing_palette} (b) and \ref{fig:foaming_palette} (b) demonstrate Strategy 2 applied to the same object and gradient. This method achieves significantly higher color counts within the palette while eliminating the need for purge towers. However, Strategy 2 sacrifices traditional toolpath structures, such as walls, skins, and infill, to prioritize gradient fidelity by only using concentric walls that produce a 100\% dense part. For foaming PLA, the diminishing visual gradient quality is attributed to two factors: the subtle color variation between unfoamed and maximally foamed states and the printer's hotend controller limitations.

\begin{figure}[h!]
    \centering
    \includegraphics[width=\linewidth]{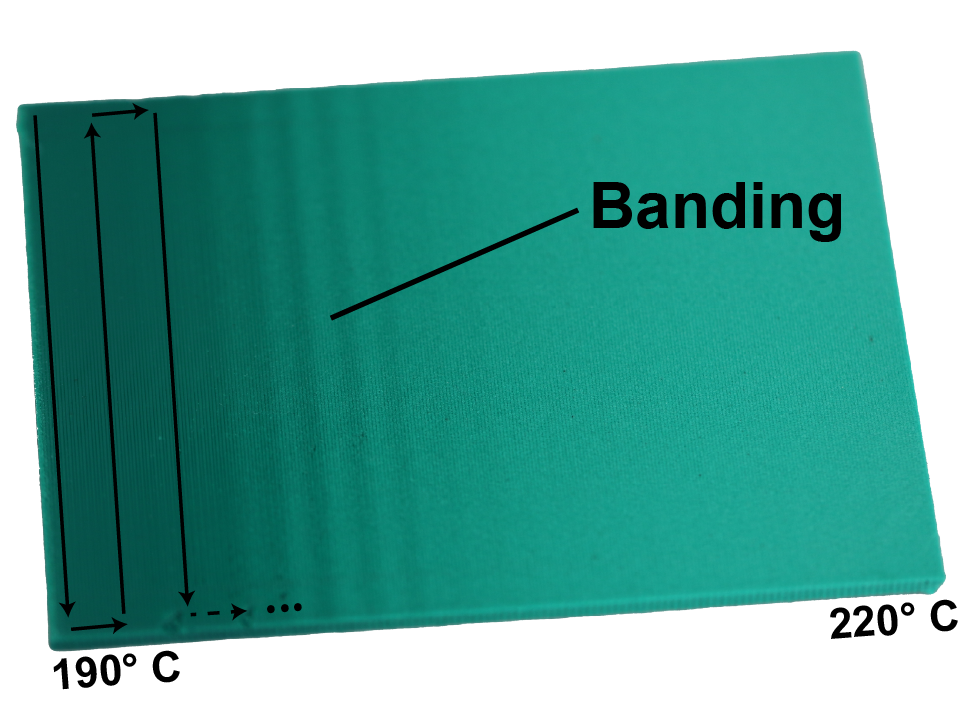}
    \caption{Color banding in a temperature-graded object printed with foaming TPU. The banding results from hotend controller overshooting.}
    \label{fig:color_banding}
\end{figure}

The challenges with hotend temperature control are further highlighted in Figure \ref{fig:color_banding}, which depicts a test object graded from 190°C (unfoamed) to 220°C (maximum foaming). Toolpaths are planned to minimize temperature changes between adjacent segments by following a serpentine pattern perpendicular to the gradient. Despite this optimization, distinct color banding is visible at the center of the object, caused by the hotend controller overshooting target temperatures. These results emphasize that typical controllers, designed for stable single-temperature operation, require retuning for dynamic grading applications. For this work, the following default PID values were used: $P=14.00$, $I=1.00$, and $D=100.00$. Further investigation is needed into optimal values when printing with continuously varying temperature.

\begin{figure}[h!]
    \centering
    \includegraphics[width=\linewidth]{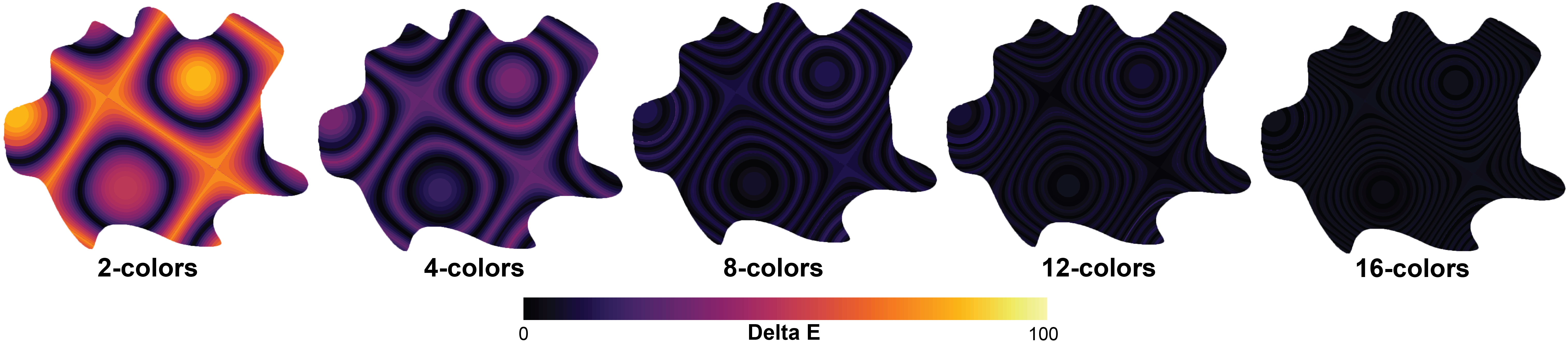}
    \caption{Pixel-wise Delta E comparison of five objects, each planned with an increasing number of colors in the palette. Lower Delta E values indicate better alignment with the reference image. The images illustrate significant improvements in gradient quality as the number of regions increases, with smoother transitions and reduced perceptual differences between colors. }
    \label{fig:delta_e_comparison}
\end{figure}

\begin{figure}[h!]
    \centering
    \includegraphics[width=\linewidth]{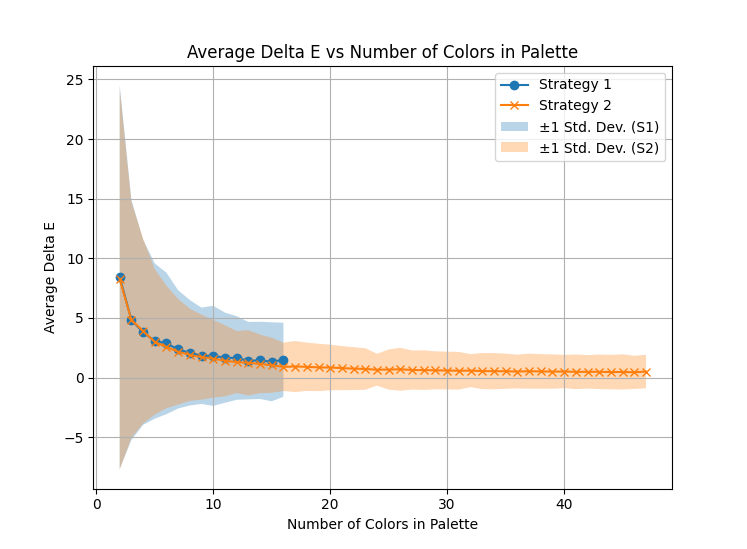}
    \caption{Comparison of the average Delta E for objects sliced with an increasing number of colors in their palette. Lower Delta E values correspond to better alignment with the reference image. The graph demonstrates rapid improvements in Delta E for both strategies as the number of regions increases, with diminishing returns observed beyond 16 regions. Strategy 2 outperforms Strategy 1 at higher color counts due to its capability to manage more regions effectively. The standard deviation of the Delta-e for each image sample is also given. }
    \label{fig:delta_e_graph}
\end{figure}

Figure \ref{fig:delta_e_graph} quantifies the gradient quality as the number of colors in the palette increases. The Figure plots the average Delta E for each printed object, where Delta E measures the perceptual difference between two colors.  Lower Delta E values indicate better alignment with the reference image, and in this context, higher accuracy gradients. Commonly used in color science, Delta E values below 1 are typically considered imperceptible to the human eye \cite{minaker_optimizing_2021}.  Figure \ref{fig:delta_e_comparison} visualizes pixel-wise Delta E for five objects planned using Strategy 1. These images illustrate substantial improvements in gradient quality as the number of regions increases. The corresponding average Delta E values for each image are plotted in Figure \ref{fig:delta_e_graph}, which also includes a comparison with Strategy 2 sampled up to 47 regions. From the graph, we observe rapid improvements in Delta E for both strategies as the number of regions increases, with diminishing returns beyond approximately 16 regions. Up to 16 colors, Strategies 1 and 2 perform comparably. However, Strategy 2 achieves superior gradient quality at higher color counts due to its ability to handle more regions.

To further understand the practical implications of color palette size, we measured the material consumption and total printing time associated with each strategy. Specifically, we quantified both additional filament usage due to purging and the increased print duration resulting from adding colors to the palette.

For Strategy 1 using the mixing system, the objects from Figure \ref{fig:mixing_palette} (without purging) requires 25.56 grams of material. Each additional color in the palette requires 1.9 grams of filament for purging. Consequently, the 16-color specimen consumed a total of 30.4 grams for purge towers, exceeding the material requirement of the primary print itself. It is important to highlight that purge tower volume and filament consumption depend significantly on both the specific printer hardware, and filament characteristics. For instance, substantially smaller purge volumes are sufficient for achieving stable foaming rates when using foaming PLA filaments, with only 0.16 grams required per additional palette color.

Printing duration also scales with palette size. For the mixing system, each additional color introduced to the palette adds approximately 15 minutes to the total print duration. The object in Figure \ref{fig:mixing_palette} takes 253 minutes to print, not including purge towers. In contrast, adding a color to the foaming filament prints increases printing time by only about 1.5 minutes per additional color, starting from a baseline of 161 minutes without purging. We note that these estimates are inherently printer-dependent and influenced by the maximum print speed achievable by the hardware.

In contrast, Strategy~2, when applied to sufficiently large objects, does not require purge towers, and incurs no additional material consumption or printing time penalties regardless of palette size. Thus, Strategy 2 provides an efficient option for maximizing palette fidelity without increasing resource consumption.
\subsection{Multi-axis Gradient Demonstration}

\begin{figure*}[h!]
    \centering
    \includegraphics[width=\linewidth]{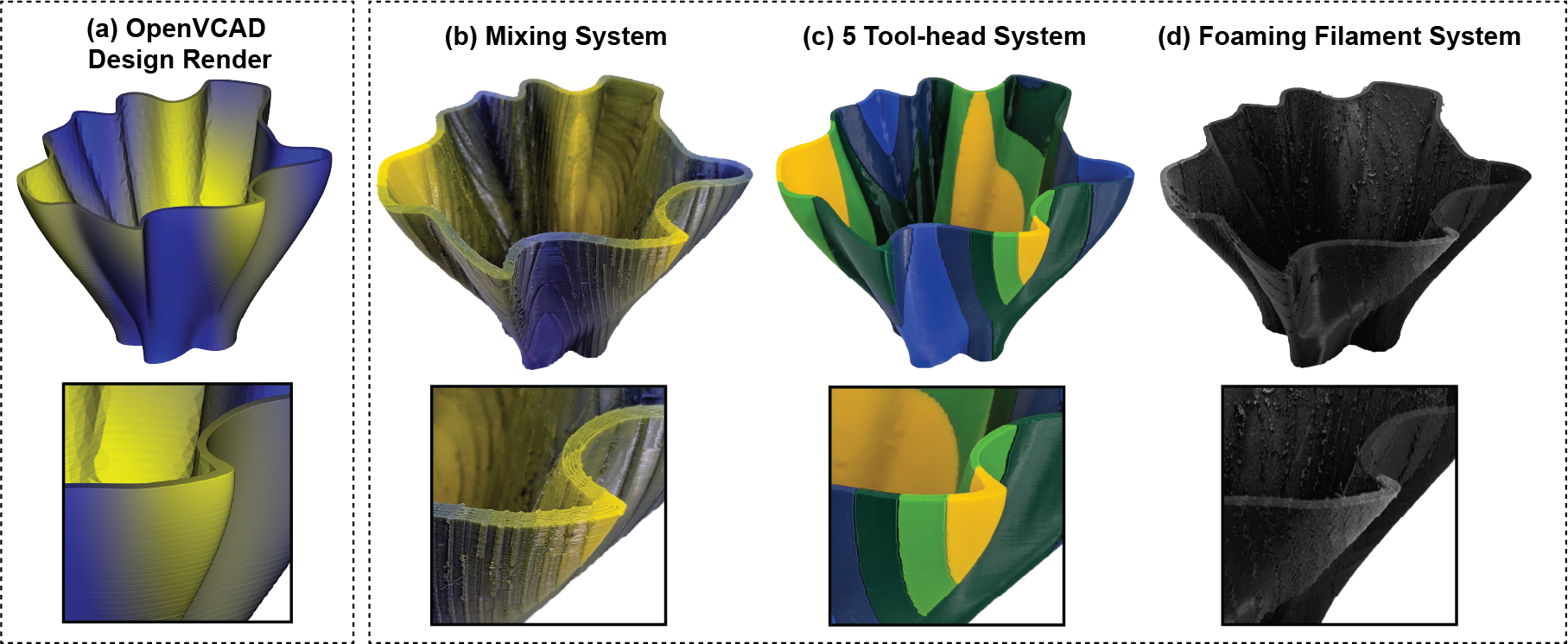}
    \caption{Demonstration of multi-axis gradient slicing versatility using a single OpenVCAD vase design with the gradient defined by Equation \ref{eq:color_distributions}. (a) OpenVCAD design preview. The design was sliced and fabricated using: (b) a filament mixing system employing a sixteen-color palette, (c) a discrete 5-toolhead MEX system transitioning between yellow and blue filaments, and (d) a single-material MEX system with temperature-responsive foaming filament exhibiting variable density. This example underscores the method's adaptability across diverse additive manufacturing systems and non-instantaneous process parameters.}
    \label{fig:vase}
\end{figure*}

To further demonstrate the versatility of the proposed method, a more complex, multi-axis gradient print was fabricated. The selected geometry, a vase shown in Figure
\ref{fig:vase}(a), exhibits intricate curvature and material variation across all three spatial dimensions. This example showcases diverse regions within the prescribed gradient (Equation\ref{eq:color_distributions}). 

Figure \ref{fig:vase}(b) shows the vase printed with a sixteen-color palette using Strategy 1 on a filament mixing system. The results clearly demonstrate the capability of the proposed slicing method to reproduce intricate, spatially varying gradients across complex surfaces. However, noticeable artifacts such as seams due to material oozing at interface boundaries were observed. These defects align with findings by Green et al., who noted similar deposition inconsistencies during travel moves \cite{green_local_2023}. Techniques aimed at improving extrusion consistency, including optimized retraction and coasting settings, might reduce such artifacts and enhance overall print quality \cite{charia_real-time_2025,jackson_multi-parameter_2022}.

Figure \ref{fig:vase}(c) demonstrates the same OpenVCAD gradient design applied to the 5-toolhead MEX system, using discrete filament colors transitioning from yellow to blue. Although this discrete approach inherently provides lower gradient resolution compared to continuous mixing systems, it illustrates the method's broad applicability and compatibility with widely available multichannel printing systems. Notably, while demonstrated here with a 5-toolhead configuration, the method readily generalizes to any multichannel G-code syntax using tool selection commands ($T0, T1, \dots, T_\mathrm{N}$) and can scale efficiently to systems with greater numbers of filament channels.

Additionally, Figure \ref{fig:vase}(d) showcases the application of the method to a single-material MEX system using temperature-responsive foaming filament. Here, nozzle temperature was spatially optimized and controlled using Strategy 1, resulting in variable density across the printed object. Although subtle, the resultant color variations correspond to the underlying material density changes, with lighter regions indicating increased foaming (lower density) and darker regions reflecting minimal foaming and greater density, consistent with the base polymer.

Figure \ref{fig:vase} collectively illustrates the versatility of the proposed slicing architecture across multiple printing systems and diverse non-instantaneous process parameters. A notable advantage of this approach is its fully automated workflow; our method, as implemented in the slicer, directly processes the OpenVCAD design alongside standard slicing parameters (e.g., layer height, printer specifications) and generates print-ready G-code without additional post-processing or manual intervention. This contrasts significantly with conventional multi-material printing workflows, which typically require manual generation and insertion of material transition commands or intermediate boundary surfaces, thus substantially simplifying and accelerating gradient fabrication.

\begin{figure}[h!]
    \centering
    \includegraphics[width=\linewidth]{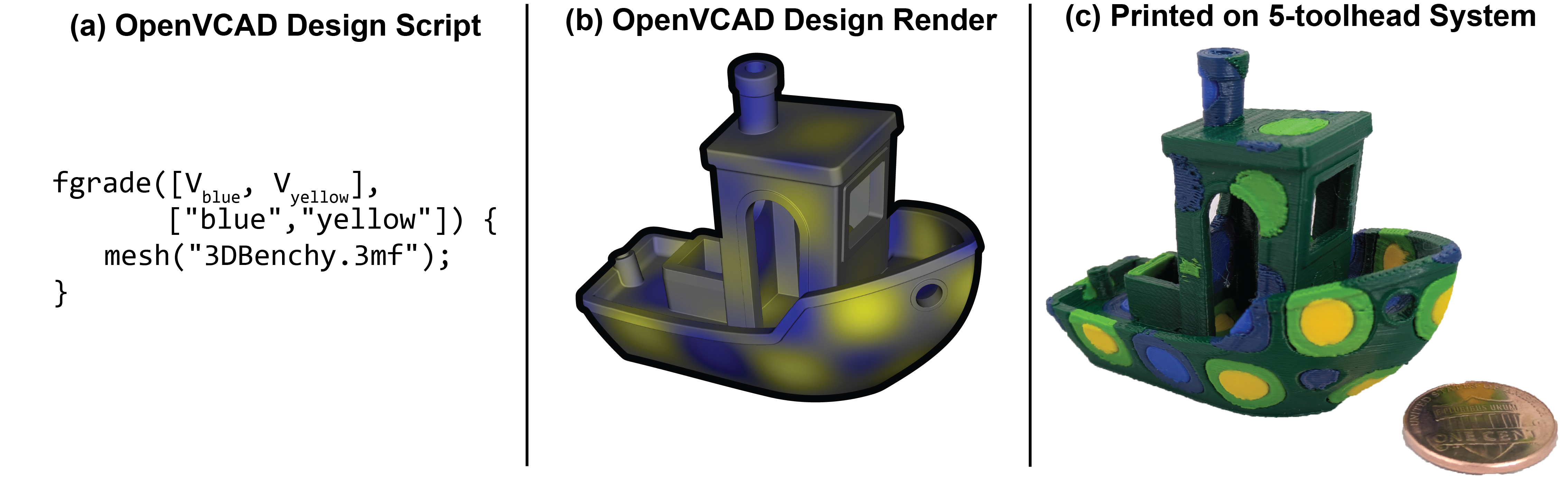}
    \caption{Demonstration of the 3D Benchy model with a spatially complex gradient defined by Equation \ref{eq:3d_gradient}. (a) and (b) The computational design illustrating the intended gradient. (c) The resulting print fabricated on a Prusa XL, showcasing an accurate gradient transition from yellow to blue filaments.}
    \label{fig:benchy}
\end{figure}

To further illustrate the versatility and effectiveness of the proposed method for the 5-toohead system, a 3D Benchy model was printed using a spatially complex gradient defined by Equation~\ref{eq:3d_gradient}~\cite{3DBenchy2015}. For this demonstration, a gradient density parameter of $\alpha = 25\,\mathrm{mm}$ was chosen, determining the scale and frequency of the gradient variation. The Benchy geometry was sliced using Strategy 1 and fabricated using the Prusa XL, employing a two-``color'' gradient transitioning from yellow to blue filaments. As shown in Figure~\ref{fig:benchy}, the printed model accurately reproduces the intended spatial gradient, clearly illustrating precise filament placement. Notably, our process directly converted a single input mesh and 3D gradient function to G-code, eliminating the need for intermediate boundary surfaces required by conventional methods, thereby significantly simplifying the workflow for multi-material gradient fabrication.

\begin{equation}
\begin{aligned}
    V_{\text{blue}} &= \frac{1}{2} + \frac{1}{2}\sin\left(\frac{2\pi x}{\alpha}\right)\cos\left(\frac{2\pi y}{\alpha}\right)\sin\left(\frac{2\pi z}{\alpha}\right), \\
    V_{\text{yellow}} &= 1 - \left[\frac{1}{2} + \frac{1}{2}\sin\left(\frac{2\pi x}{\alpha}\right)\cos\left(\frac{2\pi y}{\alpha}\right)\sin\left(\frac{2\pi z}{\alpha}\right)\right].
\end{aligned}
\label{eq:3d_gradient}
\end{equation}

\subsection{Tensile Testing}

To evaluate the mechanical performance implications of our proposed slicing strategies, we conducted tensile tests on specimens fabricated using each method. These tests isolate the impact of toolpath planning on tensile strength by using identical material conditions and varying only filament color, thereby attributing observed differences to slicing methodology alone. All specimens followed use ASTM D638 Type I geometry, with full details given in Section \ref{sec:materials_and_methods}.

Specimens were divided into five groups to facilitate comparison of slicing strategies:

\begin{enumerate}[label=\textbf{(\alph*)}]
    \item \textbf{Single Material (Baseline)}: Printed in a single filament color using PrusaSlicer. These specimens represent maximum achievable strength due to the absence of internal interfaces.

    \item \textbf{PrusaSlicer Multi-Color (Baseline)}: Three-color specimens manually segmented in CAD and sliced using PrusaSlicer. This group demonstrates typical multi-color performance achievable with conventional slicing techniques.

    \item \textbf{Strategy 1 (Without Zippering)}: Specimens printed using Strategy~1 (Section~\ref{sec:strategy_1}) segmented into three colors without applying zippering.

    \item \textbf{Strategy 1 (With Zippering, $\beta = 15\%~\text{or aprox.}~ 15$~mm)}: Strategy~1 specimens with a 15~mm zipper overlap applied at interfaces (Section~\ref{sec:strategy_1_zipper}). This test evaluates the mechanical benefit of the zippering technique. 15~mm was chosen as the overlap distance based on the findings of Sardinha et al. \cite{sardinha2024multi_extrusion}.

    \item \textbf{Strategy 2}: Specimens fabricated using Strategy~2 (Section~\ref{sec:strategy_2}) segmented into three colors. This strategy prioritizes gradient continuity at the expense of traditional wall and infill structures.
\end{enumerate}

\begin{figure}[]
    \centering
    \includegraphics[width=\linewidth]{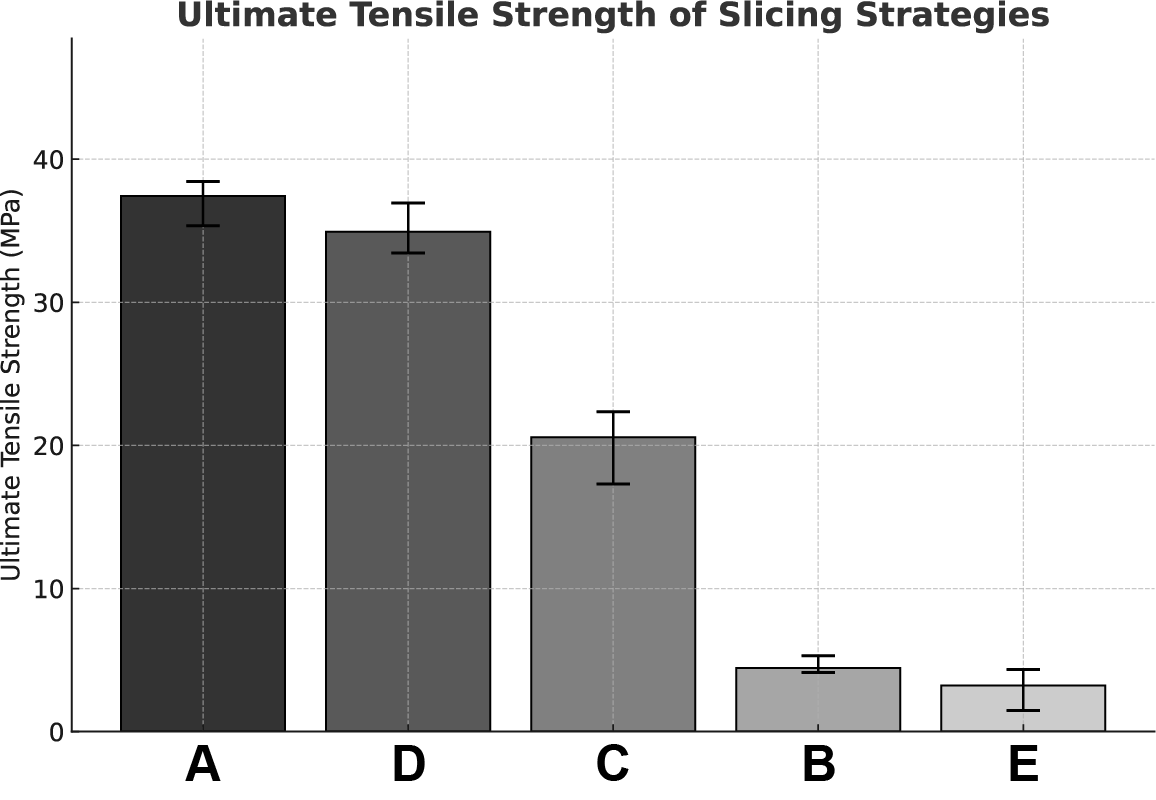}
    \caption{Ultimate tensile strength (UTS) comparison across slicing strategies. Bars represent mean values, and error bars indicate minimum and maximum values across five replicates per group.}
    \label{fig:tensile_results}
\end{figure}

\begin{figure}[]
    \centering
    \includegraphics[width=\linewidth]{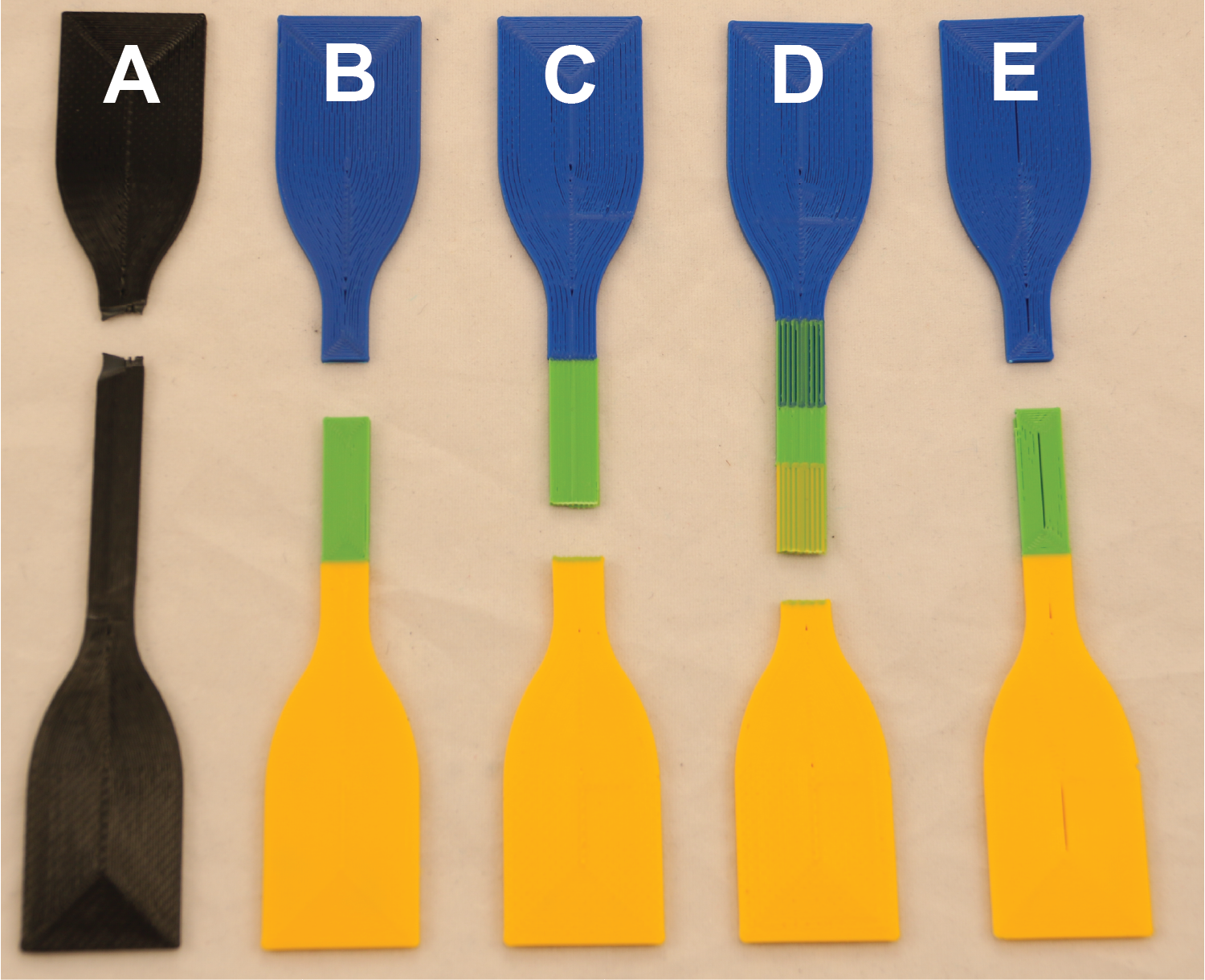}
    \caption{Fractures specimens for each of the five test groups (A-E). The color specimens always fail at the material interface. }
    \label{fig:fractures}
\end{figure}

Each group comprised five test specimens. The results, summarized in Figure~\ref{fig:tensile_results}, illustrate average ultimate tensile strength (UTS) with error bars showing the observed minimum and maximum values. Figure~\ref{fig:fractures} shows fractured specimens from each of the 5 groups. We observe that for the multi-color groups, the specimen always failed at the interface between two different filaments.

The single-material specimens exhibited the highest tensile strength, confirming that internal material interfaces negatively impact mechanical performance. Multi-color specimens sliced using conventional methods (PrusaSlicer) displayed notably reduced tensile strength, similar to those produced using Strategy~2.

Specimens sliced with Strategy~1 without zippering showed approximately half the tensile strength of single-material prints but significantly outperformed conventional multi-color methods. Introducing a 5~mm zipper overlap in Strategy~1 significantly improved tensile strength, nearly reaching single-material performance levels.

These tensile tests confirm that Strategy~1 with zippering significantly mitigates the mechanical degradation at interfaces, achieving near single-material strength. Strategy~1 without zippering still offers considerable mechanical improvement over conventional slicing techniques, whereas Strategy~2 provides high-quality gradients while exacting a mechanical cost.

\subsection{Runtime Benchmarking of the Slicer}
To assess the computational scalability of the proposed slicing methods, Table~\ref{tab:benchmarking} summarizes wall-clock slicing times on a convention desktop system (details given in Section \ref{sec:materials_and_methods}). The largest object, the multi-axis vase (500 layers, Strategy 1, 16 colors), required approximately 21 minutes (1271 s) to slice. The smaller 3D Benchy (240 layers, Strategy 2, 5 colors) took around 13 minutes (754 s). For the palette test samples (10 layers), slicing times increased nearly linearly with palette size, ranging from 33 s (4 colors) to 99 s (16 colors) using Strategy 1, with zippering approximately doubling the slicing duration (219 s). Strategy 2 exhibited a similar trend, scaling up to 424 s at 48 colors. These results indicate that slicing time primarily scales with the number of layers and the complexity introduced by increased palette size. However, even the longest observed slicing time remains small relative to typical printing durations (several hours), confirming the practical scalability of the proposed methods. Further improves to speed could be possible with the addition of parallelization to the slicer.

\begin{table*}[]
\centering
\resizebox{\linewidth}{!}{%
\begin{tabular}{|l|c|c|c|c|c|c|}
\hline
\rowcolor[HTML]{C0C0C0} 
{\color[HTML]{000000} Object} & \multicolumn{1}{l|}{\cellcolor[HTML]{C0C0C0}{\color[HTML]{000000} Object Size (mm)}} & \multicolumn{1}{l|}{\cellcolor[HTML]{C0C0C0}{\color[HTML]{000000} Layers}} & \multicolumn{1}{l|}{\cellcolor[HTML]{C0C0C0}{\color[HTML]{000000} Slicing Strategy}} & \multicolumn{1}{l|}{\cellcolor[HTML]{C0C0C0}{\color[HTML]{000000} Colors in Palette}} & \multicolumn{1}{l|}{\cellcolor[HTML]{C0C0C0}{\color[HTML]{000000} Zippering}} & \multicolumn{1}{l|}{\cellcolor[HTML]{C0C0C0}{\color[HTML]{000000} Time (s)}} \\ \hline
Vase (Fig. \ref{fig:vase}) & 135 x 175 x 100 & 500 & 1 & 16 & No & 1271 \\ \hline
3D Benchy (Fig. \ref{fig:benchy}) & 60 x 31 x 48 & 240 & 2 & 5 & NA & 754 \\ \hline
Palette (Fig. \ref{fig:mixing_palette}) & 135 x 175 x 2 & 10 & 1 & 4 & No & 33 \\ \hline
Palette (Fig. \ref{fig:mixing_palette}) & 135 x 175 x 2 & 10 & 1 & 8 & No & 54 \\ \hline
Palette (Fig. \ref{fig:mixing_palette}) & 135 x 175 x 2 & 10 & 1 & 12 & No & 76 \\ \hline
Palette (Fig. \ref{fig:mixing_palette}) & 135 x 175 x 2 & 10 & 1 & 16 & No & 99 \\ \hline
Palette (Fig. \ref{fig:mixing_palette}) & 135 x 175 x 2 & 10 & 1 & 16 & Yes & 219 \\ \hline
Palette (Fig. \ref{fig:mixing_palette}) & 135 x 175 x 2 & 10 & 2 & 12 & NA & 86 \\ \hline
Palette (Fig. \ref{fig:mixing_palette}) & 135 x 175 x 2 & 10 & 2 & 24 & NA & 174 \\ \hline
Palette (Fig. \ref{fig:mixing_palette}) & 135 x 175 x 2 & 10 & 2 & 36 & NA & 285 \\ \hline
Palette (Fig. \ref{fig:mixing_palette}) & 135 x 175 x 2 & 10 & 2 & 48 & NA & 424 \\ \hline
\end{tabular}%
}
\caption{Slicing times for Figures \ref{fig:mixing_palette}, \ref{fig:vase}, and \ref{fig:benchy}.}
\label{tab:benchmarking}
\end{table*}

\section{Discussion}
\subsection{Tensile Testing}
The results of tensile testing support prior findings: material interfaces  (indicated by materials of different colors) create weak points and interface zippering can help mitigate these weaknesses. Our findings highlight that print applications will drive design choice. If strength is the primary objective, single material or zippered multi-material with our Strategy 1 will be best. However, when strength is not a priority, Strategy 2 may produce more accurate gradients and less material waste in manufacturing.

\subsection{Summary of Contribution}
Figure~\ref{fig:big_comparison} illustrates the capabilities of our proposed slicing method across multiple gradient types, systems, and process parameters. Columns one through four represent prior work in single-axis gradients along different spatial orientations, while column five highlights our novel contribution:  slicing for arbitrary 3D, multi-axis gradients and geometries. Boxes shaded in light blue indicate gradient configurations demonstrated for the first time in this work, whereas white boxes represent gradient capabilities previously established in the literature. It is worth highlighting that these prior works required custom manually-generated G-code for each different geometry. Additionally, this research provides the first demonstration of radial and circumferential gradients specifically tailored for temperature-responsive foaming filaments.

It is important to emphasize that all examples depicted in Figure \ref{fig:big_comparison}, including previously reported gradients, were successfully fabricated using our automated approach. This underscores that our proposed method not only encompasses but substantially extends existing gradient-printing techniques. A key strength of our methodology lies in its complete automation and generalized applicability: a single OpenVCAD design can be directly sliced into ready-to-print G-code for diverse printing systems and gradient types, without the need for manual design of specific G-code or intermediate boundary-surface processing.

\subsection{Extending to Other Additive Manufacturing Processes}
Our method for gradient-informed slicing is applicable to all G-code-based system, so we focus this section on the extensions to the open-source code that would be necessary to create executable G-code for a system not currently supported. Users must adapt the output to match the system’s command syntax and select the appropriate instructions for controlling process parameters. Some parameters might include ink mixture ratios in DIW or laser power and wire-feed rates in DED and are analogous to the parameter changes shown for MEX systems in Figure~\ref{fig:printers}. Within our open-source slicer, this primarily involves modifying the \texttt{gcode\_writer.py} file, which handles commands such as linear moves (\texttt{G1}). The key function in this file, \texttt{write\_mixing\_ratios()}, is invoked whenever the slicer specifies a change in mixture, temperature, or toolhead, and must be extended to support system-specific parameters and syntax. For example, Sevcik et al. describe a two-material DIW system built on a Prusa i3 MK3S+ platform \cite{sevcik_dual_2024}. While this system uses the standard motion commands, it relies on a custom command (\texttt{M42}) to control material fractions rather than the \texttt{M165} command used in our setup. Supporting such a system requires only updating the \texttt{write\_mixing\_ratios()} function to issue \texttt{M42} commands in place of \texttt{M165}

\section{Conclusion and Future Work}
In this work, we introduced the first fully automated slicing framework for functionally graded additive manufacturing, capable of converting arbitrarily complex, three-dimensional gradient designs directly into printer-ready G-code. Our proposed method addresses a critical bottleneck, manual gradient G-code generation, significantly simplifying the workflow from volumetric gradient design to physical fabrication on any G-code–based additive manufacturing platform.

Experimental results demonstrated successful application across different material extrusion systems, including mixing hotends, multi-toolhead setups, and single-material printers using temperature-responsive foaming filaments. Tensile testing confirmed that Strategy~1, combined with automated zippering, effectively mitigates mechanical weaknesses at material interfaces, achieving tensile strengths comparable to single-material prints while printing with a gradient. While Strategy~2 prioritizes gradient continuity and visual fidelity, it does so at the expense of mechanical strength.

Future work will focus on methodological improvements and expanding practical applications. Methodologically, there is substantial room for enhancing print efficiency and quality. A deeper investigation into eliminating or significantly reducing purge towers by carefully managing color transitions and toolpath sequencing could drastically improve printing efficiency and reduce waste. Additionally, exploring more advanced zippering geometries and strategies can further strengthen material interfaces, particularly when combining dissimilar materials or materials with challenging adhesion characteristics.

From an applications perspective, our open-source framework aims to accelerate research into functionally graded materials by providing tools to explore complex gradients rapidly. Future work will leverage this capability to print graded electronics, sensors, and radio-frequency components using direct ink write, where precise spatial control of conductive and dielectric inks can yield devices with tailored electromagnetic properties. Additionally, demonstrating our methods with fundamentally different and dissimilar materials, such as elastomers (TPU) combined with rigid polymers (PLA), will further expand the application domains, enabling new designs in soft robotics and complaint mechanisms.

By releasing our slicing methods as an open-source extension to OpenVCAD, we seek to foster collaboration and accelerate progress within the additive manufacturing community, addressing critical limitations in gradient design and fabrication capabilities across diverse printing technologies and materials. By providing a robust slicing method capable of handling complex gradients and continuous material transitions, our work directly addresses a significant barrier previously limiting application-driven research in multi-material additive manufacturing.

\section{Materials and Methods}
\label{sec:materials_and_methods}
\subsection{Hardware, Materials, and Process Settings}
We apply our method to a two-material mixing hotend system to achieve functional gradients through controlled material mixtures. For these experiments, we use an inexpensive Geeetech Mizar M printer equipped with a passive mixing hotend \cite{geeetech_mizar_m}. As illustrated in Figure \ref{fig:printers}, two base filaments (blue and yellow) were mixed at varying ratios. Mixture control is accomplished through the Marlin firmware's \texttt{M165} G-code instruction, which specifies the relative proportions of each extruder feeding into a single nozzle. The command syntax follows the format: \texttt{M165 [A<factor>] [B<factor>] ...}, where each factor represents the volume fraction from the corresponding extruder. For example, the command \texttt{M165 A0.25 B0.75} sets a mixture ratio of 25\% from extruder A and 75\% from extruder B. Our method automatically inserts this command into the G-code stream prior to printing each mixture region, ensuring proper stabilization of the desired ratio. 

We also demonstrate the method's applicability to controlling temperature gradients using temperature-responsive foaming filaments on a single-material MEX system. By optimizing print order during slicing, we minimize abrupt temperature changes between adjacent regions. Temperature control is implemented using the standard G-code command \texttt{M104}, as illustrated in Figure~\ref{fig:printers}. This optimization reduces both the stabilization delay and the amount of purged material necessary for achieving consistent foaming. For these experiments, we use ColorFabb LW-PLA and varioShore TPU foaming filaments \cite{colorfabb_lwpla_black, colorfabb_varioshore_green}, printed on a standard Prusa MK4S within an enclosure \cite{prusa_mk4s_enclosure_bundle}. Since OpenVCAD represents objects as both geometry and continuous fields of base material volume fractions, a mapping was required to translate this abstract two-material gradient space into corresponding nozzle temperatures for foaming control. We employed a linear mapping in which 100\% material A was assigned to 190°C (representing the unfoamed base state) and 100\% material B was assigned to 225°C (corresponding to the maximum foaming condition). Intermediate material mixtures were linearly interpolated within this temperature range, allowing the implicit gradient design to drive spatially controlled variations in print density.

Finally, we demonstrate our method on a conventional multi-material printer equipped with five independent toolheads, each loaded with a different filament color, spanning from blue to yellow. For this experiment, we use a Prusa XL printer configured with a 5-toolhead setup~\cite{prusa_xl_toolchanger}. Material selection and switching are handled using the \texttt{T<N>} G-code command, where \texttt{N} identifies the toolhead index. For instance, issuing the \texttt{T3} command prompts the printer to dock the currently active toolhead and activate the third indexed toolhead. The Prusa XL system indexes toolheads from \texttt{T0} to \texttt{T4}. This demonstration highlights our slicing approach's general applicability to widely-used multi-channel systems.

\subsection{Foaming Filament Expansion Compensation Details}
When printing with foaming filaments, it is necessary to compensate for the increased bead width as the temperature rises. Although standard slicers, such as Prusa Slicer, provide a setting to control flow rate, they do not support dynamic control over the flow during the print. Equations~\ref{eq:flow_rate_pla} and \ref{eq:flow_rate_tpu} were used to adjust for expansion based on the nozzle temperature, $T$. Our slicer automatically controls flow rate to compensate for foaming expansion by using the \texttt{M221 T0 S<$F$>} G-code command in conjunction with each \texttt{M104} temperature update. 

\begin{equation}
\label{eq:flow_rate_pla}
\begin{aligned}
\text{F}_{\mathrm{PLA}}
  &= 100\Bigl(
       8.35479\times10^{-6}\,t^{3}
       - 5.37075\times10^{-3}\,t^{2} \\[4pt]
  &\hphantom{= 100\Bigl(}
       + 1.13374\,t
       - 77.814
     \Bigr)
\end{aligned}
\end{equation}

\begin{equation}
\label{eq:flow_rate_tpu}
\begin{aligned}
\text{F}_{\mathrm{TPU}}
  &= 100\Bigl(
       3.09637\times10^{-4}\,t^{2}
       - 1.38401\times10^{-1}\,t \\[4pt]
  &\hphantom{= 100\Bigl(}
       + 15.9560
     \Bigr)
\end{aligned}
\end{equation}

\subsection{Slicing Settings}
For all objects presented in this work, a layer height of $0.2~\mathrm{mm}$ and a nozzle diameter/bead width of $0.4~\mathrm{mm}$ were used. The \textit{Marlin} G-code syntax was used for all systems~\cite{marlin_gcode_syntax}.

\subsection{Vase Object (Figure~\ref{fig:vase}) Details}
The vase measures $135\,\mathrm{mm}$ in width, $175\,\mathrm{mm}$ in depth, and $100\,\mathrm{mm}$ in height. The model was designed as a solid object; however, for Figure~\ref{fig:vase} it was sliced with 10 perimeters and infill disabled to create a hollow structure.

\subsection{Color Palette Objects (Figures~\ref{fig:mixing_palette} and \ref{fig:foaming_palette}) Details}
The test object is a top cross-section slice of the vase, measuring $135\,\mathrm{mm}$ in width, $175\,\mathrm{mm}$ in depth, and $2\,\mathrm{mm}$ in height. Unlike the vase in Figure~\ref{fig:vase}, the color palette objects were printed with 3 perimeters and 100\% infill.

\subsection{Delta-E Calculation Details}
For the Delta-E analysis, the object generated using Strategy~2 with 48 colors was selected as the reference image because it provided the highest-quality print. In principle, the OpenVCAD render could be used as the reference; however, developing an accurate color mapping from render to printed part is outside the scope of this work. For each pixel, Delta-E was computed using aligned, top-down photographs of the prints. A fixed camera and jig ensured pixel-perfect alignment between the reference image and the samples.

\subsection{Tensile Testing Details}
All tensile specimens followed the ASTM D638 Type~I geometry, measuring $138\,\mathrm{mm}$ in length, $22.8\,\mathrm{mm}$ in gauge width, and $1.5\,\mathrm{mm}$ in thickness (8 layers at $0.2\,\mathrm{mm}$ layer height). ColorFabb High-Speed PLA filament was used, with a $0.4\,\mathrm{mm}$ nozzle and concentric infill at 100\% density.

\subsection{Slicer Benchmarking Details}
Benchmark times in Table~\ref{tab:benchmarking} were measured on a desktop system with an AMD~Ryzen~7~7700X CPU, 65~GB of RAM, running Windows~11. The slicer implementation uses a single thread in Python~3.12. All values are wall-clock times representing the total duration to convert an OpenVCAD design into sliced G-code.

\section*{Data and material availability}
All data needed to evaluate the conclusions in the paper are present
in the paper and/or the Supplementary Materials as well as provided
on the VCAD-Slicer GitHub repository (\href{https://github.com/MacCurdyLab/VCAD-Slicer}{https://github.com/MacCurdyLab/VCAD-Slicer}) at commit hash \texttt{c663b3ab0c9ea06db10e4966e661fac9be357ef9}. Additional data related to this paper may be requested from the corresponding author.

\section*{Research Support}
This material is based upon work supported by the Charles Stark Draper Laboratory, Inc. under Contract No. N00030-24-C-6001. Any opinions, findings and conclusions or recommendations expressed in this material are those of the author(s) and do not necessarily reflect the views of Strategic Systems Programs.

\section*{CRediT authorship contribution statement}
\textbf{Charles Wade:} Writing - review \& editing, Writing - original  draft, Validation, Methodology, Investigation, Formal analysis, Data curation, Conceptualization, Software. \textbf{Devon Beck}: Writing - review \& editing, Conceptualization, Supervision. \textbf{Robert MacCurdy}: Writing - review \& editing, Conceptualization, Methodology, Project administration, Supervision, Resources

\section*{Declaration of competing interest}
The authors have no competing interests.

\appendix
\section{Procedure for Compensating for Dead Volume}
This appendix outlines procedures for determining the necessary look-ahead distance to compensate for dead volume effects during gradient-based 3D printing. Accurate compensation is essential for mixing systems, where residual melt-chamber volume affects material mixtures, as well as for systems using temperature-responsive foaming filaments, where delayed thermal responses can cause unintended over- or under-foaming. Two calculation methods are provided: the first for systems where the melt chamber geometry and volume are explicitly known, allowing direct computation; and the second for commercial or otherwise closed-source systems where the exact melt chamber volume is unknown and must be empirically determined.

\subsection{Calculating Look-ahead Distance for a Known Melt Chamber Volume}
To ensure the entire melt chamber volume is purged, a sufficient length of material must be extruded. The required extrusion distance is derived by considering the cross-sectional geometry of the deposited filament. For this purpose, a capsule model is employed, where the filament's cross section is approximated as a rectangular prism with two semicircular ends. The rectangular portion has a width of $(w - h)$ and a height of $h$, while the semicircular ends each have a radius of $h/2$. This model provides an accurate representation of the extruded filament's geometry on the build surface \cite{percoco_analytical_2021}.

Given:
\begin{itemize}
    \item $h$: Layer height.
    \item $w$: Bead width.
    \item $V_{\text{melt}}$: Melt chamber volume to be purged.
\end{itemize}

The total toolpath length $L$ required to extrude volume $V_{\text{melt}}$ is computed by dividing the purge volume by the cross-sectional area $A$. Using the capsule model, the cross-sectional area $A$ is:
\[
A = h(w - h) + \frac{\pi h^2}{4}.
\]

The extrusion length $L$ needed to dispense $V_{\text{melt}}$ is therefore:
\begin{equation}
    \label{eq:ExtrusionLength}
    L = \frac{V_{\text{melt}}}{h(w - h) + \frac{\pi h^2}{4}}.
\end{equation}

This toolpath length $L$ ensures that the melt chamber volume $V_{\text{melt}}$ is fully purged at the specified layer height $h$ and bead width $w$. This value can be used as the look-ahead distance for Strategy 2 or as the minimum purge tower size for Strategy 1.

\subsection{Calculating Look-ahead Distance for an Unknown Melt Chamber Volume}
For cases where the melt chamber volume is unknown, the look-ahead distance can be determined experimentally using visual inspection and iteration. We employed the following procedure:
\begin{enumerate}
    \item \textbf{Initial Test Print}: Print the object shown in Figure \ref{fig:lookadhead}(a) with zero look-ahead distance, where the mixture ratio change command is issued precisely at the midpoint of the object.
    \item \textbf{Identify the Error Distance}: Using visual inspection, determine the x-location $x_{\text{error}}$ at which the observed change in mixture ratio begins. The error distance $e_x$ is the horizontal offset between the midpoint and $x_{\text{error}}$:
    \[
    e_x = |x_{\text{midpoint}} - x_{\text{error}}|.
    \]
    \item \textbf{Calculate the Number of Incorrect Segments}: Given the bead width $w$, calculate the total number of incorrectly colored segments:
    \[
    s_{\#} = \text{ceil}\left(\frac{e_x}{w}\right).
    \]
    \item \textbf{Determine the Look-ahead Distance}: Compute the total incorrect extrusion distance as:
    \[
    L = s_{\#} \times L_y,
    \]
    where $L_y$ is the extrusion length for a single segment along the y-direction.
    \item \textbf{Iterative Refinement}: Repeat this process with updated look-ahead distances until the gradient's position aligns with the design intent. Each iteration should reduce the observed error distance $e_x$. If greater resolution is needed, decrease the object's length, $L_y$.
\end{enumerate}

This iterative approach enables the determination of an accurate look-ahead distance for printers with unknown melt chamber volumes. By systematically reducing the error distance, the gradient can be calibrated for a printer and ensure all objects have their gradients aligned with the OpenVCAD design.

\section{OpenVCAD Code for Objects}
This appendix gives the OpenVCAD code used to generate the objects in Figure \ref{fig:big_comparison}.
\lstdefinestyle{mystyle}{
    backgroundcolor=\color{gray!10},
    basicstyle=\ttfamily\small,
    frame=single,
    breaklines=true,
    tabsize=4,
    keywordstyle=\color{blue},
    stringstyle=\color{teal},
    commentstyle=\color{gray}
}

\begin{lstlisting}[style=mystyle, language=, caption={OpenVCAD Code for Figure~\ref{fig:big_comparison} Column 1}]
fgrade(["z/70", "1-z/70)"],
       ["blue", "yellow"]) {
    cylinder(15, 70);
}
\end{lstlisting}

\begin{lstlisting}[style=mystyle, language=, caption={OpenVCAD Code for Figure~\ref{fig:big_comparison} Column 2}]
fgrade(["y/75 + 0.5", "0.5 - y/75"], 
       ["blue",       "yellow"]) {
    rectprism(150,75,2.5);
}
\end{lstlisting}

\begin{lstlisting}[style=mystyle, language=, caption={OpenVCAD Code for Figure~\ref{fig:big_comparison} Column 3}]
fgrade(["abs(phi)/pi", "1 - abs(phi)/pi"],
       ["blue",         "yellow"]) {
    difference() {
       cylinder(50, 15);
       cylinder(15, 15);
    }
}
\end{lstlisting}

\begin{lstlisting}[style=mystyle, language=, caption={OpenVCAD Code for Figure~\ref{fig:big_comparison} Column 5}]
fgrade(["(rho-15)/35", "1 - ((rho-15)/35)"], 
       ["blue",        "yellow"]) {
    difference() {
        cylinder(50, 10);
        cylinder(15, 10);
    }
}
\end{lstlisting}

\begin{lstlisting}[style=mystyle, language=, caption={OpenVCAD Code for Figure~\ref{fig:big_comparison} Column 6}]
fgrade(["(1+sin(0.025*x+0.0375*y)*
            cos(0.0375*x-0.025*y))/2",
        "1-(1+sin(0.025*x+0.0375*y)*
            cos(0.0375*x-0.025*y))/2"],
       ["yellow","blue"]){
     mesh("Vase.stl");
}
\end{lstlisting}

\bibliographystyle{elsarticle-num} 
\bibliography{references}

\begin{thebibliography}{10}
\expandafter\ifx\csname url\endcsname\relax
  \def\url#1{\texttt{#1}}\fi
\expandafter\ifx\csname urlprefix\endcsname\relax\def\urlprefix{URL }\fi
\expandafter\ifx\csname href\endcsname\relax
  \def\href#1#2{#2} \def\path#1{#1}\fi

\bibitem{prusa_mmu3}
Prusa3D, Original prusa mmu3,
  \url{https://www.prusa3d.com/category/original-prusa-mmu3/}, accessed:
  2024-12-11 (2024).

\bibitem{bambu_ams}
B.~Lab, Ams -- multi-color and multi-material printing,
  \url{https://us.store.bambulab.com/products/ams-multicolor-printing},
  accessed: 2024-12-11 (2024).

\bibitem{li_review_2020}
Y.~Li, Z.~Feng, L.~Hao, L.~Huang, C.~Xin, Y.~Wang, E.~Bilotti, K.~Essa,
  H.~Zhang, Z.~Li, F.~Yan, T.~Peijs, A {Review} on {Functionally} {Graded}
  {Materials} and {Structures} via {Additive} {Manufacturing}: {From}
  {Multi}‐{Scale} {Design} to {Versatile} {Functional} {Properties}, Advanced
  Materials Technologies 5~(6) (2020) 1900981.
\newblock \href {https://doi.org/10.1002/admt.201900981}
  {\path{doi:10.1002/admt.201900981}}.

\bibitem{wade_openvcad_2024}
C.~Wade, G.~Williams, S.~Connelly, B.~Kopec, R.~MacCurdy, {OpenVCAD}: {An} open
  source volumetric multi-material geometry compiler, Additive Manufacturing 79
  (2024) 103912.
\newblock \href {https://doi.org/https://doi.org/10.1016/j.addma.2023.103912}
  {\path{doi:https://doi.org/10.1016/j.addma.2023.103912}}.

\bibitem{leoni_functionally_2023}
F.~Leoni, P.~Dal~Fabbro, S.~Rosso, L.~Grigolato, R.~Meneghello, G.~Concheri,
  G.~Savio, Functionally {Graded} {Additive} {Manufacturing}: {Bridging} the
  {Gap} between {Design} and {Material} {Extrusion}, Applied Sciences 13~(3)
  (2023) 1467.
\newblock \href {https://doi.org/10.3390/app13031467}
  {\path{doi:10.3390/app13031467}}.

\bibitem{borish_automated_2022}
M.~Borish, B.~T. Gibson, C.~Adkins, P.~Mhatre, Automated {Process} {Planning}
  for {Embossing} and {Functionally} {Grading} {Materials} via
  {Site}-{Specific} {Control} in {Large}-{Format} {Metal}-{Based} {Additive}
  {Manufacturing}, Materials 15~(12) (2022) 4152.
\newblock \href {https://doi.org/10.3390/ma15124152}
  {\path{doi:10.3390/ma15124152}}.

\bibitem{gibson_beyond_2019}
B.~T. Gibson, B.~S. Richardson, T.~W. Sundermann, L.~J. Love, Beyond the
  {Toolpath}: {Site}-{Specific} {Melt} {Pool} {Size} {Control} {Enables}
  {Printing} of {Extra}-{Toolpath} {Geometry} in {Laser} {Wire}-{Based}
  {Directed} {Energy} {Deposition}, Applied Sciences 9~(20) (2019) 4355.
\newblock \href {https://doi.org/10.3390/app9204355}
  {\path{doi:10.3390/app9204355}}.

\bibitem{green_local_2023}
J.~T. Green, I.~A. Rybak, J.~J. Slager, M.~Lopez, Z.~Chanoi, C.~M. Stewart,
  R.~V. Gonzalez, Local composition control using an active-mixing hotend in
  fused filament fabrication, Additive Manufacturing Letters (2023)
  100177Publisher: Elsevier.
\newblock \href {https://doi.org/https://doi.org/10.1016/j.addlet.2023.100177}
  {\path{doi:https://doi.org/10.1016/j.addlet.2023.100177}}.

\bibitem{geng_effects_2019}
P.~Geng, J.~Zhao, W.~Wu, W.~Ye, Y.~Wang, S.~Wang, S.~Zhang, Effects of
  extrusion speed and printing speed on the {3D} printing stability of extruded
  {PEEK} filament, Journal of Manufacturing Processes 37 (2019) 266--273.
\newblock \href {https://doi.org/10.1016/j.jmapro.2018.11.023}
  {\path{doi:10.1016/j.jmapro.2018.11.023}}.

\bibitem{ozdemir_xpandables_2023}
M.~Ozdemir, Z.~Doubrovski, Xpandables: {Single}-filament {Multi}-property {3D}
  {Printing} by {Programmable} {Foaming}, in: Extended {Abstracts} of the 2023
  {CHI} {Conference} on {Human} {Factors} in {Computing} {Systems}, ACM,
  Hamburg Germany, 2023, pp. 1--7.
\newblock \href {https://doi.org/10.1145/3544549.3585731}
  {\path{doi:10.1145/3544549.3585731}}.

\bibitem{lee_influence_2019}
C.-Y. Lee, C.-Y. Liu, The influence of forced-air cooling on a {3D} printed
  {PLA} part manufactured by fused filament fabrication, Additive Manufacturing
  25 (2019) 196--203.
\newblock \href {https://doi.org/10.1016/j.addma.2018.11.012}
  {\path{doi:10.1016/j.addma.2018.11.012}}.

\bibitem{prusa_xl_toolchanger}
All3DP, Toolchanger game changer: Hands-on with the prusa xl, accessed:
  2024-12-11 (2023).

\bibitem{loskot_influence_2023}
J.~Loskot, D.~Jezbera, R.~Loskot, D.~Bušovský, A.~Barylski, K.~Glowka,
  P.~Duda, K.~Aniołek, K.~Voglová, M.~Zubko, Influence of print speed on the
  microstructure, morphology, and mechanical properties of {3D}-printed {PETG}
  products, Polymer Testing 123 (2023) 108055.
\newblock \href {https://doi.org/10.1016/j.polymertesting.2023.108055}
  {\path{doi:10.1016/j.polymertesting.2023.108055}}.

\bibitem{duncan_lowloss_2023}
B.~Duncan, R.~D. Weeks, B.~Barclay, D.~Beck, P.~Bluem, R.~Rojas, M.~Plaut,
  J.~Russo, S.~G.~M. Uzel, J.~A. Lewis, T.~Fedynyshyn, Low‐{Loss} {Graded}
  {Dielectrics} via {Active} {Mixing} of {Nanocomposite} {Inks} during {3D}
  {Printing}, Advanced Materials Technologies 8~(3) (2023) 2201496.
\newblock \href {https://doi.org/10.1002/admt.202201496}
  {\path{doi:10.1002/admt.202201496}}.

\bibitem{lipton_3d_2016}
J.~I. Lipton, H.~Lipson, {3D} {Printing} {Variable} {Stiffness} {Foams} {Using}
  {Viscous} {Thread} {Instability}, Scientific Reports 6~(1) (2016) 29996.
\newblock \href {https://doi.org/10.1038/srep29996}
  {\path{doi:10.1038/srep29996}}.

\bibitem{emery_foams_2024}
B.~Emery, K.~L. Snapp, D.~Revier, V.~Sarkar, M.~Nakura, K.~A. Brown, J.~I.
  Lipton, Foams with {3D} {Spatially} {Programmed} {Mechanics} {Enabled} by
  {Autonomous} {Active} {Learning} on {Viscous} {Thread} {Printing}, Advanced
  Science 11~(44) (2024) 2408062.
\newblock \href {https://doi.org/10.1002/advs.202408062}
  {\path{doi:10.1002/advs.202408062}}.

\bibitem{shaukat_review_2022}
U.~Shaukat, E.~Rossegger, S.~Schlögl, A {Review} of {Multi}-{Material} {3D}
  {Printing} of {Functional} {Materials} via {Vat} {Photopolymerization},
  Polymers 14~(12) (2022) 2449.
\newblock \href {https://doi.org/10.3390/polym14122449}
  {\path{doi:10.3390/polym14122449}}.

\bibitem{shaukat_dual-vat_2024}
U.~Shaukat, A.~Thalhamer, E.~Rossegger, S.~Schlögl, Dual-vat
  photopolymerization {3D} printing of vitrimers, Additive Manufacturing 79
  (2024) 103930.
\newblock \href {https://doi.org/10.1016/j.addma.2023.103930}
  {\path{doi:10.1016/j.addma.2023.103930}}.

\bibitem{yuan_effect_2022}
W.~Yuan, X.~Zhao, S.~Li, Y.~Zhu, Effect of laser scanning speed on
  microstructure and mechanical properties of {SLM} porous
  {Ti}-{5Al}-{5V}-{5Mo}-{3Cr}-{1Fe} alloy, Frontiers in Materials 9 (2022)
  973829.
\newblock \href {https://doi.org/10.3389/fmats.2022.973829}
  {\path{doi:10.3389/fmats.2022.973829}}.

\bibitem{green_hotend_nodate}
J.~T. Green, R.~V. Gonzalez, Hotend for additive manufacturing with an actuated
  rod in a heated chamber.

\bibitem{garland_design_2015}
A.~Garland, G.~Fadel, Design and {Manufacturing} {Functionally} {Gradient}
  {Material} {Objects} {With} an {Off} the {Shelf} {Three}-{Dimensional}
  {Printer}: {Challenges} and {Solutions}, Journal of Mechanical Design
  137~(11) (2015) 111407.
\newblock \href {https://doi.org/10.1115/1.4031097}
  {\path{doi:10.1115/1.4031097}}.

\bibitem{fayolle_modeling_2021}
P.-A. Fayolle, L.~McLoughlin, M.~Sanchez, G.~Pasko, A.~Pasko, Modeling and
  {Visualization} of {Multi}-material {Volumes}, Scientific Visualization
  13~(2) (2021).
\newblock \href {https://doi.org/10.26583/sv.13.2.09}
  {\path{doi:10.26583/sv.13.2.09}}.

\bibitem{khan_review_2024}
I.~Khan, I.~Barsoum, M.~Abas, A.~Al~Rashid, M.~Koç, M.~Tariq, A review of
  extrusion-based additive manufacturing of multi-materials-based polymeric
  laminated structures, Composite Structures 349-350 (2024) 118490.
\newblock \href {https://doi.org/10.1016/j.compstruct.2024.118490}
  {\path{doi:10.1016/j.compstruct.2024.118490}}.

\bibitem{e3d_toolchanger_review}
D.~Sher, Review: E3d motion system and toolchanger multitool and multi-material
  3d printer, accessed: 2024-12-11 (2021).

\bibitem{abilgaziyev_design_2015}
A.~Abilgaziyev, T.~Kulzhan, N.~Raissov, M.~H. Ali, W.~Match, N.~Mir-Nasiri,
  Design and development of multi-nozzle extrusion system for {3D} printer, in:
  2015 {International} {Conference} on {Informatics}, {Electronics} \& {Vision}
  ({ICIEV}), IEEE, Fukuoka, Japan, 2015, pp. 1--5.
\newblock \href {https://doi.org/10.1109/ICIEV.2015.7333982}
  {\path{doi:10.1109/ICIEV.2015.7333982}}.

\bibitem{ali_multi-nozzle_2016}
M.~H. Ali, N.~Mir-Nasiri, W.~L. Ko, Multi-nozzle extrusion system for {3D}
  printer and its control mechanism, The International Journal of Advanced
  Manufacturing Technology 86~(1-4) (2016) 999--1010.
\newblock \href {https://doi.org/10.1007/s00170-015-8205-9}
  {\path{doi:10.1007/s00170-015-8205-9}}.

\bibitem{mosaic_palette_3}
M.~Manufacturing, Palette 3,
  \url{https://www.mosaicmfg.com/products/palette-3}, accessed: 2024-12-11
  (2024).

\bibitem{littler_automated_2022}
E.~Littler, B.~Zhu, W.~Jarosz, Automated {Filament} {Inking} for {Multi}-color
  {FFF} {3D} {Printing}, in: Proceedings of the 35th {Annual} {ACM} {Symposium}
  on {User} {Interface} {Software} and {Technology}, ACM, Bend OR USA, 2022,
  pp. 1--13.
\newblock \href {https://doi.org/10.1145/3526113.3545654}
  {\path{doi:10.1145/3526113.3545654}}.

\bibitem{davinci}
D.~Schaffhauser, New da vinci 3d printer creates full-color objects, accessed:
  2024-12-11 (Aug. 2017).

\bibitem{reiner_dualcolor_2014}
T.~Reiner, N.~Carr, R.~Měch, O.~Št'ava, C.~Dachsbacher, G.~Miller,
  Dual‐color mixing for fused deposition modeling printers, Computer Graphics
  Forum 33~(2) (2014) 479--486.
\newblock \href {https://doi.org/10.1111/cgf.12319}
  {\path{doi:10.1111/cgf.12319}}.

\bibitem{kuipers_3d_2017}
T.~Kuipers, E.~Doubrovski, J.~Verlinden, {3D} hatching: linear halftoning for
  dual extrusion fused deposition modeling, in: Proceedings of the 1st {Annual}
  {ACM} {Symposium} on {Computational} {Fabrication}, ACM, Cambridge
  Massachusetts, 2017, pp. 1--7.
\newblock \href {https://doi.org/10.1145/3083157.3083163}
  {\path{doi:10.1145/3083157.3083163}}.

\bibitem{kuipers_hatching_2018}
T.~Kuipers, W.~Elkhuizen, J.~Verlinden, E.~Doubrovski, Hatching for {3D}
  prints: {Line}-based halftoning for dual extrusion fused deposition modeling,
  Computers \& Graphics 74 (2018) 23--32.
\newblock \href {https://doi.org/10.1016/j.cag.2018.04.006}
  {\path{doi:10.1016/j.cag.2018.04.006}}.

\bibitem{song_colored_2019}
H.~Song, J.~Martínez, P.~Bedell, N.~Vennin, S.~Lefebvre, Colored {Fused}
  {Filament} {Fabrication}, ACM Transactions on Graphics 38~(5) (2019) 1--11.
\newblock \href {https://doi.org/10.1145/3183793} {\path{doi:10.1145/3183793}}.

\bibitem{takahashi_programmable_2020}
H.~Takahashi, P.~Punpongsanon, J.~Kim, Programmable {Filament}: {Printed}
  {Filaments} for {Multi}-material {3D} {Printing}, in: Proceedings of the 33rd
  {Annual} {ACM} {Symposium} on {User} {Interface} {Software} and {Technology},
  ACM, Virtual Event USA, 2020, pp. 1209--1221.
\newblock \href {https://doi.org/10.1145/3379337.3415863}
  {\path{doi:10.1145/3379337.3415863}}.

\bibitem{ahn_3d_2024}
S.-J. Ahn, H.~Lee, K.-J. Cho, {3D} printing with a {3D} printed digital
  material filament for programming functional gradients, Nature Communications
  15~(1) (2024) 3605.
\newblock \href {https://doi.org/10.1038/s41467-024-47480-5}
  {\path{doi:10.1038/s41467-024-47480-5}}.

\bibitem{tiwary_overview_2021}
V.~K. Tiwary, A.~P., V.~R. Malik, An overview on joining/welding as
  post-processing technique to circumvent the build volume limitation of an
  {FDM}-{3D} printer, Rapid Prototyping Journal 27~(4) (2021) 808--821.
\newblock \href {https://doi.org/10.1108/RPJ-10-2020-0265}
  {\path{doi:10.1108/RPJ-10-2020-0265}}.

\bibitem{dairabayeva_investigation_2023}
D.~Dairabayeva, A.~Perveen, D.~Talamona, Investigation on the mechanical
  performance of mono-material vs multi-material interface geometries using
  fused filament fabrication, Rapid Prototyping Journal 29~(11) (2023) 40--52.
\newblock \href {https://doi.org/10.1108/RPJ-07-2022-0221}
  {\path{doi:10.1108/RPJ-07-2022-0221}}.

\bibitem{sardinha2024multi_extrusion}
M.~Sardinha, L.~Reis, J.~J. Netto, N.~Frutuoso, M.~Leite, Insights on
  multi-extrusion filament fabrication: Exploring the impact of seam overlap
  and ironing process, in: Proceedings of the 35th Annual International Solid
  Freeform Fabrication Symposium – An Additive Manufacturing Conference,
  IDMEC-IST, Universidade de Lisboa, Lisboa, Portugal, 2024, pp. 1143--1152.

\bibitem{han_design_2017}
S.~Han, Y.~Xiao, T.~Qi, Z.~Li, Q.~Zeng, Design and {Analysis} of {Fused}
  {Deposition} {Modeling} {3D} {Printer} {Nozzle} for {Color} {Mixing},
  Advances in Materials Science and Engineering 2017 (2017) 1--12.
\newblock \href {https://doi.org/10.1155/2017/2095137}
  {\path{doi:10.1155/2017/2095137}}.

\bibitem{kennedy_printing_2020}
Z.~C. Kennedy, J.~F. Christ, Printing polymer blends through in situ active
  mixing during fused filament fabrication, Additive Manufacturing 36 (2020)
  101233.
\newblock \href {https://doi.org/10.1016/j.addma.2020.101233}
  {\path{doi:10.1016/j.addma.2020.101233}}.

\bibitem{adapa_design_2023}
S.~K. Adapa, {Jagadish}, Design and fabrication of internal mixer and filament
  extruder for extraction of hybrid filament composite for {FDM} applications,
  International Journal on Interactive Design and Manufacturing (IJIDeM) (Oct.
  2023).
\newblock \href {https://doi.org/10.1007/s12008-023-01521-3}
  {\path{doi:10.1007/s12008-023-01521-3}}.

\bibitem{teng_single-nozzle_2024}
T.~Teng, Y.~Zhi, M.~Akbarzadeh, Single-{Nozzle} {Multi}-{Filament} {System}
  with {Active} {Mixing} for {High}-{Fidelity} {Multimaterial} {Additive}
  {Manufacturing} (2024).
\newblock \href {https://doi.org/10.2139/ssrn.4773564}
  {\path{doi:10.2139/ssrn.4773564}}.

\bibitem{ober_active_2015}
T.~J. Ober, D.~Foresti, J.~A. Lewis, Active mixing of complex fluids at the
  microscale, Proceedings of the National Academy of Sciences 112~(40) (2015)
  12293--12298.
\newblock \href {https://doi.org/10.1073/pnas.1509224112}
  {\path{doi:10.1073/pnas.1509224112}}.

\bibitem{serex_microfluidics_2018}
L.~Serex, A.~Bertsch, P.~Renaud, Microfluidics: {A} {New} {Layer} of {Control}
  for {Extrusion}-{Based} {3D} {Printing}, Micromachines 9~(2) (2018) 86.
\newblock \href {https://doi.org/10.3390/mi9020086}
  {\path{doi:10.3390/mi9020086}}.

\bibitem{golobic_active_2019}
A.~M. Golobic, M.~D. Durban, S.~E. Fisher, M.~D. Grapes, J.~M. Ortega, C.~M.
  Spadaccini, E.~B. Duoss, A.~E. Gash, K.~T. Sullivan, Active {Mixing} of
  {Reactive} {Materials} for {3D} {Printing}, Advanced Engineering Materials
  21~(8) (2019) 1900147.
\newblock \href {https://doi.org/10.1002/adem.201900147}
  {\path{doi:10.1002/adem.201900147}}.

\bibitem{ortega_active_2019}
J.~M. Ortega, M.~Golobic, J.~D. Sain, J.~M. Lenhardt, A.~S. Wu, S.~E. Fisher,
  L.~X. Perez~Perez, A.~W. Jaycox, J.~E. Smay, E.~B. Duoss, T.~S. Wilson,
  Active {Mixing} of {Disparate} {Inks} for {Multimaterial} {3D} {Printing},
  Advanced Materials Technologies 4~(7) (2019) 1800717.
\newblock \href {https://doi.org/10.1002/admt.201800717}
  {\path{doi:10.1002/admt.201800717}}.

\bibitem{liao_active_2024}
E.~Liao, Y.~Zhang, Y.~Su, C.~Zhang, C.~Geng, C.~Li, X.~Liu, Y.~Liu, A.~Lu,
  Active mixing of two-component viscoelastic silicone ink at molecular level
  for spatiotemporally controlled {3D}/{4D} printing of cellular silicones,
  Polymer 302 (2024) 127053.
\newblock \href {https://doi.org/10.1016/j.polymer.2024.127053}
  {\path{doi:10.1016/j.polymer.2024.127053}}.

\bibitem{pelz_multi-material_2021}
J.~S. Pelz, N.~Ku, W.~T. Shoulders, M.~A. Meyers, L.~R. Vargas-Gonzalez,
  Multi-material additive manufacturing of functionally graded carbide ceramics
  via active, in-line mixing, Additive Manufacturing 37 (2021) 101647.
\newblock \href {https://doi.org/10.1016/j.addma.2020.101647}
  {\path{doi:10.1016/j.addma.2020.101647}}.

\bibitem{li_fabricating_2018}
W.~Li, A.~J. Martin, B.~Kroehler, A.~Henderson, T.~Huang, J.~Watts, G.~E.
  Hilmas, M.~C. Leu, Fabricating {Functionally} {Graded} {Materials} by
  {Ceramic} {On}-{Demand} {Extrusion} with {Dynamic} {Mixing} (2018).
\newblock \href {https://doi.org/10.26153/TSW/17109}
  {\path{doi:10.26153/TSW/17109}}.

\bibitem{damanpack_porous_2021}
A.~R. Damanpack, A.~Sousa, M.~Bodaghi, Porous {PLAs} with {Controllable}
  {Density} by {FDM} {3D} {Printing} and {Chemical} {Foaming} {Agent},
  Micromachines 12~(8) (2021) 866.
\newblock \href {https://doi.org/10.3390/mi12080866}
  {\path{doi:10.3390/mi12080866}}.

\bibitem{laureijs_investigating_nodate}
N.~Laureijs, Investigating {Foaming} {Materials} in {3D} {Printing}: {An}
  {Application} for {Motorcycle} {Helmets}.

\bibitem{tammaro_microfoamed_2022}
D.~Tammaro, M.~M. Villone, P.~L. Maffettone, Microfoamed {Strands} by {3D}
  {Foam} {Printing}, Polymers 14~(15) (2022) 3214.
\newblock \href {https://doi.org/10.3390/polym14153214}
  {\path{doi:10.3390/polym14153214}}.

\bibitem{lalegani_dezaki_soft_2023}
M.~Lalegani~Dezaki, M.~Bodaghi, A.~Serjouei, S.~Afazov, A.~Zolfagharian, Soft
  {Pneumatic} {Actuators} with {Controllable} {Stiffness} by {Bio}‐{Inspired}
  {Lattice} {Chambers} and {Fused} {Deposition} {Modeling} {3D} {Printing},
  Advanced Engineering Materials 25~(6) (2023) 2200797.
\newblock \href {https://doi.org/10.1002/adem.202200797}
  {\path{doi:10.1002/adem.202200797}}.

\bibitem{loh_overview_2018}
G.~H. Loh, E.~Pei, D.~Harrison, M.~D. Monzón, An overview of functionally
  graded additive manufacturing, Additive Manufacturing 23 (2018) 34--44.
\newblock \href {https://doi.org/10.1016/j.addma.2018.06.023}
  {\path{doi:10.1016/j.addma.2018.06.023}}.

\bibitem{steuben_implicit_2016}
J.~C. Steuben, A.~P. Iliopoulos, J.~G. Michopoulos, Implicit slicing for
  functionally tailored additive manufacturing, Computer-Aided Design 77 (2016)
  107--119.
\newblock \href {https://doi.org/10.1016/j.cad.2016.04.003}
  {\path{doi:10.1016/j.cad.2016.04.003}}.

\bibitem{xia_stress-based_2020}
L.~Xia, S.~Lin, G.~Ma, Stress-based tool-path planning methodology for fused
  filament fabrication, Additive Manufacturing 32 (2020) 101020.
\newblock \href {https://doi.org/10.1016/j.addma.2019.101020}
  {\path{doi:10.1016/j.addma.2019.101020}}.

\bibitem{xiong_process_2019}
Y.~Xiong, S.-I. Park, S.~Padmanathan, A.~G. Dharmawan, S.~Foong, D.~W. Rosen,
  G.~S. Soh, Process planning for adaptive contour parallel toolpath in
  additive manufacturing with variable bead width, The International Journal of
  Advanced Manufacturing Technology 105~(10) (2019) 4159--4170.
\newblock \href {https://doi.org/10.1007/s00170-019-03954-1}
  {\path{doi:10.1007/s00170-019-03954-1}}.

\bibitem{chen_field-based_2022}
X.~Chen, G.~Fang, W.-H. Liao, C.~C. Wang, Field-{Based} {Toolpath} {Generation}
  for {3D} {Printing} {Continuous} {Fibre} {Reinforced} {Thermoplastic}
  {Composites}, Additive Manufacturing 49 (2022) 102470.
\newblock \href {https://doi.org/10.1016/j.addma.2021.102470}
  {\path{doi:10.1016/j.addma.2021.102470}}.

\bibitem{liu_stress_2024}
G.~Liu, W.~Huang, Y.~Wang, H.~Ren, G.~Zhang, L.~Zhou, Y.~Xiong, Stress
  field-aware infill toolpath generation for additive manufacturing of
  continuous fiber reinforced polymer composites, Materials \& Design 239
  (2024) 112756.
\newblock \href {https://doi.org/10.1016/j.matdes.2024.112756}
  {\path{doi:10.1016/j.matdes.2024.112756}}.

\bibitem{sales_function-aware_2021}
E.~Sales, T.-H. Kwok, Y.~Chen, Function-aware slicing using principal stress
  line for toolpath planning in additive manufacturing, Journal of
  Manufacturing Processes 64 (2021) 1420--1433.
\newblock \href {https://doi.org/10.1016/j.jmapro.2021.02.050}
  {\path{doi:10.1016/j.jmapro.2021.02.050}}.

\bibitem{roschli_motion_2023}
A.~C. Roschli, Motion and {Path} {Planning} for {Additive} {Manufacturing}, 1st
  Edition, Additive {Manufacturing} {Materials} and {Technologies} {Series},
  Elsevier, San Diego, 2023.

\bibitem{maple_geometric_2003}
C.~Maple, Geometric design and space planning using the marching squares and
  marching cube algorithms, in: 2003 {International} {Conference} on
  {Geometric} {Modeling} and {Graphics}, 2003. {Proceedings}, IEEE Comput. Soc,
  London, UK, 2003, pp. 90--95.
\newblock \href {https://doi.org/10.1109/GMAG.2003.1219671}
  {\path{doi:10.1109/GMAG.2003.1219671}}.

\bibitem{ju_dual_2002}
T.~Ju, F.~Losasso, S.~Schaefer, J.~Warren, Dual contouring of hermite data, ACM
  Transactions on Graphics 21~(3) (2002) 339--346.
\newblock \href {https://doi.org/10.1145/566654.566586}
  {\path{doi:10.1145/566654.566586}}.

\bibitem{lorensen_marching_1987}
W.~E. Lorensen, H.~E. Cline, Marching cubes: {A} high resolution {3D} surface
  construction algorithm, ACM SIGGRAPH Computer Graphics 21~(4) (1987)
  163--169.
\newblock \href {https://doi.org/10.1145/37402.37422}
  {\path{doi:10.1145/37402.37422}}.

\bibitem{wang_universal_2024}
S.-L. Wang, L.-C. Zhang, C.~Cai, M.-K. Tang, S.~Chen, J.~Huang, Y.-S. Shi,
  Universal and efficient hybrid modeling and direct slicing method for
  additive manufacturing processes, Advances in Manufacturing 12~(2) (2024)
  300--316.
\newblock \href {https://doi.org/10.1007/s40436-023-00468-8}
  {\path{doi:10.1007/s40436-023-00468-8}}.

\bibitem{cgal_arrangement_on_surface_2}
Cgal: Arrangement\_on\_surface\_2 reference,
  \url{https://doc.cgal.org/latest/Arrangement_on_surface_2/index.html},
  accessed: 2024-12-11 (2024).

\bibitem{agarwal_sharir_2000}
P.~K. Agarwal, M.~Sharir, Arrangements and their applications, in: J.-R. Sack,
  J.~Urrutia (Eds.), Handbook of Computational Geometry, Elsevier Science
  Publishers B.V. North-Holland, Amsterdam, 2000.

\bibitem{sardinha_influence_2020}
M.~Sardinha, N.~Frutuoso, C.~M. Vicente, R.~Ribeiro, M.~Leite, L.~Reis,
  Influence of seams in the mechanical properties of {PLA} produced with
  multiple extrusion modules, Procedia Structural Integrity 28 (2020) 358--363.
\newblock \href {https://doi.org/10.1016/j.prostr.2020.10.042}
  {\path{doi:10.1016/j.prostr.2020.10.042}}.

\bibitem{taiwo_investigating_2024}
E.~T. Taiwo, P.~O‘Dowd, B.~Hicks, Investigating the influence of
  inter-printhead bond strategy on tensile strength in multi-printhead {PLA}
  parts for collective additive manufacturing, Procedia CIRP 130 (2024)
  648--655.
\newblock \href {https://doi.org/10.1016/j.procir.2024.10.143}
  {\path{doi:10.1016/j.procir.2024.10.143}}.

\bibitem{minaker_optimizing_2021}
S.~A. Minaker, R.~H. Mason, D.~R. Chow, Optimizing {Color} {Performance} of the
  {Ngenuity} 3-{Dimensional} {Visualization} {System}, Ophthalmology Science
  1~(3) (2021) 100054.
\newblock \href {https://doi.org/10.1016/j.xops.2021.100054}
  {\path{doi:10.1016/j.xops.2021.100054}}.

\bibitem{charia_real-time_2025}
O.~Charia, H.~Rajani, I.~Ferrer~Real, M.~Domingo-Espin, N.~Gracias, Real-{Time}
  {Stringing} {Detection} for {Additive} {Manufacturing}, Journal of
  Manufacturing and Materials Processing 9~(3) (2025) 74.
\newblock \href {https://doi.org/10.3390/jmmp9030074}
  {\path{doi:10.3390/jmmp9030074}}.

\bibitem{jackson_multi-parameter_2022}
B.~Jackson, K.~Fouladi, B.~Eslami, Multi-{Parameter} {Optimization} of {3D}
  {Printing} {Condition} for {Enhanced} {Quality} and {Strength}, Polymers
  14~(8) (2022) 1586.
\newblock \href {https://doi.org/10.3390/polym14081586}
  {\path{doi:10.3390/polym14081586}}.

\bibitem{3DBenchy2015}
C.~Tools, {3DBenchy}: The jolly 3d printing torture-test,
  \url{https://www.3dbenchy.com}, 3D printer calibration model by Daniel
  Nor{\'e}e; accessed April 19, 2025 (2015).

\bibitem{sevcik_dual_2024}
M.~J. Sevcik, J.~Golson, G.~Bjerke, I.~Snyder, G.~Taylor, F.~Wilson, G.~I.
  Rabinowitz, D.~J. Kline, M.~D. Grapes, K.~T. Sullivan, J.~L. Belof,
  V.~Eliasson, Dual feed progressive cavity pump extrusion system for
  functionally graded direct ink write {3D} printing, HardwareX 17 (2024)
  e00515.
\newblock \href {https://doi.org/10.1016/j.ohx.2024.e00515}
  {\path{doi:10.1016/j.ohx.2024.e00515}}.

\bibitem{geeetech_mizar_m}
{Geeetech}, Geeetech mizar m dual extruder multi color 3d printer, archived on
  July 6, 2024 (n.d.).

\bibitem{colorfabb_lwpla_black}
{ColorFabb}, Lw-pla black, \url{https://colorfabb.us/lw-pla-black}, accessed:
  2025-04-20 (n.d.).

\bibitem{colorfabb_varioshore_green}
{ColorFabb}, Varioshore tpu green,
  \url{https://colorfabb.us/varioshore-tpu-green}, accessed: 2025-04-20 (n.d.).

\bibitem{prusa_mk4s_enclosure_bundle}
{Prusa Research}, Original prusa mk4s enclosure bundle,
  \url{https://www.prusa3d.com/en/product/original-prusa-mk4s-enclosure-bundle-4/},
  accessed: 2025-04-20 (n.d.).

\bibitem{marlin_gcode_syntax}
{Marlin Firmware Project}, \href{https://marlinfw.org/meta/gcode/}{Marlin
  Firmware G-code Documentation} (2025).
\newline\urlprefix\url{https://marlinfw.org/meta/gcode/}

\bibitem{percoco_analytical_2021}
G.~Percoco, L.~Arleo, G.~Stano, F.~Bottiglione, Analytical model to predict the
  extrusion force as a function of the layer height, in extrusion based {3D}
  printing, Additive Manufacturing 38 (2021) 101791.
\newblock \href {https://doi.org/10.1016/j.addma.2020.101791}
  {\path{doi:10.1016/j.addma.2020.101791}}.

\end{thebibliography}

\end{document}